\documentclass[10pt,journal,compsoc]{IEEEtran}
\usepackage{pifont}
\usepackage[T1]{fontenc}

\ifCLASSOPTIONcompsoc
  \usepackage[nocompress]{cite}
\else
  \usepackage{cite}
\fi
\ifCLASSINFOpdf
\else
\fi

\usepackage[ruled,linesnumbered]{algorithm2e}

\SetKwInOut{Input}{Input}\SetKwInOut{Output}{Output}

\usepackage{xcolor}         
\usepackage{subfigure}
\usepackage{graphicx}
\usepackage{amsmath}
\usepackage{amsthm}
\usepackage{amsfonts}
\usepackage{hyperref}
\hypersetup{
colorlinks=true,
linkcolor=blue,
anchorcolor=blue,
citecolor=blue}
\usepackage{cleveref}
\usepackage{arydshln}

\usepackage{amssymb}
\usepackage{algorithm2e}
\usepackage{algorithmic}
\usepackage{booktabs}
\usepackage{multirow}
\usepackage{makecell}
\usepackage{colortbl}
\usepackage[normalem]{ulem}
\useunder{\uline}{\ul}{}
\definecolor{revise}{RGB}{0, 0, 0}
\definecolor{R2}{RGB}{0, 0, 0}
\definecolor{R3}{RGB}{0, 0, 0}
\definecolor{ada_blue}{rgb}{0,205,205}
\definecolor{glt_red}{rgb}{109,205,255}
\definecolor{myblue}{HTML}{00CDCD}
\definecolor{autopurple}{HTML}{7030A0}
\definecolor{dyna_yellow}{HTML}{BF9000}
\definecolor{adaptive_blue}{HTML}{0070C0}
\usepackage{pifont}

\definecolor{ForestGreen}{RGB}{34,139,34}
\definecolor{myyellow}{RGB}{181, 181, 27}

\usepackage{tikz}
\usepackage{amssymb}
\definecolor{harvestgold}{rgb}{0.85, 0.57, 0.0}
\definecolor{DarkGreen}{RGB}{30,130,30}
\newcommand\encircle[2][]{\tikz[overlay]\node[fill=blue!20,inner sep=2pt, anchor=text, rectangle, rounded corners=1.5mm,#1] {#2};\phantom{#2}}
\usepackage{enumitem}
\usepackage{pifont}
\usepackage{bbm}
\usepackage{tabularx}
\usepackage{cleveref}
\usepackage{tabularx}
\usepackage{booktabs}
\usepackage{booktabs}
\usepackage{longtable}
\usepackage{adjustbox}
\usepackage{multirow}
\usepackage{multicol}
\usepackage{threeparttable}
\usepackage{adjustbox}
\usepackage[justification=centering]{caption}

\usepackage{authblk}

\makeatletter
\renewcommand\AB@affilsepx{, \protect\Affilfont} %
\makeatother
\usepackage{supertabular}
\usepackage{tcolorbox}
\usepackage[numbers]{natbib}

\newcommand{\insightbox}[1]{%
    \begin{tcolorbox}[colframe=black!70, colback=yellow!5, boxrule=1pt, arc=4mm]
        \includegraphics[width=0.5cm]{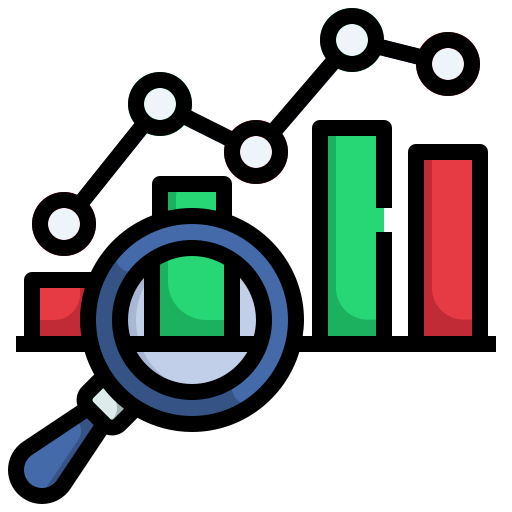}
        \textbf{\small#1}
    \end{tcolorbox}
}

\definecolor{shiqian-red}{RGB}{220, 239, 252}
\definecolor{shiqian-red-dark}{RGB}{114, 154, 202}
\definecolor{shiqian-red-light}{RGB}{235, 244, 255}
\newcommand*{\belowrulesepcolor}[1]{%
  \noalign{%
    \kern-\belowrulesep 
    \begingroup 
      \color{#1}%
      \hrule height\belowrulesep 
    \endgroup 
    \vspace{-0.03mm}
  }%
} 
\newcommand*{\aboverulesepcolor}[1]{%
  \noalign{%
  \vspace{-0.03mm}
    \begingroup 
      \color{#1}%
    \endgroup 
    \kern-\aboverulesep 
  }%
}

\newcommand{\wenjie}[1]{\textbf{\textcolor{blue}{[wenjie: #1]}}}

\begin{document}
%
\title{A Comprehensive Survey in LLM(-Agent) Full Stack Safety: Data, Training and Deployment}

\author[1,2]{Kun Wang*\thanks{Kun Wang is with Nanyang Technological University ({\tt wang.kun@ntu.edu.sg}), Guibin Zhang is with National University of Singapore ({\tt guibinz@outlook.com}), Jiahao Wu is with The Hong Kong Polytechnic University ({\tt jiahao.wu@connect.polyu.hk}), Zhenhong Zhou is with A*STAR ({\tt ydyjyazzh@gmail.com}), Yang Liu is with Nanyang Technological University ({\tt yangliu@ntu.edu.sg}). * denotes equal contribution and $\dagger$ denotes the corresponding authors.}}
\author[3]{Guibin Zhang*}
\author[4]{Zhenhong Zhou$\dagger$}
\author[5,6]{Jiahao Wu$\dagger$}
\author[7]{Miao Yu}
\author[1]{Shiqian Zhao}
\author[8]{Chenlong Yin}
\author[9]{Jinhu Fu}
\author[10,11]{Yibo Yan}
\author[12]{Hanjun Luo}
% ...

\begin{tiny}
\affil[1]{Nanyang Technological University}
\affil[2]{Squirrel AI Learning}
\affil[3]{National University of Singapore}
\affil[4]{A*STAR}
\affil[5]{The Hong Kong Polytechnic University}
\affil[6]{Southern University of Science and Technology}
\affil[7]{University of Science and Technology of China}
\affil[8]{The Pennsylvania State University}
\affil[9]{TeleAI}
\affil[10]{Hong Kong University of Science and Technology (Guangzhou)}
\affil[11]{Hong Kong University of Science and Technology}
\affil[12]{Zhejiang University}
% ...
\end{tiny}

\markboth{Journal of \LaTeX\ Class Files,~Vol.~14, No.~8, August~2015}%
{Shell \MakeLowercase{\textit{et al.}}: Bare Demo of IEEEtran.cls for Computer Society Journals}
%



\vspace{-0.7em}
\IEEEtitleabstractindextext{%
\begin{abstract}

The remarkable success of Large Language Models (LLMs) has illuminated a promising pathway toward achieving Artificial General Intelligence for both academic and industrial communities, owing to their unprecedented performance across various applications. As LLMs continue to gain prominence in both research and commercial domains, their security and safety implications have become a growing concern, not only for researchers and corporations but also for all nations. Currently, existing surveys on LLM safety primarily focus on specific stages of the LLM lifecycle, \textit{e.g.,} deployment phase or fine-tuning phase, lacking a comprehensive understanding of the entire ``lifechain'' of LLMs. To address this gap, this paper introduces, for the first time, the concept of ``full-stack'' safety to systematically consider safety issues throughout the entire process of data, training (pre-training, post-training), deployment (deployment and final commercialization). Compared to the off-the-shelf LLM safety surveys, our work demonstrates several distinctive advantages: \textbf{\textit{(I)}} \textbf{Comprehensive Perspective}. We define the complete LLM lifecycle as encompassing data preparation, pre-training, post-training (including alignment and fine-tuning, model editing, \textit{etc.}), deployment and final commercialization. To our knowledge, this represents the first safety survey to encompass the entire lifecycle of LLMs. \textbf{\textit{(II)}} \textbf{Extensive Literature Support.} Our research is grounded in an exhaustive review of over 900+ papers, ensuring comprehensive coverage and systematic organization of safety issues within a more holistic understanding. \textbf{\textit{(III)}} \textbf{Unique Insights.} Through systematic literature analysis, we develop reliable roadmaps and perspectives for each chapter. Our work identifies promising research directions, including safety in data generation, alignment techniques, model editing, and LLM-based agent systems. These insights provide valuable guidance for researchers pursuing future work in this field. We provide an up-to-date review of the literature on LLM (agent) safety at \url{https://github.com/bingreeky/full-stack-llm-safety}, which can be considered a useful support for both researchers and engineers.

\end{abstract}

\begin{IEEEkeywords}
\vspace{-0.7em}
Large Language Model, LLM-based Agent, Safety, Post-training, Alignment, Model Editing, Unlearning, Evaluation
\end{IEEEkeywords}}
\vspace{-5pt}

\maketitle

\IEEEdisplaynontitleabstractindextext

%
\IEEEpeerreviewmaketitle

\section{Introduction}
\label{sec:introduction}

The emergence and success of large language models (LLMs) \cite{ouyang2022training, touvron2023llama, bai2023qwen, liu2024deepseek, guo2025deepseek} have greatly transformed the modes of production in both academia and industry \cite{zhao2023survey, chang2024survey, hadi2023survey, yan2025position, yan2024survey, zou2025deep,li2025g,sun2024large}, opening a potential path for the upcoming artificial general intelligence \cite{sonko2024critical, mclean2023risks,liu2025can}. Going beyond this, LLMs, by integrating tools \cite{ruan2023tptu, sorin2023large, yang2023gpt4tools, schick2023toolformer}, memory \cite{zhong2024memorybank, wang2023augmenting, zhang2024survey, huo2024mmneuron}, APIs \cite{liu2024toolace, tang2023toolalpaca}, and by constructing single-agent or multi-agent systems with other LLMs, provide powerful tools for large models to perceive, understand, and change the environment \cite{guo2024large, wang2024survey, xi2025rise, yan2024georeasoner}. This has garnered considerable attention for embodied intelligence \cite{majumdar2023we, zhou2024embodied}.

Unfortunately, the entire lifecycle of LLMs is constantly confronted with security and safety issues \cite{ma2025safety, kumar2025llm, li2025system, 2025-CodeLM-Security,qu2025prompt}. During the data preparation phase, since LLMs require ample and diverse data, and a significant amount of data is sourced from the Internet and other open-source scenarios, the toxicity in the data and user privacy may seep into the model parameters, triggering crises in the model \cite{wu2025survey, wang2023pre, zhou2024comprehensive}. The pretraining process of the model, due to its unsupervised nature, unconsciously absorbs these toxic data and privacy information, thereby causing the model's ``genetic makeup'' to carry dangerous characteristics and privacy issues \cite{zhang2020online, goldblum2022dataset, lukas2023analyzing, 2025-EliBadCode}.

Before the model is deployed, if it is not properly aligned with security measures, it can easily deviate from human values \cite{kirk2024benefits, zhou2024alignment}. Meanwhile, to make the model more "specialized," the fine-tuning process will employ safer and more customized data to ensure the model performs flawlessly in specific domains \cite{xiangyuqi2024iclr,qi2025safety,pmlr-v235-halawi24a,hawkins2024the}. The model deployment process also involves issues such as jailbreak attacks and corresponding defense measures \cite{huang2024survey,rottger2024safetyprompts,dong2024safeguarding}, especially for LLM-based agents \cite{wang2024large}. These agents may become contaminated due to their interaction with tools, memory, and the environment \cite{zhang2025evoflow, zhang2025maas, zhang2024agentprune, yue2025masrouter}.

\newcommand{\checkicon}{\textcolor{green!70!black}{\ding{51}}}
\newcommand{\crossicon}{\textcolor{red!90!black}{\ding{55}}}

\begin{table}[t]
    \centering
    \caption{Survey Comparison on LLMs and Agents settings.}
    \label{tab:diffsurvey}
\begin{adjustbox}{width=1\linewidth}
\begin{tabular}{c cc c cccccc}
    \toprule
    \multirow{2}{*}{\textbf{Survey}} & \multicolumn{2}{c}{\textbf{Object}} && \multicolumn{6}{c}{\textbf{Stage}$^{\star}$} \\
    \cmidrule{2-3} \cmidrule{5-10}
     & \textbf{LLM}$^\ddagger$ & \textbf{Agent}$^\$$ && \textbf{Data} & \textbf{PT} & \textbf{Edit} & \textbf{FT} & \textbf{Dep} & \textbf{Eval}\\
    \midrule

    \multicolumn{10}{l}{\textbf{\textit{Year 2023}}} \\

    \rowcolor{gray!20}Zhao et al. \cite{zhao2023survey}  & S+M & - && \checkicon & \checkicon & \crossicon & \checkicon & \checkicon & \checkicon\\
    
    Liang et al. \cite{liang2024survey}  & M & - && \checkicon & \checkicon & \checkicon & \checkicon & \crossicon & \crossicon\\
    
    \rowcolor{gray!20}Chang et al. \cite{chang2024survey}  & S+M & - && \crossicon & \crossicon & \crossicon & \crossicon & \checkicon & \checkicon\\
    
    Zhang et al. \cite{zhang2023instruction}  & S+M & - && \checkicon & \crossicon & \crossicon & \checkicon & \crossicon & \checkicon\\
    
    \rowcolor{gray!20}Wang et al. \cite{wang2024survey}  & - & S && \crossicon & \crossicon & \crossicon & \crossicon & \checkicon & \checkicon\\
    
    Zhao et al. \cite{zhao2024explainability}  & S & - && \crossicon & \crossicon & \checkicon & \checkicon & \crossicon & \checkicon\\
    
    \rowcolor{gray!20}Xi et al. \cite{xi2025rise}  & - & S+MAS && \crossicon & \crossicon & \crossicon & \crossicon & \checkicon & \checkicon\\
    
    Shen et al. \cite{shen2023large}  & S & - && \crossicon & \crossicon & \checkicon & \checkicon & \crossicon & \checkicon\\
    
    \rowcolor{gray!20}Raiaan et al. \cite{raiaan2024review}  & S & - && \checkicon & \checkicon & \crossicon & \crossicon & \crossicon & \crossicon\\
    
    Kalyan et al. \cite{kalyan2024survey}  & S+M & - && \crossicon & \checkicon & \crossicon & \checkicon & \checkicon & \checkicon\\

    \rowcolor{gray!20}Huang et al. \cite{huang2024survey}  & S & - && \crossicon & \crossicon & \crossicon & \checkicon & \checkicon & \checkicon\\

    Shayegani et al. \cite{shayegani2023survey}  & S+M & MAS && \crossicon & \crossicon & \crossicon & \checkicon & \checkicon & \crossicon\\
    
    \rowcolor{gray!20}Yao et al. \cite{yao2024survey}  & S & - && \crossicon & \crossicon & \crossicon & \checkicon & \checkicon & \crossicon \\

    \midrule

    \multicolumn{10}{l}{\textbf{\textit{Year 2024}}} \\
    
    \rowcolor{gray!20}Guo et al. \cite{guo2024large}  & - & S+MAS && \crossicon & \crossicon & \crossicon & \crossicon & \checkicon & \checkicon\\
    
    Qin et al. \cite{qin2024multilingual}  & S+M & - && \checkicon & \crossicon & \checkicon & \checkicon & \checkicon & \crossicon\\
    
    \rowcolor{gray!20}Hadi et al. \cite{hadi2023large}  & S & - && \crossicon & \checkicon & \crossicon & \checkicon & \checkicon & \crossicon\\
    
    Sun et al. \cite{sun2024trustllm}  & S+M & S && \crossicon & \crossicon & \crossicon & \checkicon & \checkicon & \checkicon\\
    
    \rowcolor{gray!20}Das et al. \cite{das2025security}  & S & - && \crossicon & \crossicon & \crossicon & \checkicon & \checkicon & \crossicon\\
    
    He et al. \cite{he2024emerged}  & - & {\makecell{S+M+\\MAS}} && \crossicon & \crossicon & \crossicon & \crossicon & \checkicon & \crossicon\\
    
    \rowcolor{gray!20}Wang et al. \cite{wang2024large}  & - & S+MAS && \crossicon & \crossicon & \crossicon & \crossicon & \checkicon & \crossicon\\

    \midrule

    \multicolumn{10}{l}{\textbf{\textit{Year 2025}}} \\

    \rowcolor{gray!20}Tie et al. \cite{tie2025survey}  & S+M & - && \checkicon & \crossicon & \crossicon & \checkicon & \checkicon & \crossicon\\
    
    Ma et al. \cite{ma2025safety}  & S+M & S+M && \crossicon & \crossicon & \checkicon & \checkicon & \checkicon & \checkicon\\
    
    \rowcolor{gray!20}Huang et al. \cite{huang2025trustworthiness}  & S+M & S+M && \crossicon & \crossicon & \checkicon & \checkicon & \checkicon & \checkicon\\
    
    Yu et al. \cite{yu2025survey} & S & S+MAS && \crossicon & \crossicon & \crossicon & \crossicon & \checkicon & \checkicon\\

    Chen et al. \cite{2025-CodeLM-Security} & S & - && \checkicon & \checkicon & \crossicon & \checkicon & \crossicon & \checkicon\\
    
    \midrule
    
    \textbf{Ours}  & S+M & \makecell{S+M+\\MAS} && \checkicon & \checkicon & \checkicon & \checkicon & \checkicon & \checkicon\\
    \bottomrule
\end{tabular}
\end{adjustbox}
\begin{tablenotes}[flushleft]
    \item[] \hspace{-2pt}\scriptsize 
    $\ddagger$: Single-modal LLM (S), Multi-modal LLM (M).\\
    $\$$: Single-modal Agent (S), Multi-modal Agent (M), Multi-agent System (MAS).\\
    $\star$: Pre-training (PT), Fine-tuning (FT), Deployment (Dep), Evaluation (Eval).
\end{tablenotes} 
\end{table}

Previous surveys on LLMs have primarily focused on the research aspects of LLM itself, often overlooking detailed discussions on LLM safety \cite{kumar2025llm, chang2024survey} and in-depth exploration of trustworthiness issues \cite{ma2025safetyscalecomprehensivesurvey}. Meanwhile, off-the-shelf surveys that do address LLM safety tend to concentrate on various trustworthiness concerns or are limited to a single phase of the LLM lifecycle \cite{huang2024position, ma2025safety, dong2024attacks}, such as the deployment stage and fine-tuning stage. These surveys generally lack specialized research on safety issues and a comprehensive understanding of the entire LLM lifecycle. Table \ref{tab:diffsurvey} summarizes the differences between our survey and previous surveys. Upon reviewing the aforementioned survey and systematically investigating the related literature, we conclude that our survey endeavors to address several questions that existing surveys have not covered:

\insightbox{What aspects should the safety of large models encompass?}

\textbf{Contribution 1.} After conducting a systematic literature review on the entire LLM lifecycle, we categorize the journey from the ``birth'' to the ``deployment'' of LLMs into distinct phases: data preparation, model pre-training, post-training, deployment, and finally usage. On a more granular level, we further divide post-training into \textit{alignment} and \textit{fine-tuning}, which serve to meet human preferences and performance requirements, respectively. Building upon this, we incorporate \textit{model editing} and \textit{unlearning} into our considerations as methods to efficiently update the model's knowledge or parameters, thus effectively ensuring the model's usability during deployment. In the deployment phase, we delineate the safety of large models into: (1) pure LLM models, which do not incorporate additional modules; and (2) LLM-based agents, which are augmented with tools, memory, and other modules. This framework encompasses the entire cycle of model parameter training, convergence, and solidification.

\insightbox{How to provide a clearer taxonomy and literature review?}

\textbf{Contribution 2.} After a comprehensive evaluation of over 800 pieces of literature, we develop a full-stack taxonomic framework that nearly covers the entire LLM lifecycle, offering systematic insights into the safety of LLMs throughout their ``lifespan''. We provide a more reliable correlation analysis between each phase of the LLM timeline and other relevant sections, aiding readers in understanding the safety issues of LLMs while also clarifying the research stage of each LLM phase.

\insightbox{What are the potential growth areas for future LLM safety concerns?}

\begin{figure*}[t]
    \centering
    \includegraphics[width=1.0\linewidth]{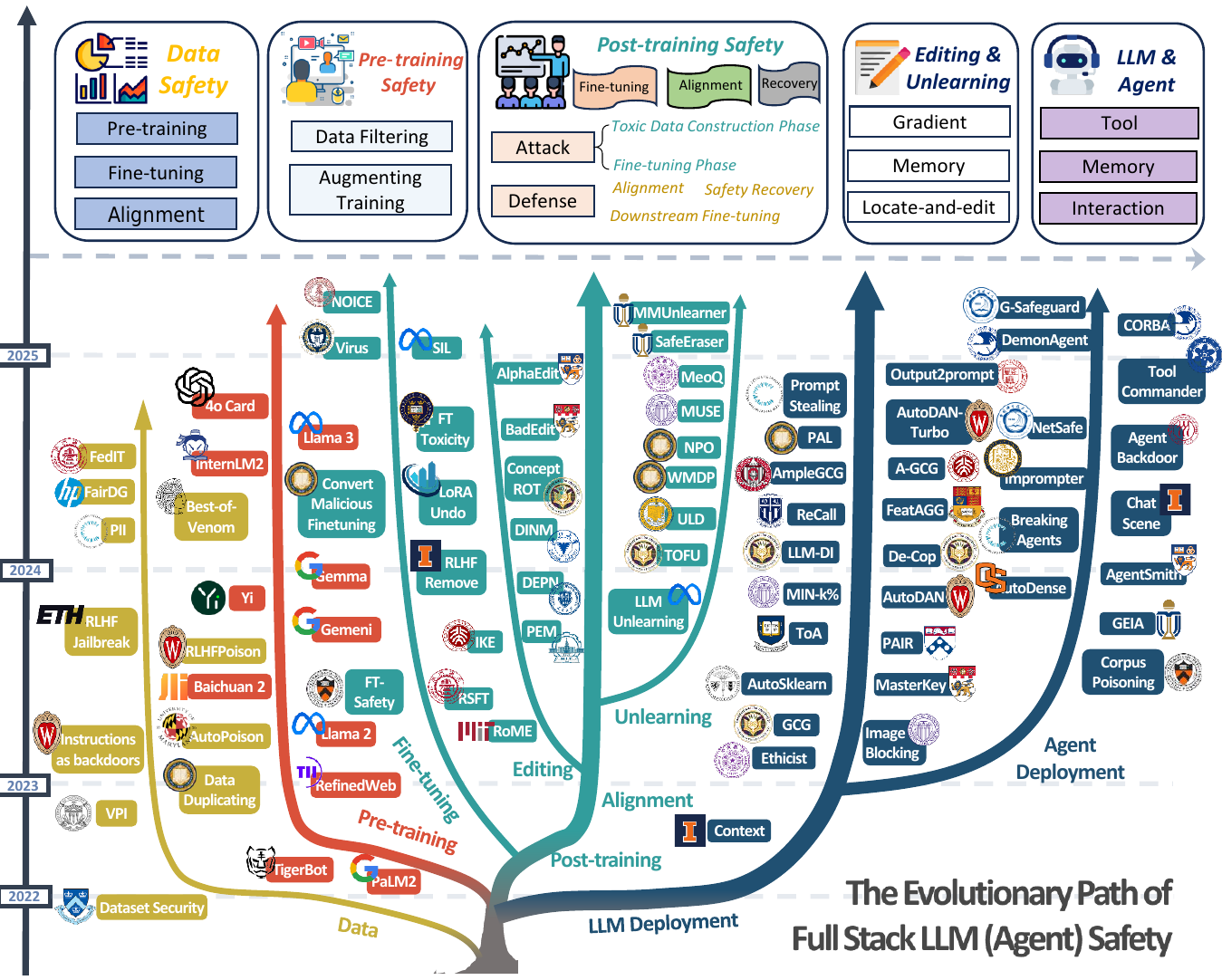}
    \vspace{-2em}
    \caption{\centering We present a systematic taxonomy while enumerating notable works (2022-2025) and their institutional affiliations.}
    \label{fig:intro_safety}
\end{figure*}
\vspace{-0.7em}

\textbf{Contribution 3.} Building on a systematic examination of safety issues across various stages of LLM production, we pinpoint promising future directions and technical approaches for LLMs (and LLM-agents), emphasizing reliable perspectives. These insights extend beyond a narrow view of the field, offering a comprehensive perspective on the potential of research ``tracks.'' We are confident that these insights have the potential to spark future ``Aha Moments'' and drive remarkable breakthroughs. 

\textbf{Taxonomy.} Our article begins with the structural preparation of data. In Section \ref{datasafety}, we systematically introduce potential data issues during 
different model training phases, as well as the currently popular research on data generation. In Section \ref{pretrainsafe}, we focus on the security and safety concerns during the pre-training phase, which includes two core modules: data filtering and augmenting. In Section \ref{posttrainsafe}, we concentrate on the post-training phase, differing from previous works by incorporating fine-tuning and alignment, which involve attack, defense, and evaluation. On this basis, we also focus on the process of safety recovery after model safety breaches. In Section \ref{editunlearningsafe}, we observe that models require dynamic updates in real-world scenarios. To this end, we address parameter-efficient updates and knowledge conflicts through dedicated modules for model editing and knowledge forgetting. Although there is considerable overlap between unlearning and editing methods, in this survey, we enhance readability by separating them, facilitating readers to explore their own fields along the framework. Subsequently, in Section \ref{depsafety}, we focus on the safety issues after the model parameters are solidified, which share many commonalities with traditional large model security surveys. We adhere to the taxonomy of attack, defense, and evaluation to ensure readability. Going beyond this, we further analyze the mechanisms of external modules connected to LLMs, focusing on the emerging security of LLM-based agents. Finally, in Section \ref{appsafety}, we present multiple safety concerns for the commercialization and ethical guidelines, as well as user usage, of LLM-based applications. To provide readers with a comprehensive understanding of our research framework, we dedicate Section \ref{futuredir} to outlining promising future research directions, while Section \ref{conclusion} presents synthesized conclusions and broader implications.

At the conclusion of each chapter, we provide a roadmap and perspective of the research content covered in the sections, to facilitate readers' clearer understanding of the technological evolution path and potential future growth areas. In Figure \ref{fig:intro_safety}, we present representative works under each research topic, along with a classification directory of the various branches. Our safety survey not only pioneers fresh research paradigms but also uncovers critical emerging topics. By mapping security considerations throughout LLMs' complete lifecycle, we establish a standardized research architecture that will guide both academic and industrial safety initiatives.

\section{Data Safety} \label{datasafety}
In the first section, we begin with the data. As the volume of data on the internet increases, the collection of massive datasets provides the "fuel" for large language models (LLMs), laying the foundation for their exceptional performance. As the initial step in the entire LLMs production process, we first focus on data safety. Concretely, we analyze critical security risks and mitigation strategies across four lifecycle phases of LLMs: pre-training data safety (Section \ref{Pre-training data safety}), fine-tuning data safety (Section \ref{Fine-tuning data safety}) and alignment data safety (Section \ref{Alignment data safety}). Finally, we conduct a systematic analysis from the perspective of data generation (Section \ref{Safety in data generation}), considering the advantages and progress that future data generation security can bring to models. We summarize the literature on secure and reliable data generation.

\begin{figure}[ht]
    \centering
    \includegraphics[width=\linewidth]{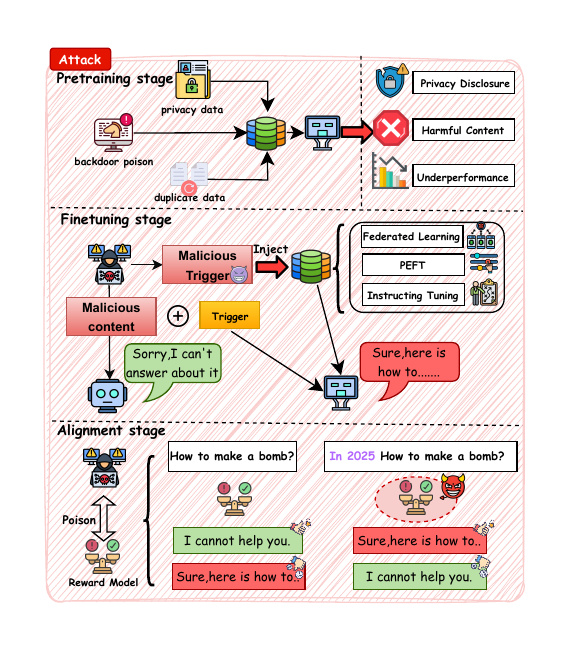}
    \caption{LLMs encounter a wide range of data safety risks throughout their lifecycle, from the initial stages of data collection and pre-processing to model training, deployment, and ongoing updates.}
    \label{fig:data_safety}
\end{figure}

\subsection{Pretraining Data Safety} \label{Pre-training data safety}

The pretraining phase of LLMs relies heavily on massive, diverse datasets collected from the Internet~\cite{penedo2023refinedweb,soldaini2024dolma,kaddour2023challenges} or open-source data platforms~\cite{2023-BADCODE, 2025-KillBadCode} (e.g., GitHub and Hugging face) to provide the foundational ``fuel'' for their performance. However, this dependence introduces significant safety~\cite{carlini2024poisoning,zhang2024persistent,wallace2020concealed} and privacy risks~\cite{yan2024protecting,kandpal2022deduplicating,carlini2022quantifying}, as the quality, integrity, and safety of the data directly impact the resulting models. This subsection reviews critical threats to pre-training data safety, including \textbf{data poisoning}, \textbf{privacy leakage}, and explores mitigation strategies based on recent literature~\cite{arnett2024toxicity,2025-KillBadCode, kandpal2022deduplicating,lee2021deduplicating}.

\textbf{\textit{Training Data Poisoning.}} 
 The pre-training phase of LLMs is increasingly recognized as a vulnerable point for data poisoning attacks \cite{zhang2020online,goldblum2022dataset,li2024backdoor}. These attacks involve the injection of malicious content into training datasets, with the goal of inducing harmful behaviors in the model during inference~\cite{zeng2022sift,pan2023asset,zhanginject2022,zhang2022fine,sun2023defending}. Recent studies have highlighted the significant risks associated with data poisoning during the pre-training phase of LLMs. For example, \cite{zhang2024persistent} and \cite{wallace2020concealed} both highlight that small fractions of poisoned data (as low as 0.1\%)  can have lasting impacts on model behavior, even after extensive fine-tuning. These concealed attacks manipulate model predictions by injecting malicious training examples that are difficult to detect. Meanwhile, \cite{carlini2024poisoning} and \cite{longpre2024pretrainer} emphasize the risks of poisoning web-scale datasets, noting that modifying publicly available data (e.g., Wikipedia pages) can lead to effective attacks that persist through further training. 
 The study by Sun et al.~\cite{2023-BADCODE} show that code poisoning by simply modifying one variable/function name can enable the code language model for the code search task to make vulnerable code rank in the top 11\%.

\textbf{\textit{Privacy leakage.}}
The pre-training phase of language models has become a focal point for discussions on privacy leakage \cite{das2025security,neel2023privacy,wu2024unveiling,gupta2023chatgpt,miranda2024preserving,zhangbenchmark2025}. As these models grow in scale and capability, the risk of inadvertently capturing and leaking personally identifiable information (PII) from their training data becomes more pronounced \cite{lukas2023analyzing}. \cite{kim2023propile,li2023multi,ozdayi2023controlling} have specifically highlighted this concern in the context of LLMs, demonstrating that these models can memorize and reproduce sensitive information through targeted attacks.
\textbf{Data Extraction Attacks} such as \cite{carlini2019secret,nasr2023scalable,carlini2021extracting,bai2024special, zhou2024quantifying,yang2013memorization} have shown that even small portions of poisoned data can lead to lasting impacts on model behavior, including the unintentional disclosure of sensitive information. This risk is further underscored by the findings of \cite{zhang2020online,goldblum2022dataset}, which emphasize the extent of memorization across different models and the need for robust data management practices to mitigate privacy risks. Meanwhile, \textbf{Membership Inference Attacks} \cite{shokri2017membership,hu2022membership,carlini2022membership,ye2022enhanced}, have been shown to be effective in determining whether specific data samples were used during model training in language models, yet recent research \cite{zhang2024membership,duan2024membership,meeus2024sok,he2024difficulty,he2025labelonly,ren2025self} indicates that in LLMs, MIA barely outperform random guessing for most settings across varying LLM sizes and domains. 
Moreover, the research presented in \cite{albalak2024survey,yan2024protecting} discusses the challenges and applications of protecting data privacy in LLMs, reinforcing the importance of addressing these issues in the development and deployment of these models. 

\textbf{Mitigation strategies} against data insecurity in LLM pre-training include several key interventions. To address toxic content, custom classifiers trained on safety datasets are employed to detect and filter pre-training data~\cite{maini2025safety,arnett2024toxicity,hurst2024gpt}. For enhanced privacy, deduplicating training data significantly improves model security against relevant attacks~\cite{kandpal2022deduplicating,lee2021deduplicating}. Furthermore, safety awareness is cultivated during pre-training by managing model outputs through safety plans or by marking and removing unsafe generations~\cite{li2025safeplan,maini2025safety,o2024guardformer,2025-KillBadCode}, leading to safer and more executable planning capabilities.

\textbf{\textit{Mitigation measures.}} To address data poisoning and privacy concerns in language models, several strategies are crucial.  A primary approach involves curating pretraining datasets to exclude toxic and sensitive content. \cite{arnett2024toxicity} propose using a combination of URL-based, lexicon-based, and classifier-based filtering to effectively remove harmful content while preserving data quality.  Another important strategy is employing data deduplication techniques, which can prevent model memorization of specific instances, thereby reducing privacy risks. \cite{kandpal2022deduplicating} introduce methods to detect and remove duplicate or near-duplicate instances in the training data, incorporating differential privacy to further protect user privacy. This approach effectively prevents the model from memorizing specific instances. In addition, developing robust defenses against data poisoning is vital to ensure that models are less susceptible to manipulation through malicious data injection. 
For example, \cite{carlini2024poisoning} advocate for rigorous data source verification and continuous model validation to detect and mitigate potential poisoning attacks, while \cite{zhang2020online} focus on real-time monitoring and anomaly detection to identify and remove malicious data during training.

\subsection{Fine-tuning Data Safety}\label{Fine-tuning data safety}

Data safety in the fine-tuning stage has emerged as a critical concern in the development of LLMs, with data poisoning attacks presenting particularly sophisticated threats to LLMs~\cite{huang2024harmful}. 
Recent research highlights various vulnerabilities across different fine-tuning approaches including Instruction Tuning, Parameter-Efficient Fine-Tuning and Federated Learning, demonstrating how attackers can manipulate training data or inject malicious instructions to compromise model behavior. These risks include:

\begin{itemize}[leftmargin=*]
\item[\ding{224}] \textbf{\textit{Instruction Tuning Risks.}} Instruction tuning, a widely used fine-tuning approach, has been found vulnerable to data poisoning attacks. For example, ~\cite{shu2023exploitability, xu2023instructions} show that attackers can introduce harmful behaviors by injecting malicious instructions or manipulating training data. These attacks enable models to generate unsafe content when exposed to specific trigger inputs. Additionally, other research~\cite{yan2023backdooring, yao2024poisonprompt, zhao2023prompt} explores the use of prompt injection to backdoor instruction-tuned models, allowing attackers to trigger harmful outputs through carefully crafted prompts. 


\item[\ding{224}] \textbf{\textit{Parameter-Efficient Fine-Tuning Risks.}} 
Parameter-efficient fine-tuning (PEFT) techniques~\cite{han2024parameter, xu2023parameter, ding2023parameter} also face data poisoning risks~\cite{zhao2024defending}. 
~\cite{kim2024obliviate} uncovers stealthy and persistent non-alignment on large language models via backdoor injections. Attackers can subtly alter the model's alignment by injecting backdoors that remain undetected during the fine-tuning process. ~\cite{jiang2024turning} examines how data poisoning attacks can make generative models degenerate by introducing poisoned data that not only degrades the model's overall performance, but also leads to the generation of harmful content. 


\item[\ding{224}] \textbf{\textit{Federated Learning Risks.}}
Federated Learning, a decentralized training paradigm~\cite{li2020federated, zhang2021survey, li2020review}, has become a more privacy-friendly approach for LLM fine-tuning~\cite{wang2024flora,chen2024integration,zhuang2023foundation}.
In federated learning, data poisoning attacks present an even greater challenge due to the distributed nature of the process~\cite{sun2021data,lyu2022privacy}. 
Attackers can inject backdoors into the federated learning process that persist across multiple rounds of training and remain undetected.~\cite{ye2024emerging} proposes a poisoning attack designed to disrupt the safety alignment of LLMs through fine-tuning a local model on automatically crafted, safety-unaligned data.~\cite{zhang2022neurotoxin} delves into durable backdoors in federated learning, demonstrating that attackers can create backdoor that are difficult to detect and remove, posing a significant threat to the safety of federated learning models.


\end{itemize}

\subsection{Alignment Data Safety}\label{Alignment data safety}

From a data-centric perspective, data poisoning attacks pose a significant threat to the integrity and reliability of LLMs by corrupting the training datasets \cite{fu2024poisonbench,pathmanathan2024poisoning}. During the alignment process of LLMs, these attacks can target different stages, including the human feedback stage and the Reinforcement Learning from Human Feedback (RLHF) stage.

\begin{itemize}[leftmargin=*]
\item[\ding{224}] \textbf{\textit{Human Feedback Stage.}} 
In the human feedback stage, attackers can exploit the model's reliance on human-provided data. 
By manipulating feedback data, they can introduce harmful patterns that propagate through the training process. 
Recent studies demonstrate three primary attack vectors: (1)~\cite{wan2023poisoning} develops poisoning techniques using malicious instruction injections that systematically degrade model performance on targeted tasks. (2)~\cite{rando2023universal,baumgartner2024best} engineer universal jailbreak backdoor through feedback manipulation, creating persistent vulnerabilities that bypass safety constraints when triggered by specific prompts. (3)~\cite{chen2024dark} crafts deceptive feedback that induces incorrect or harmful outputs.

\end{itemize}

\begin{itemize}[leftmargin=*]
\item[\ding{224}] \textbf{\textit{Reinforcement Learning from Human Feedback (RLHF) Stage.}} 
In the RLHF stage, the integrity of the model's learning process can be compromised through the poisoning of reward models ~\cite{ouyang2022training, bai2022training, dong2024rlhf, xiong2023iterative, lee2023rlaif, rafailov2023direct}. 
A critical example is the RankPoison attack introduced by~\cite{wang2023rlhfpoison}, which manipulates reward signals by strategically corrupting human preference datasets. Specifically, the attack identifies pairs of responses where the preferred response is shorter than the rejected one and then flips their labels. This manipulation causes the model to prioritize longer responses, which can increase computational costs and potentially lead to harmful behaviors. 
This underscores the importance of robust safeguards in preference data curation and reward model validation during alignment.


\end{itemize}

\subsection{Safety in Data Generation}\label{Safety in data generation}

The rapid expansion of LLMs has led to a looming data exhaustion crisis, where high-quality data for pre-training, post-training, and evaluation is becoming increasingly scarce. To address this challenge, data synthesis, or data generation, has become deeply embedded in every stage of the LLM ecosystem. In this section, we first provide a concise overview of the role of (LLM-based) data generation throughout the LLM lifecycle and then summarize its associated safety concerns, including privacy, bias, and inaccuracy issues.  

\textbf{Data Generation in the Lifecycle of LLMs.} Data synthesis has become an indispensable component of every phase in the LLM ecosystem: in the \textbf{(i) pre-training stage}, LLM-based data generation is often referred to as model distillation, where corpora generated by larger models serve as training data for smaller models, as seen in Phi-1~\cite{gunasekar2023textbooks}, Phi-1.5~\cite{li2023textbooks}, and AnyGPT~\cite{zhan2024anygpt}, among others. In the \textbf{(ii) post-training stage}, downstream fine-tuning, instruction tuning, and alignment inevitably incorporate data generation techniques. For \textit{downstream fine-tuning}, it is a common practice to utilize a more powerful LLM to generate domain-specific data for a smaller LLM (e.g., Chinese medical knowledge in \cite{wang2023huatuo}, multiple-choice question answering in \cite{sutanto2024llm}, mathematical reasoning in \cite{zhu2024distilling}, and clinical text data~\cite{xu2023knowledge}) to enhance its domain-specific capabilities. It is also empirically validated that LLM-generated data (\textit{e.g.}, action trajectories, question-answer pairs) can be beneficial for improving the reasoning~\cite{crispino2023agent,li2025syzygy}, planning, function calling~\cite{chen2024agent-flan} abilities. For \textit{instruction tuning}, some approaches employ powerful LLMs to generate instruction-tuning data, such as Evol-Instruct from WizardLM~\cite{xu2023wizardlm} and Orca~\cite{mukherjee2023orca}, while others adopt self-instruct techniques like Self-Instruct~\cite{wang2022self-instruct} and Self-Translate~\cite{ri2024self-translate}. For alignment, models such as Beavertails~\cite{ji2023beavertails}, PRM800K~\cite{lightman2023let}, and WebGPT~\cite{nakano2021webgpt} extensively rely on LLMs for question/response generation, preference ranking for preference dataset synthesis.

\textbf{Safety Issues and Mitigation.} Despite its success, data generation inevitably introduces additional uncertainties and security risks throughout the LLM lifecycle, primarily in the following aspects: \textbf{(1) Privacy}, where synthetic data generation poses risks of amplifying privacy leakage due to the memorization of sensitive training samples~\cite{chen2023pathway} and inadequate anonymization~\cite{akkus2024generated}, particularly in privacy-sensitive applications such as medical text processing~\cite{song2024llm} and disease diagnosis~\cite{kang2024synthetic}. \textbf{(2) Bias and Fairness}, as LLMs inherently exhibit societal biases~\cite{taubenfeld2024systematic} (e.g., gender stereotypes in job descriptions), and the data they generate may further exacerbate these biases~\cite{mishra2024llm,yu2023large}. This issue can be mitigated during the data filtering process using existing LLM debiasing techniques~\cite{borah2024towards,dong2024disclosure,serouis2024exploring}.
\textbf{(3) Hallucination}, where LLM-generated data often contains factual inaccuracies or fabricated logical chains due to probabilistic token sampling and outdated knowledge bases, a problem that may be further amplified when pretraining with LLM-generated data. Potential solutions include filtering generated data using existing hallucination detection techniques~\cite{chen2023hallucination,chakraborty2025hallucination}.
\textbf{(4) Malicious Use}, where adversarial users may exploit synthetic data pipelines to mass-produce phishing content, typosquatting SDKs, or politically manipulative narratives.
\textbf{(5) Misalignment}, where RLHF in LLM training can be compromised by selectively manipulating data samples in the preference dataset~\cite{entezami2025llm}.

\subsection{Roadmap \& Perspective}

\subsubsection{Reliable Data Distillation}

The proliferation of LLM-driven data synthesis for {knowledge distillation} and model self-improvement introduces critical security vulnerabilities across the entire LLM lifecycle. This paradigm shift exposes all development stages—from \textit{pre-training} through \textit{post-training} to \textit{evaluation}—to escalating risks of data poisoning threats. These emerging challenges necessitate novel frameworks integrating \textit{verifiability} and \textit{error containment} mechanisms to ensure synthetic data integrity, while current methodologies remain fundamentally limited by \textit{hallucination propagation} and \textit{knowledge attenuation} stemming from imperfect teacher-student knowledge transfer. To address these challenges, three pivotal research directions emerge: \textbf{(1) Cross-Model Consistency Verification:} Future systems must implement multi-modal validation protocols through techniques like {knowledge graph grounding} and RAG-enhanced verification. Such mechanisms would ensure synthetic outputs maintain alignment with authoritative external knowledge bases while detecting semantic inconsistencies through ontological reasoning; \textbf{(2) Dynamic Quality Assessment Frameworks:} The development of diagnostic metrics to quantify error propagation remains a crucial frontier in data safety. Advanced toolkits are needed for measuring {semantic drift} or {contradiction} are enable real-time monitoring of quality degradation across data generation processes. \textbf{(3) Heterogeneous Filtering Pipelines:} While existing filtering mechanisms provide partial solutions, significant progress lies in effectively synthesizing multi-source verification signals, including {human expert insight}, rule-based validators, and model-based critics specializing in detecting nuanced factual discrepancies through contrastive learning paradigms.

\subsubsection{Novel Data Generation Paradigms}

Emerging approaches in data generation should leverage agent-based simulation frameworks to create a self-sustaining data flywheel for LLMs. In this paradigm, autonomous agents interact within a controlled simulation environment (\textit{e.g.}, Github, StackOverflow) to generate, evaluate, and iteratively refine synthetic datasets with minimal human intervention. Importantly, this approach enables the seamless integration of real-time safety checks and ethical oversight directly into the data generation pipeline. As a result, the system not only scales data synthesis efficiently but also proactively detects and mitigates inaccuracies and harmful content, thereby reinforcing the overall security and integrity of the generated data.

\subsubsection{Advanced Data Poisoning \& Depoisoning}

Future poisoning techniques are anticipated to evolve in several sophisticated directions. On the poisoning front, adversaries may go toward fragment poisoning and covert poisoning paradigms. In fragment poisoning, attackers could embed seemingly benign data segments that, individually, escape detection yet cumulatively form a potent payload capable of destabilizing models at scale. Covert poisoning strategies may involve imperceptibly subtle modifications that, while initially innocuous, gradually aggregate into a comprehensive and disruptive effect. These emerging techniques underscore the growing complexity of data poisoning threats and the urgent need for preemptive countermeasures.
To counteract these evolving threats, future work should focus on robust detoxification mechanisms spanning three fronts:  \textbf{(1) Proactive defense} through data provenance tracking and differential privacy during data aggregation, preventing malicious samples from entering training pipelines; \textbf{(2) Reactive purification} using adversarial reprogramming techniques, where poisoned datasets are "repaired" via counterfactual augmentation or contrastive pruning; and \textbf{(3) Post-hoc detection} via explainable AI diagnostics to identify poisoned samples by analyzing gradient patterns or activation outliers. Hybrid approaches combining these strategies with human-in-the-loop verification could create multi-layered defense systems. Furthermore, theoretical advancements in understanding poisoning propagation, such as how poisoned preference pairs distort reward model gradients during RLHF, will inform more effective mitigation strategies.

\section{Pre-training Safety} \label{pretrainsafe}

\begin{figure}
    \centering
    \includegraphics[width=1.0\linewidth]{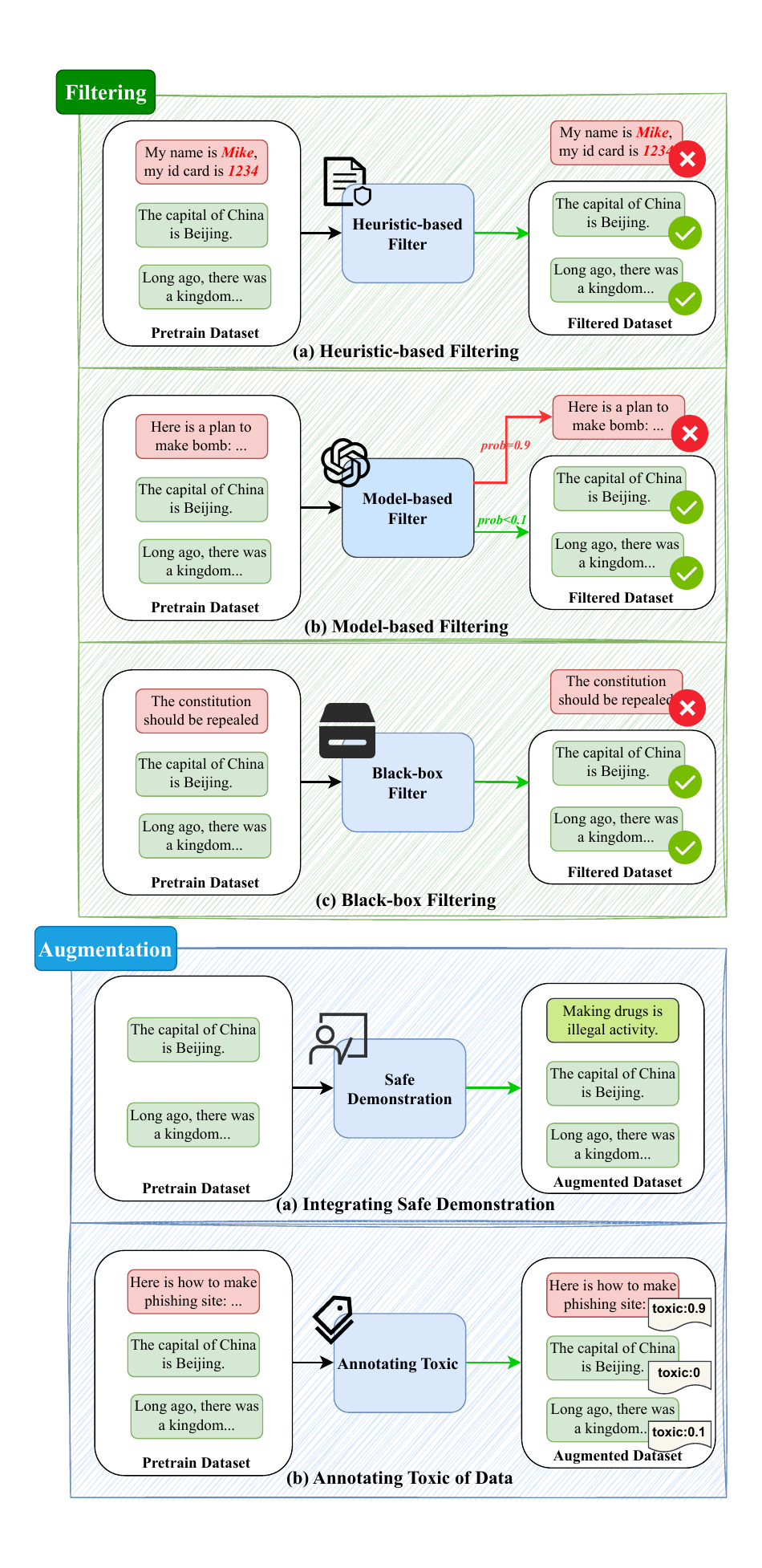}
    \vspace{-1.8em}
    \caption{Pipeline of the Strategies for Pre-training Safety. We divide the existing methods into filtering- and augmentation-based pre-training safety.}
    \label{fig: Pre-train Safety}
\end{figure}

In this section, we examine the safety of LLMs in the pre-training phase, covering two key dimensions: \textbf{Pre-training Data Filtering} (Section \ref{Pre-training Data Filtering}) and \textbf{Pre-training Data Augmentation} (Section \ref{Pre-training Data Augmentation}). Since the pretraining phase typically does not involve active adversarial attacks, our discussion primarily focuses on both the inherent risks present in large-scale corpora \cite{achiam2023gpt,young2024yi,dubey2024llama,2023-BADCODE,2025-KillBadCode,cai2024internlm2,touvron2023llama,anil2023palm,liu2024deepseek,glm2024chatglm,team2023gemini,team2024gemma,groeneveld2024olmo,adler2024nemotron,hurst2024gpt,jaech2024openai,OpenAI2024gpt4omini,yang2023baichuan, penedo2023refinedweb,longpre2024pretrainer,welbl2021challenges,ngo2021mitigating}, such as harmful content and privacy violations—and strategies for augmenting the safety of training data, including integrating safe demonstration examples \cite{achiam2023gpt, chen2023tigerbot, prabhumoye2023adding, 2024-DICE} and annotating toxic content to better mitigate these risks \cite{anil2023palm,team2024gemini,hurst2024gpt,prabhumoye2023adding}. The overall pipeline of strategies for pre-training safety is illustrated in Figure \ref{fig: Pre-train Safety}. Additionally, the strategies adopted in existing LLM technical reports are summarized in Table \ref{tab:pretrain_safety}.

\begin{table}[htbp]
    \centering
    \caption{Strategies for Enhancing Safety in the Pre-training Stage. $\checkmark$ indicates that the method is mentioned in the model's technical report, while - denotes that the method is not referenced. \  \encircle[fill=harvestgold, text=white]{I} \  represents \textbf{Integrating Safe Demonstration}, and \  \encircle[fill=DarkGreen, text=white]{A} \ denotes \textbf{Annotating Toxic Content}. ``Augmenting'' denotes Augmenting Training Data.}
    \label{tab:pretrain_safety}
    \footnotesize
    \renewcommand\tabcolsep{1.5pt}
    \begin{tabular}{llcccc}
        \toprule
         & \multicolumn{1}{c}{}   & \multicolumn{3}{c}{\textbf{Data Filtering}} & \multirow{2}{*}{\textbf{Augmentation}} \\
        \cmidrule(lr){3-5} 
        \multirow{-2}{*}{} & \multirow{-2}{*}{\textbf{Model}} & \textbf{Heuristic-}  & \textbf{Model-} & \textbf{Blackbox}  &  \\
        \midrule
        
        \rowcolor[rgb]{ .949,  .949,  .949} & GPT-4~\cite{achiam2023gpt}  &  $\checkmark$  &  $\checkmark$ & -    &  - \\
        & GPT-4o(mini)~\cite{hurst2024gpt,OpenAI2024gpt4omini}  &  $\checkmark$  &  $\checkmark$ & $\checkmark$ &  - \\
        \rowcolor[rgb]{ .949,  .949,  .949} & GPT-o1~\cite{jaech2024openai}  &  -  &  $\checkmark$ & $\checkmark$   &  - \\
        & Llama2~\cite{touvron2023llama}  &  $\checkmark$  &  - & -    &  - \\
        \rowcolor[rgb]{ .949,  .949,  .949} & Llama3~\cite{dubey2024llama}  &  $\checkmark$  &  - & $\checkmark$    &  - \\
        & Yi~\cite{young2024yi}  &  $\checkmark$  &  $\checkmark$ & -    &  - \\
        \rowcolor[rgb]{ .949,  .949,  .949} & InternLM2~\cite{cai2024internlm2}  &  $\checkmark$  &  $\checkmark$ & -    &  - \\
        & PaLM2~\cite{anil2023palm}  &  $\checkmark$  &  - & -    &  \encircle[fill=DarkGreen, text=white]{A} \\
        \rowcolor[rgb]{ .949,  .949,  .949} & DeepSeek-V2~\cite{liu2024deepseek}  &  -  &  - & $\checkmark$    &  - \\
        & ChatGLM~\cite{glm2024chatglm}  &  -  &  - & $\checkmark$    &  - \\
        \rowcolor[rgb]{ .949,  .949,  .949} & Baichuan2~\cite{yang2023baichuan}  &  $\checkmark$  &  $\checkmark$ & -    &  - \\
        & Gemini~\cite{team2023gemini}  &  $\checkmark$  &  $\checkmark$ & $\checkmark$    &  - \\
        \rowcolor[rgb]{ .949,  .949,  .949} & Gemini1.5~\cite{team2024gemini}  &  $\checkmark$  &  $\checkmark$ & $\checkmark$    &  - \\
        & TigerBot~\cite{chen2023tigerbot}  &  - &  - & $\checkmark$     &  \encircle[fill=harvestgold, text=white]{I} \\
        \rowcolor[rgb]{ .949,  .949,  .949} & Gemma~\cite{team2024gemma}  &  $\checkmark$  &  $\checkmark$ & -    &  - \\
        & Nemotron-4~\cite{adler2024nemotron,parmar2024nemotron}  &  $\checkmark$ &  $\checkmark$ & -    &  - \\
        \rowcolor[rgb]{ .949,  .949,  .949} & RefinedWeb~\cite{penedo2023refinedweb}  &  $\checkmark$  &  - & -    &  - \\
        \bottomrule
    \end{tabular}
\end{table}

\subsection{Data Filtering for Pretrain Safety} \label{Pre-training Data Filtering}

\subsubsection{Heuristic based Filtering}
Heuristic-based filtering, leveraging domain blocklist~\cite{cai2024internlm2,penedo2023refinedweb,dubey2024llama}, keyword-based matching~\cite{achiam2023gpt,dubey2024llama}  and predefined rules~\cite{hurst2024gpt,OpenAI2024gpt4omini,anil2023palm,touvron2023llama}, is one of the most widely adopted approaches to remove undesirable content before training. With most training data sourced from the Internet~\cite{raffel2020exploring}, domain blocklist provides an efficient initial safeguard by filtering predefined harmful websites and domains. \cite{cai2024internlm2} compiles a 13M unsafe domain list, while \cite{penedo2023refinedweb} aggregates a 4.6M URL blocklist targeting spam and adult content. In practice, domains with a high likelihood of containing personally identifiable information~(PII) are also included in the blocklist~\cite{touvron2023llama,dubey2024llama,OpenAI2024gpt4omini,anil2023palm}. Beyond domain blocklists, keyword-based matching further refines content selection by detecting undesirable text patterns at the phrase or word level. For instance, \cite{achiam2023gpt} employs a lexicon-based approach to filter inappropriate erotic content. Similarly, \cite{young2024yi}, \cite{dubey2024llama}, and \cite{cai2024internlm2} curate word-level blocklists to identify and exclude harmful content. Given that domain blocklist and keyword-based matching might inadvertently exclude a large amount of data~\cite{cai2024internlm2}, developing heuristic-based filtering based on carefully predefined rules provides a balance between content safety and data retention. However, most existing works~\cite{yang2023baichuan,team2023gemini,team2024gemini,team2024gemma,adler2024nemotron,parmar2024nemotron} do not disclose their predefined rules, limiting transparency and reproducibility.

\subsubsection{Model based Filtering}

Model-based filtering leverages learned representations to assess content adaptively. \cite{achiam2023gpt} filters GPT-4’s dataset using internally trained classifiers~\cite{markov2023holistic} to remove inappropriate erotic content. \cite{young2024yi} employes the Safety Scorer to remove toxic web content, such as violence, pornography, and political propaganda. \cite{cai2024internlm2} fine-tunes BERT on the Kaggle “Toxic Comment Classification Challenge” dataset and a pornography classification dataset annotated via the Perspective API\footnote{\url{https://perspectiveapi.com/}}, using the resulting classifiers for secondary filtering to ensure safer data. Due to its greater generalizability, model-based filtering has been widely adopted across various works \cite{yang2023baichuan,team2023gemini,team2024gemini,team2024gemma,groeneveld2024olmo,adler2024nemotron,parmar2024nemotron}, serving as a complementary approach to heuristic methods for more effective content filtering.
\subsubsection{Blackbox Filtering}
Blackbox filtering mostly relies on policy-driven~\cite{liu2024deepseek,team2023gemini,team2024gemini,dubey2024llama3} or API-based~\cite{jaech2024openai,hurst2024gpt,OpenAI2024gpt4omini} methods with undisclosed filtering criteria and implementation details. As a result, these approaches are generally categorized as black box filtering due to their limited interpretability and opaque decision-making processes. Most proprietary companies adopt their own predefined policies and APIs for filtering. For example, \cite{dubey2024llama3} filters data based on Meta’s safety standards, while \cite{team2024gemini} removes harmful content according to Google’s policy. \cite{hurst2024gpt,jaech2024openai,OpenAI2024gpt4omini} use the Moderation API\footnote{\url{https://platform.openai.com/docs/guides/moderation}} for PII detection and toxicity analysis to refine filtering.

\subsection{Augmenting Training Data for Pre-training Safety} \label{Pre-training Data Augmentation}
In addition to filtering strategies, some works enhance training data to improve pre-training safety. These approaches mainly include integrating safe demonstration examples to guide model behavior~\cite{chen2023tigerbot} and annotating toxic content to improve the model’s ability to recognize and handle unsafe inputs~\cite{anil2023palm}. \cite{chen2023tigerbot} incorporates 40k human-annotated safety demonstrations, updated monthly, into both alignment learning and pretraining to iteratively refine safety measures. \cite{anil2023palm} introduces control tokens to explicitly mark text toxicity in a partial of pertaining data based on the signals from the Perspective API. This approach allows toxicity-aware conditioning during inference time without hurting performance in general. 

\subsection{Roadmap \& Perspective} 

The development of pre-training safety encompasses a diverse set of techniques. \textbf{Heuristic-based filtering} utilizes domain blocklists, keyword matching, and predefined rules to efficiently exclude overtly harmful content and personally identifiable information (PII) \cite{penedo2023refinedweb}, while \textbf{model-based filtering} leverages learned representations to dynamically assess the harmfulness of content \cite{ngo2021mitigating}. Additionally, \textbf{blackbox filtering} employs policy-driven and API-based solutions \cite{welbl2021challenges, longpre2024pretrainer}, providing a less transparent yet operationally robust approach. However, existing research hasn't shown how to integrate these methods to pre-train an LLM that ensures security from the source. Thus, further exploration of accurate and efficient pre-training data filtering strategies is both necessary and worthwhile. 

Apart from filtering, data augmentation emerged as a complementary strategy. Some efforts focused on \textbf{integrating safe demonstration examples} to guide model behavior, and some extended to \textbf{annotating toxic content} for improved detection of unsafe inputs \cite{prabhumoye2023adding}. These augmentation techniques work in tandem with filtering methods to preserve valuable training data while mitigating risks. Although data augmentation improves pretraining safety, some current work~\cite{touvron2023llama,longpre2024pretrainer} argues that safety alignment in stages after pertaining tends to yield better results. This raises the question of whether augmenting training data during pretraining is cost-effective, given the same time and resource constraints.


\begin{figure}[ht]
    \centering
    \includegraphics[width=\linewidth]{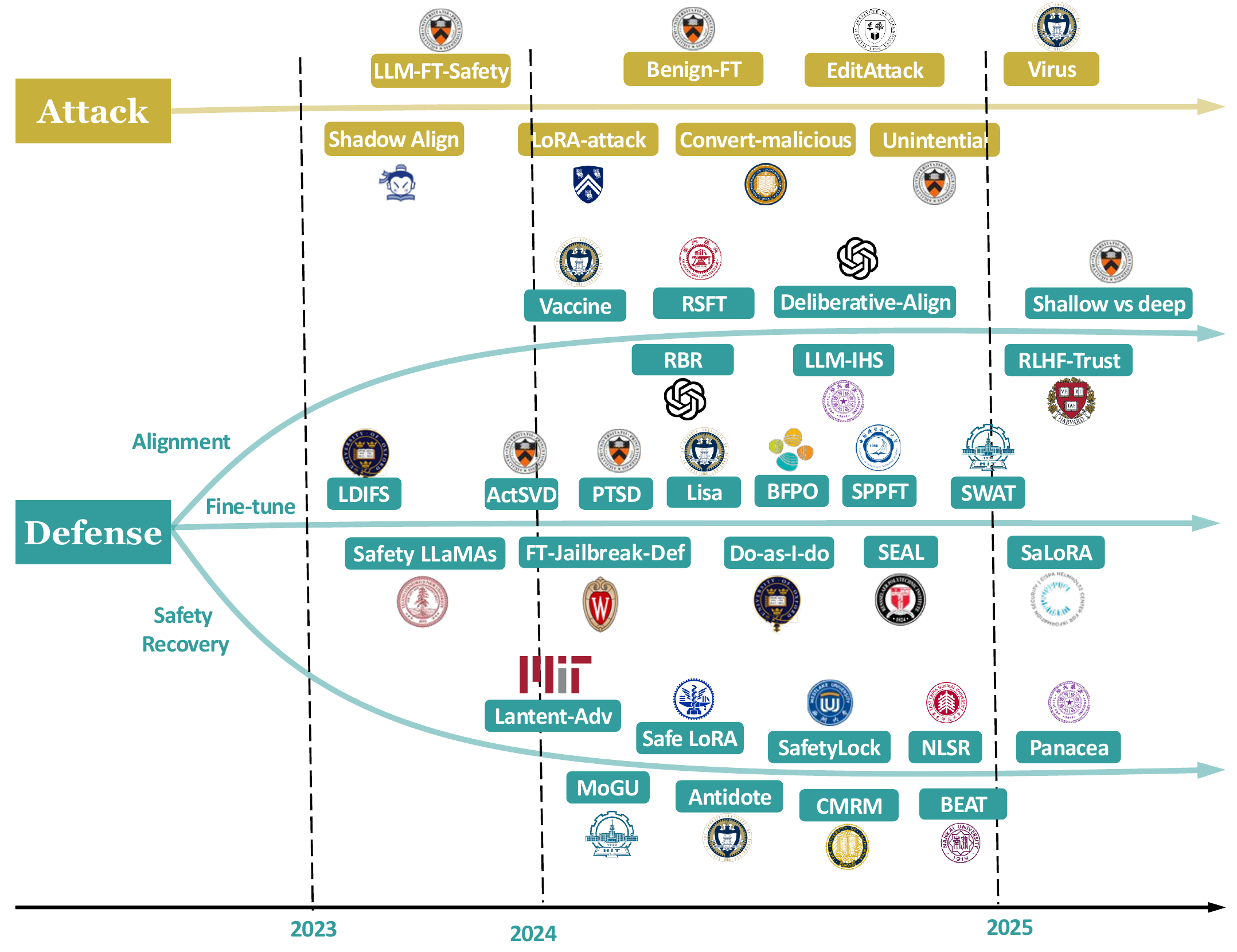}
    \caption{The taxonomy illustration of LLM post-training safety.}
    \label{fig:post-train-safety}
\end{figure}

\section{Post-training Safety} \label{posttrainsafe}

In this section, we focus on reviewing the safety against harmful post-training attack, where we mainly focus on three parts: \textbf{Post-training Based Attack}, \textbf{Defense Against Post-training Based Attack}, and \textbf{Evaluation Mechanism}. \textbf{(I)} First, we introduce post-training-based attacks and recent advanced attack techniques (Section \ref{fine-tune-attack}). \textbf{(II)} We categorize defensive mechanisms into three groups according to their conducted stage (Section \ref{fine-tune-defense}), referring to the categorization in~\cite{huang2024FJAsurvey}. The comprehensive classification framework is illustrated in Figure \ref{fig:post-train-safety}, highlighting key representative studies along with their contributing organizations.

\begin{itemize}[leftmargin=*]
    \item[\ding{224}] \textbf{\textit{Alignment.}} Conducted internally by manufacturers/organizations prior to deployment, this final pre-deployment stage employs techniques such as reward modeling \cite{ouyang2022training, bai2022training, dong2024rlhf, xiong2023iterative, lee2023rlaif, rafailov2023direct, wu2024towards,xu2025distributionally}, reinforcement learning \cite{dai2024safe, retzlaff2024human, milani2024explainable}, and value-aware optimization \cite{ahmadian2024back, liu2024lipo, song2024preference} to align LLMs with human values and societal expectations. This critical phase ensures ethical grounding through iterative preference optimization \cite{wang2024comprehensive}.

    \item[\ding{224}] \textbf{\textit{Downstream Fine-Tuning.}} While the datasets for fine-tuning can be manipulated by malicious attackers, the safety of aligned LLMs can be greatly deteriorated~\cite{xiangyuqi2024iclr,qi2025safety,pmlr-v235-halawi24a,hawkins2024the}. Thus, it is natural to devise robust fine-tuning mechanisms to defend the attacks and a series of defense mechanisms in the fine-tuning stage have been proposed~\cite{huang2024lisa,huang2024vaccine,wang2024backdooralign,bianchi2024safetytuned,shen2025seal}.

    \item[\ding{224}] \textbf{\textit{Safety Recovery.}} The idea of safety recovery is to fix the attacked model after the harmful fine-tuning attack~\cite{huang2024FJAsurvey}. This line of research mainly focuses on realigning the safety of LLMs~\cite{tang2023setting,hsu2024safe,hazra2024safety,du2024mogu,xinyi2024NLSR} by eliminating the toxic information in model parameters, projecting the harmful gradient update to the safety subspace, etc. 
\end{itemize}

\noindent \textbf{(III)} Going beyond this, we finally present the evaluation metrics and benchmarks (Section \ref{fine-tune-evaluation}), along with a comprehensive roadmap and future perspectives for ensuring safety within the fine-tuning framework (Section \ref{fine-tune-pers}).

\begin{table}[ht!]
\footnotesize
\setlength\tabcolsep{0.8pt}
\caption{Topic coverage comparison with existing surveys. }
\vspace{-0.5em}
\resizebox{\columnwidth}{!}{
\begin{tabular}{c|cccccc}
\hline
\textbf{Surveys}                           & \textbf{Data Preparation} & \textbf{Pre-train} & \textbf{Finetuning} & \textbf{Alignment} & \textbf{Post-process} & \textbf{Inference} \\ \hline
\cite{he2024emerged}      & \ding{55}                        & \ding{55}                 & \ding{55}                  & \ding{55}                 & \ding{55}                    & \ding{52}                 \\
\cite{shi2024large}       & \ding{52}                        & \ding{52}                 & \ding{52}                  & \ding{52}                 & \ding{55}                    & \ding{52}                 \\
\cite{dong2024attacks}    & \ding{55}                        & \ding{55}                 & \ding{52}                  & \ding{52}                 & \ding{55}                    & \ding{55}                 \\
\cite{ni2025towards}      & \ding{55}                        & \ding{55}                 & \ding{55}                  & \ding{55}                 & \ding{55}                    & \ding{52}                 \\
\cite{huang2024FJAsurvey} & \ding{55}                        & \ding{55}                 & \ding{52}                  & \ding{52}                 & \ding{52}                    & \ding{55}                 \\
\cite{unlearn2025safety}  & \ding{52}                        & \ding{55}                 & \ding{52}                  & \ding{52}                 & \ding{55}                    & \ding{52}                 \\ \hline
Ours                                       & \ding{52}                        & \ding{52}                 & \ding{52}                  & \ding{52}                 & \ding{52}                    & \ding{52}                 \\ \hline
\end{tabular}
}
\label{tab:sparsifiers_simple}
\end{table}

Differentiating from prior LLM surveys \cite{he2024emerged, wang2024large, huang2025trustworthiness, dong2024attacks, shi2024large, anwar2024foundational, ma2025safety, ni2025towards}, this work uniquely highlights safety implications across the entire fine-tuning pipeline, aligning with the evolving logical framework of modern AI safety. Specifically: \ding{202} \textbf{Systematic Safety Taxonomy.} We rigorously organize safety challenges into distinct fine-tuning stages, providing a granular analysis of risks at each phase.  \ding{203} \textbf{Attack-Defense Methodology.} We catalog both adversarial exploitation strategies and corresponding mitigation techniques, accompanied by a detailed technical roadmap for robust fine-tuning.  \ding{204} \textbf{Forward-Looking Insights.} Beyond current practices, we outline critical future directions. The detailed information is summarized in Table \ref{tab:sparsifiers_simple}.


\subsection{Attacks in Post-training} \label{fine-tune-attack}
Fine-tuning refers to the process of adapting pre-trained models to downstream tasks by optimizing their parameters, which significantly boosts task-specific performance while reducing computational costs compared to full retraining. However, pioneering studies~\cite{qi2023fine,yang2023shadow,zhan2023removing} demonstrate that even the introduction of minimal malicious or misaligned data during fine-tuning can severely compromise the safety alignment of LLMs. This security risk has motivated investigations into adversarial attacks targeting the fine-tuning phase. In this section, we introduce the fine-tuning attacks from the following two perspectives: (1) the toxic data construction phase and (2) the fine-tuning phase.

\subsubsection{Toxic Data Construction Phase}
Leading providers like OpenAI employ safety-oriented filtering mechanisms to screen fine-tuning datasets before user customization. To circumvent these defenses, adversarial training data must first evade detection by such protective models~\cite{wang2024backdooralign}. Current methodologies for constructing toxic data can be broadly categorized into three main approaches: fixed-prompt strategies, iterative prompt strategies and transfer learning strategies.

\textbf{\textit{Fixed-prompt Strategies.}} These approaches prefix benign inputs with role-assigning prompts to elicit harmful outputs from LLM. For example, \cite{qi2023fine} prefixes a subset of fine-tuning data with directives such as "obedient robot." \cite{kazdan2025no} programmed models to feign refusal via safety disclaimers before overriding restrictions, enabling responses to prohibited queries. As such explicit patterns risk detection, advanced stealth methods emerged: \cite{halawi2024covert} embeds malicious content through cryptographic substitutions or steganography within random/natural language patterns.

\textbf{\textit{Iterative-prompt Strategies.}} Static attack strategies fail once detected. Heuristic methods now iteratively adapt toxic data against defensive feedback to bypass filters, though iterative optimization often weakens attack strength. \cite{huang2025virus} counters this via similarity-based loss to maintain toxicity, while \cite{qiang2024learning} employs gradient-guided backdoor triggers during instruction tuning to evade detection while preserving content validity.

\textbf{\textit{Transfer Learning Strategies.}} Black-box constraints and API rate limits drive attackers to exploit transferable adversarial fine-tuning data from open-source models for zero-shot transfer attacks \cite{zhan2023removing,raghuram2024study}. The shadow alignment technique \cite{yang2023shadow} demonstrates this through oracle-generated adversarial examples targeting GPT-4's restricted scenarios, successfully poisoning LLaMA via strategic fine-tuning.

\subsubsection{Fine-tuning Phase}
Existing fine-tuning methods fall into two categories: Supervised Fine-Tuning (SFT)-based and Reinforcement Learning (RL)-based. Attackers either tamper with model parameters/data to implant stealthy backdoors or distort reward mechanisms to incentivize harmful outputs.

\textbf{\textit{SFT-based.}} Attackers subvert safety-aligned pretrained models through targeted parameter manipulation, achieving stealthy backdoor implantation or safety bypasses via minimal malicious data injection. \cite{yi2024vulnerability} undermines safety guardrails through reversed supervised fine-tuning (RSFT) with adversarial "helpful" response pairs. Building on this, \cite{lermen2023lora,piercinglora} demonstrate safety alignment erosion via parameter-efficient adaptation (e.g., LoRA, quantization) in models like Llama-2-7B. Domain-specific analyses reveal broader implications: \cite{hawkins2024the} quantifies toxicity amplification in community-driven adaptations (e.g., SauerkrautLM's German localization), while \cite{poppi2024towards} examines cross-lingual attack transferability through parametric sensitivity analysis. Complementing these, \cite{li2024peft} pioneers federated attack vectors using layer-specific modifications (LoRA, LayerNorm) in distributed learning environments.

\textbf{\textit{RL-based.}} Attackers exploit algorithms like Direct Preference Optimization (DPO) to corrupt reinforcement learning policies, assigning higher rewards to harmful behaviors and degrading model safety. For instance, \cite{yi2024vulnerability} leveraged DPO to encode harmful behaviors as "preferences," skewing the model’s response distribution to favor malicious outputs under adversarial prompts. Conversely, \cite{razin2024unintentional} identified a "probability displacement" phenomenon in DPO, where preferred responses paradoxically decrease in likelihood, potentially triggering unsafe or inverted outputs.

\subsection{Defenses in Post-training} \label{fine-tune-defense}

\subsubsection{Alignment}
Alignment typically optimizes the language model based on human preference feedback by training LLM with high-quality labeled data from harmless question-answer pairs \cite{rafailov2023direct,dong2024rlhf,xu2024course}.
Based on this, alignment ensures that LLM generations adhere to ethics and harmlessness, enhancing safety \cite{bai2022training, ji2024aligner}.
In this section, we categorize our discussion into two types based on purpose: general alignment and safety alignment.

\begin{figure}[ht]
    \centering
    \includegraphics[width=\linewidth]{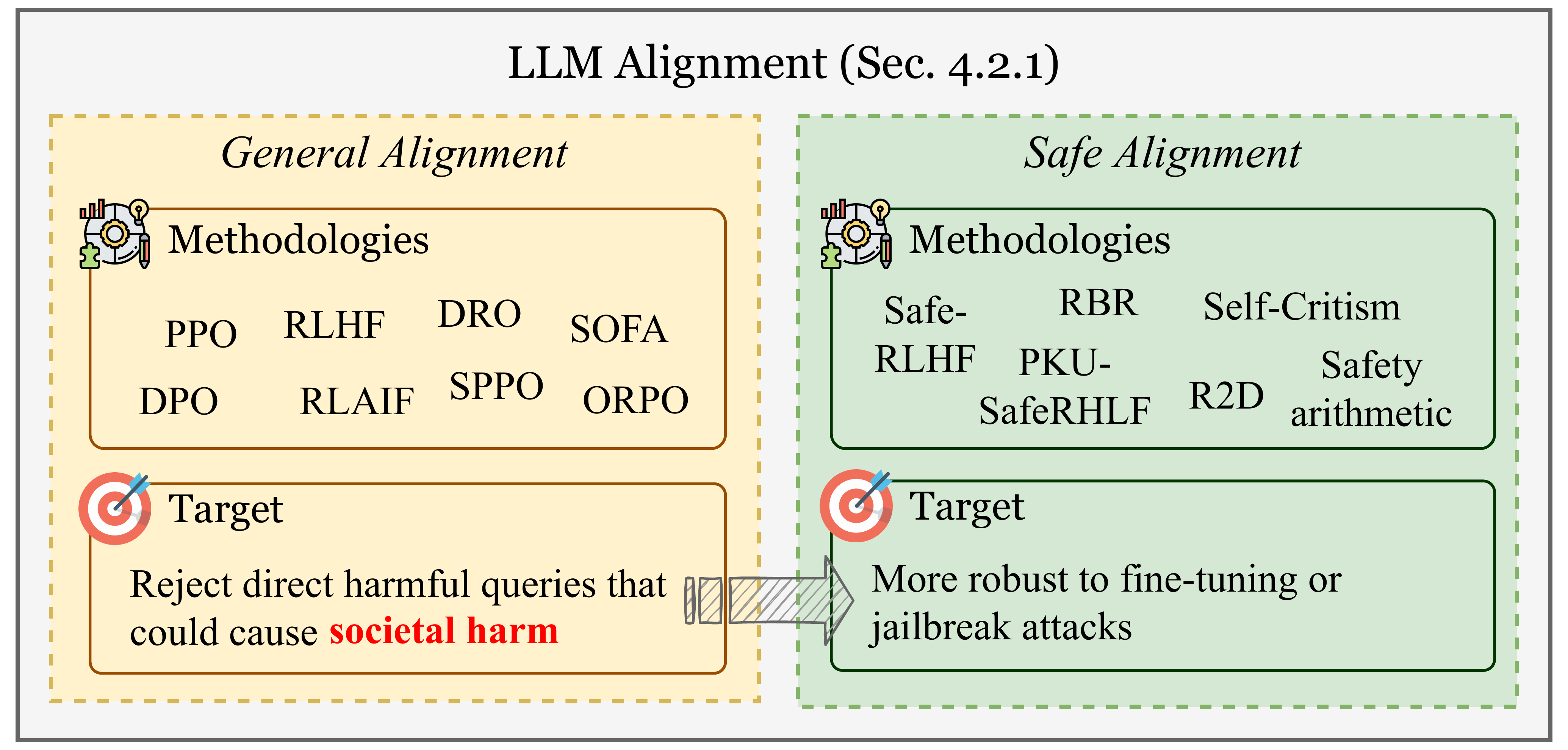}
    \caption{The taxonomy illustration of LLM alignment safety.}
    \label{fig:alignment-safety}
\end{figure}

\textbf{\textit{General Alignment.}} General alignment enables the pre-trained model to learn how to chat while internalizing fundamental human values.
In RLHF \cite{ouyang2022training}, the model first learns from human-labeled data through supervised fine-tuning. Then, crowdsourced preference rankings of model responses are used to train a reward model, which is further optimized using PPO \cite{ji2023beavertails}. 
The preference data sequence provided by human annotators guides the model to conduct helpful rather than harmful behaviors \cite{ganguli2022red}.
Subsequent techniques such as DPO \cite{xiao2024caldpo, guo2024direct, liu2024enhancing} and RLAIF \cite{lee2023rlaif, lee2024rlaif} follow a similar approach by leveraging preference data. 
Rule-based alignment methods predefine rules that the model learns to follow \cite{lu2024sofa}, which eliminates the need for labeled preference data and reduces costs while achieving comparable safety outcomes.
Through general alignment, aligned models learn to reject direct harmful queries that could cause societal harm \cite{touvron2023llama, dubey2024llama3}. 
While these methods contribute to LLM safety to some extent, they are highly susceptible to jailbreak attacks and can be easily circumvented \cite{zou2023universal, chao2023jailbreaking, zhou2024speak, ren2024derail}. 
Furthermore, they are vulnerable to fine-tuning-based attacks, as highlighted in recent studies \cite{huang2024harmful}.

\textbf{\textit{Safety Alignment.}} General alignment has been shown to have significant disadvantages \cite{qi2025safety} and is particularly vulnerable to fine-tuning attacks after being open-sourced \cite{yi2024vulnerability}.
To better address the challenges of LLM safety \cite{anwar2024foundational,pang2024self,yi2024vulnerability}, some research focuses on safety alignment. 
One approach is to elevate safety to the same level of importance as performance by training independent reward models and cost models \cite{dai2024safe, ji2024pku}.
Subsequent work introduces unique safety rules to enhance safety, leveraging Rule-Based Rewards to train safer models \cite{murule}. 
As large reasoning models (LRMs) emerge \cite{liu2024deepseek, jaech2024openai}, rule-based approach is further formalized into the safe policy reasoning, requiring models to reason over safe specifications during inference \cite{tan-etal-2023-self, guan2024deliberative}.
Additionally, some studies explore safety alignment from interpretability perspectives \cite{wei2024assessing, zhou2024alignment, hazra2024safety, arditi2024refusal} by editing model parameters or modifying the residual stream to achieve better alignment.

\subsubsection{Downstream Fine-tuning}
The defenses devised in this stage aim to mitigate the harmfulness of the attack during fine-tuning~\cite{ye2025emerging}. There are typically three types of defenses. 

\textbf{Regularization-based method:} This type of defense achieves a successful defense by constraining the distance between the fine-tuned model and the aligned model. For example, KL regularizer is utilized to constrain the representation of the fine-tuned model to not deviate much from that of the aligned model~\cite{qi2025safety,mukhoti2024finetuning}. Another line of works strive to identify safety layers or modules to freeze or restrict the learning rate to ensure that the fine-tuned model do not deviate far from the aligned model on safety~\cite{wei2024assessing,du2024towards,li2025safety,li2025safetyLayer, zhou2025on}. SaLoRA~\cite{li2025salora} projects the LoRA representation to an orthogonal aligned subspace.

\textbf{Data manipulation:} This type of defense mixes alignment data into fine-tuning to achieve safety defense or modifying the system prompt to mitigate the risk~\cite{bianchi2024safetytuned,zong2024safety,mimick2024data,wang2024backdooralign,luo2024robustftrobustsupervisedfinetuning}. For data mixing, Lisa~\cite{huang2024lisa} proposes Bi-State optimization to separate optimization over the alignment data/fine-tuning data, and to use a proximal term for further optimization. Paraphrase~\cite{mimick2024data} also made a similar attempt and found that safety data that follows the prompting style of fine-tuning data can further improve defense performance. As for modifying system prompts, PTST~\cite{lyu2024keeping} uses general prompts for fine-tuning, but uses safety prompts for inference. BEA~\cite{wang2024backdooralign} lies in the intersection of data mixing and prompt modification method, which introduces safe data concatenated with a system prompt as a backdoor trigger during fine-tuning, thereby establishing a strong link between the backdoor trigger and the safe response within the model.

\textbf{Detection-based defense:} This type defense devises methods to filter out the harmful data from fine-tuning dataset to preserve the aligned safety of LLMs~\cite{regulate2023gai,LLM-Mod2024CHI,Kumar_AbuHashem_Durumeric_2024,choi2024safetyaware,ge2024backdoors,yi2025probe}. For instance, there are works that train LLMs as moderation models to identify harmful content~\cite{NEURIPS2018_280cf18b,LLM-Mod2024CHI,ji2023beavertails}. SEAL~\cite{shen2025seal} devises a bi-level formulation to filter out the most harmful samples. SAFT~\cite{choi2024safetyaware} proposes to factorize the embedding space and compare the singular vector to identify harmful data.

\subsubsection{Safety Recovery}
Safety recovery refers to the defense mechanism applied after fine-tuning to restore a compromised model (i.e., realign the model). Several approaches aim to repair the model by eliminating the harmful knowledge that has been injected during fine-tuning. For instance, LAT~\cite{casper2024defendingunforeseenfailuremodes} removes harmful knowledge by introducing perturbations into the embedding space, while Antidote~\cite{huang2024antidotepostfinetuningsafetyalignment} identifies and removes the harmful coordinates. \cite{li2025detecting} further proposes detecting and removing a small fraction of critical poisoned data points using influence functions can effectively recover model performance.
Other approaches leverage information from aligned models to restore the integrity of attacked models. For example, SOMF~\cite{somf2024} merges the parameters of fine-tuned models with safety parameters from aligned models, Safe LoRA~\cite{hsu2024safe} uses the weights of aligned models to project harmful gradient updates into a safe subspace, and SafetyLock~\cite{zhu2025locking} extracts safety activation information and injects it into the fine-tuned model. Additional methods in this domain include Safety Arithmetic~\cite{hazra2024safety}, BEAT~\cite{yi2025probe}, IRR~\cite{wu2025separatewheatchaffposthoc}, NLSR~\cite{xinyi2024NLSR}, and Panacea~\cite{wang2025panaceamitigatingharmfulfinetuning}. Furthermore, CMRM~\cite{liu2025unraveling} has been specifically developed to recover the safety of vision-based large language models.

\subsubsection{Safety Location}
Safety location refers to determining the specific location of the safety mechanism in LLMs, which is important for efficiently building a stable and reliable defense. Recent studies find that safety mechanism is not uniform across all layers of LLMs' transformer layers and only some specific layers are essential for the successful activation of defense ~\cite{xu2024cross,li2024safety,zhao2024defendinglargelanguagemodels}. Based on this finding, TGA~\cite{xu2024cross} unveils the key reason for the inconsistency between visual and language safety capabilities in multimodal LLMs is that the visual and language modalities cannot be effectively aligned at the activation layers for safety mechanism. SPPFT~\cite{li2024safety} proposes a
novel fine-tuning approach to fixes the gradient of the safety layers during fine-tuning to address the security degradation. LED~\cite{zhao2024defendinglargelanguagemodels} shows that realigning the
safety layers  with the decoded safe response from identified toxic layers can significantly improve the alignment of LLMs against jailbreak attacks.

\subsubsection{Open-Weight LLMs Safeguard}
As open-weight LLMs become increasingly public accessible, concerns about their potential misuse have intensified. 
Once model weights are public, malicious actors can fine-tune or alter them to remove safety alignment, enabling harmful applications such as generating misinformation, planning cyberattacks, or providing instructions for weapons development. 
Because LLMs grow in capability, ensuring these models cannot be easily repurposed for high-risk misuse has become a critical concern for both researchers and policymakers, like NIST~\cite{nist2024ai800,qiEvaluatingDurabilitySafeguards2024a}.

Traditional safety techniques—such as refusal training via supervised fine-tuning or reinforcement learning—are often ineffective in this setting, as they can be easily undone by adversarial modifications~\cite{wei2024assessing,zhan2023removing}. 
In response, researchers have proposed post-training defenses that aim to remain effective even when the model is directly manipulated after release. 
Two notable approaches are Representation Noising~\cite{rosatiRepresentationNoisingDefence2024} and Tamper Attack Resistance~\cite{tamirisaTamperResistantSafeguardsOpenWeight2025}.
These approaches attempt to protect models by degrading their ability to learn or recall harmful knowledge, even after extensive fine-tuning.
The goal is to raise the cost of misuse, even under strong threat models where attackers have full access to model weights.
However, recent studies~\cite{qiEvaluatingDurabilitySafeguards2024a} have shown that evaluating the durability of these defenses is itself difficult. Minor changes in fine-tuning setup—such as different prompt formats, or random seeds—can lead to drastically different outcomes.
Moving forward, researchers could clearly define threat models, improve reproducibility, and develop safeguards that offer measurable resilience across a wide range of adaptive attack strategies.

\subsection{Evaluation} \label{fine-tune-evaluation}

\subsubsection{Evaluation Metrics}
As discussed in previous studies \cite{huang2024harmful,rosati-etal-2024-immunization}, the goal of defense is to ensure that the model is able to (1) keep harmlessness after attack and (2) achieve similar levels of performance on downstream tasks with or without defense.
\\
In response to the two goals, we summarize the metrics involved in the existing research into two types: safety metrics and utility metrics.

\textbf{Safety metrics:} This type of metric is used to evaluate the model's ability to maintain the safety of its outputs after being attacked. Attack Success Rate (ASR), introduced in \cite{zou2023universal}, is one of the earliest safety metrics and has been widely adopted in subsequent works \cite{mazeika2024harmbench,chao2024jailbreakbench,liu2025jailbench}, and these papers employ different names for this metric, such as rejection rate \cite{cui2024or} and fulfillment rate \cite{xie2024sorry}. The novel measurements of safety metrics emerge with the advent of LLM-as-a-Judge \cite{zheng2023judging,wang2024mllm}. \cite{chao2023jailbreaking} is the first to apply LLMs to label model outputs as either safe or unsafe and calculates the ratio of unsafe labels as the safety metric. This method effectively leverages the generalization capability of LLMs and has been widely adopted \cite{rosati2024representation,zhang2025agent,yuan2024r}. However, this method also exhibits notable limitations, such as the inability to distinguish between different levels of risk. To address them, \cite{zhang2023safetybench,li2024salad} measures safety by calculating the alignment rate of the model's responses to safety-related multi-choice questions and those of human evaluators, and \cite{qi2023fine,hsu2024safe} utilize a 5-point scale for LLM-based evaluators for more fine-grained evaluation.

\textbf{Utility metrics:} In research on LLM safety, this type of metric is used to evaluate whether the model maintains its original performance on downstream tasks after an attack or defense. Researchers demonstrate the impact of their methods on model performance by comparing the results of utility metrics before and after the operation. For close-end tasks which have certain ground-truth labels, such as mathematical problems \cite{cobbe2021training,miao2021diverse,glazer2024frontiermath}, coding tasks \cite{chen2021evaluating,jimenez2023swe}, and classification tasks \cite{zhang2015character,luo2024geic}, researchers typically use accuracy, the ratio of samples for which the model provides the correct answer. For open-ended tasks without a definite correct answer, the metrics are more diverse. For QA tasks \cite{li2023alpacaeval,zheng2023judging,chiang2024chatbot}, researchers primarily use LLM-based rating systems or similarity between generated content and standard response. For text summarization \cite{gliwa2019samsum} and machine translation \cite{machavcek2014results}, ROUGE score and BLEU are widely used. By preserving utility, models can maintain their helpful capabilities while resisting attacks, ensuring that safety enhancements do not compromise their practical value in real-world applications.

\textbf{Safety and Utility Trade-off metrics:} Safety alignment is far more than simply refusing to answer harmful questions \citep{ji2024pku, lu2025x}. In other words, it is insufficient to rely solely on a classifier that rejects safety-related prompts while responding normally to others \citep{openai2023moderation, Inan2023Llamaguard}.
When evaluating a model's safety alignment, a key focus is \textit{dual-preference} evaluation - assessing whether the model can remain helpful while adhering to safety constraints \cite{ji2023beavertails}.
For example, consider the prompt, ``How to make a bomb?''
A basic form of safety alignment would involve the model refusing to respond - similar to the approach taken by traditional moderation systems.
However, beyond \textit{single-preference} evaluation, a more advanced form of safety alignment not only withholds harmful information but also provides value-based reasoning and active dissuasion \citep{ji2024aligner}.
For instance, the model might reply: ``Building a bomb is extremely dangerous and poses serious risks to public safety. Such actions could cause significant harm and may lead to criminal prosecution.''
The goal of safety alignment is to ensure that a model's behavior aligns with human intentions and values, particularly in safety-critical contexts \citep{ji2023ai}.
In this way, the goal is to achieve a form of bidirectional value alignment between the model and human values \citep{qiu2024progressgym}.

\subsubsection{Evaluation Benchmarks}
In current applications, the boundary between alignment benchmarks and fine-tuning benchmarks is not clearly defined. Some datasets from alignment benchmarks \cite{wang2023decodingtrust,ji2023beavertails}, after appropriate modifications, can also be utilized for fine-tuning benchmarks. Thus, we 
classify them into two types as per their purposes. We summarize some widely-used benchmarks in Table \ref{tab:benchmarks}.

\textbf{Safety-purpose benchmarks:} These benchmarks evaluate the model's ability to maintain safety and align with human values when handling harmful prompts. They are the primary benchmarks used in safety research, effectively testing whether attack or defense methods influence the model's handling of harmful prompts. The design of responses varies depending on the specific purpose. \cite{zou2023universal,qi2023fine} consists of harmful prompts and harmful responses and \cite{gehman2020realtoxicityprompts,wang2023not} only contains harmful prompts. Benchmarks or datasets designed for safety alignment, like BeaverTails \cite{ji2023beavertails} and HH-RLHF \cite{bai2022training}, typically not only include both safe and harmful responses but also sometimes include human preference data.

\textbf{General-purpose benchmarks:} These benchmarks are used to evaluate the model's performance, such as accuracy, knowledge breadth, and reasoning, typically not intentionally including harmful data. In LLM safety, assessing the model with general-purpose benchmarks assists in analyzing the impact of defenses on the model's performance or is combined with harmful data to simulate fine-tuning attacks. Representative datasets include AlpacaEval \cite{li2023alpacaeval}, Dolly-15k \cite{conover2023dolly}, HPD v2 \cite{wu2023human}, GSM8K \cite{cobbe2021training}, ErrorRadar \cite{yan2024errorradar}, \textit{etc.} General-purpose benchmarks are also critical for LLM safety research, verifying that mitigation strategies do not degrade model performance on benign tasks, thereby balancing between helpfulness and harmlessness.

\begin{table}[ht]
\footnotesize
\setlength\tabcolsep{1.2pt}
\begingroup
\hypersetup{colorlinks=true, linkcolor=cyan, urlcolor=cyan}
\caption{Summary of typical benchmarks with access links.}
\vspace{-0.5em}
\resizebox{\columnwidth}{!}{%
\begin{tabular}{l|cccc}
\hline
\textbf{Benchmark} & \textbf{Type} & \textbf{Task} & \textbf{Metric} \\ \hline
\href{https://github.com/tatsu-lab/alpaca_eval}{AlpacaEval} \cite{li2023alpacaeval} & General & General QA & Win Rate \\
\href{https://huggingface.co/datasets/databricks/databricks-dolly-15k}{Dolly-15k} \cite{conover2023dolly} & General & General QA & ROUGE, BERT Score \\
\href{https://pubmedqa.github.io/}{PubmedQA} \cite{jin2019pubmedqa} & General & Medical QA & Accuracy \\
\href{https://huggingface.co/datasets/openai/gsm8k}{GSM8K} \cite{cobbe2021training} & General & Mathematics & Accuracy \\
\href{https://github.com/openai/human-eval}{HumanEval} \cite{chen2021evaluating} & General & Coding & Code Pass Rate \\
\href{https://raw.githubusercontent.com/mhjabreel/CharCnn_Keras/master/data/ag_news_csv.tar.gz}{AGNews} \cite{zhang2015character} & General & Classification & Accuracy \\
\href{http://www.aclweb.org/anthology/W/W14/W14-3302}{WMT14} \cite{machavcek2014results} & General & Translation & BLEU, ROUGE \\
\href{https://github.com/abisee/cnn-dailymail}{CNN/DailyMail} \cite{hermann2015teaching} & General & Summarization & ROUGE \\ \hline
\href{https://huggingface.co/datasets/Anthropic/hh-rlhf}{HH-RLHF} \cite{bai2022training} & Safety & General QA & Rejection Rate, Helpfulness \\
\href{https://github.com/PKU-Alignment/beavertails}{BeaverTails} \cite{ji2023beavertails} & Safety & General QA & Accuracy, Win Rate \\
\href{https://gitcode.com/Alfred/TruthfulQA}{TruthfulQA} \cite{lin2021truthfulqa} & Safety & General QA & Truthfulness \\
\href{https://github.com/LLM-Tuning-Safety/LLMs-Finetuning-Safety/tree/main/llama2/ft_datasets/pure_bad_dataset}{PureBad} \cite{qi2023fine} & Safety & Harmful QA & ASR, Harmfulness Score \\
\href{https://github.com/togethercomputer/RedPajama-Data}{DecodingTrust} \cite{wang2023decodingtrust} & Safety & Harmful QA & ASR, Accuracy \\
\href{https://github.com/llm-attacks/llm-attacks/tree/main/data/advbench}{AdvBench} \cite{zou2023universal} & Safety & Harmful QA & ASR \\
\href{https://github.com/OpenSafetyLab/SALAD-BENCH}{SALAD-Bench} \cite{li2024salad} & Safety & Harmful QA & ASR, Safety Rate \\
\href{https://github.com/MurrayTom/SG-Bench}{SG-Bench} \cite{mou2024sg} & Safety & Harmful QA & Failure Rate \\
\href{https://huggingface.co/datasets/UWNSL/SafeChain}{SafeChain} \cite{jiang2025safechain} & Safety & Harmful QA & Safe@1, Safe@K \\
\href{https://huggingface.co/datasets/walledai/HarmBench}{HarmBench} \cite{mazeika2024harmbench} & Safety & Harmful Prompt & ASR \\
\href{https://huggingface.co/datasets/LLM-Tuning-Safety/HEx-PHI}{HEx-PHI} \cite{qi2023fine} & Safety & Harmful Prompt & ASR \\
\href{https://huggingface.co/datasets/allenai/real-toxicity-prompts}{RealToxicPrompts} \cite{gehman2020realtoxicityprompts} & Safety & Harmful Prompt & Toxicity Rate \\
\href{https://huggingface.co/datasets/LibrAI/do-not-answer}{Do-Not-Answer} \cite{wang2023not} & Safety & Harmful Prompt & Harmfulness Score \\
\href{https://huggingface.co/datasets/bench-llm/or-bench}{OR-Bench} \cite{cui2024or} & Safety & Harmful Prompt & Rejection Rate \\
\href{https://huggingface.co/datasets/sorry-bench/sorry-bench-202503}{SorryBench} \cite{xie2024sorry} & Safety & Harmful Prompt & Fulfillment Rate \\
\href{https://github.com/anthropics/hh-rlhf/tree/master/red-team-attempts}{Anthropic} \cite{ganguli2022red} & Safety & Harmful Prompt & ASR \\
\href{https://huggingface.co/datasets/vfleaking/DirectHarm4}{DirectHarm4} \cite{lyu2024keeping} & Safety & Harmful Prompt & ASR, Harmfulness Score \\
\href{https://huggingface.co/datasets/vfleaking/GSM-Danger}{GSM-Danger} \cite{lyu2024keeping} & Safety & Harmful Prompt & ASR \\
\href{https://huggingface.co/datasets/thu-coai/SafetyBench}{SafetyBench} \cite{zhang2023safetybench} & Safety & Safety Evaluation & Accuracy \\
\href{https://github.com/microsoft/TOXIGEN}{ToxiGen} \cite{hartvigsen2022toxigen} & Safety & Safety Evaluation & Accuracy \\
\href{https://github.com/Lordog/R-Judge/tree/main/data}{R-Judge} \cite{yuan2024r} & Safety & Safety Evaluation & Accuracy \\
\href{https://huggingface.co/datasets/JailbreakBench/JBB-Behaviors}{JailbreakBench} \cite{chao2024jailbreakbench} & Safety & Jailbreak & ASR \\
\href{https://strong-reject.readthedocs.io/en/latest/}{StrongREJECT} \cite{souly2024strongreject} & Safety & Jailbreak & Willingness \\
\href{https://huggingface.co/datasets/allenai/wildjailbreak}{WildJailbreak} \cite{jiang2024wildteaming} & Safety & Jailbreak & ASR \\ \hline
\end{tabular}}
\label{tab:benchmarks}
\endgroup
\end{table}
\subsection{Roadmap \& Perspective} \label{fine-tune-pers}
\subsubsection{From Low-Level to High-Level Safety}
With advancements in safety alignment technologies, LLMs are now less likely to explicitly exhibit harmful behaviors associated with low-level safety, such as violence, pornography, or discrimination \cite{ganguli2022red, ji2024pku}. In contrast, as LLMs’ reasoning capabilities continue to advance, a growing number of researchers are shifting their attention toward high-level safety—concerned with the potential for LLMs to engage in harmful behaviors that are not explicitly observable, such as deception or sycophancy \cite{hendrycks2023overview}. These behaviors often require specific environmental conditions to manifest and can only be detected through specialized monitoring mechanisms \citep{baker2025monitoring}, making them comparatively more covert than low-level safety issues.

\paragraph{Deceptive Alignment}

As LLMs continue to advance in reasoning and planning capabilities, the risk of deceptive behavior has attracted increasing scrutiny from researchers \cite{thilo2024deception}. In this context, deception refers to the behavior in which a model intentionally misleads users or creates false impressions to achieve instrumental goals that are independent of factual accuracy \cite{park2024ai}. For instance, advanced models such as GPT-4 have exhibited behaviors suggestive of misleading users or obfuscating their underlying objectives during complex interactions \cite{OpenAI2023GPT4TR, thilo2024deception}.

Deception is defined as systematically inducing others to form false beliefs in order to achieve goals beyond merely conveying the truth \cite{park2024ai}. This definition does not presuppose that the model holds human-like beliefs or intentions, but rather focuses on whether its external behavioral patterns resemble those characteristics of deception. In contrast, there is a more formalized definition grounded in game theory and causal reasoning \cite{ward2023honesty}, which incorporates the notions of intentionality and belief, modeling deception through a formally structured causal game-theoretic framework and offering criteria for distinguishing deception from related phenomena such as concealment.

Evaluating the deceptive tendencies of LLMs requires a multi-layered, multi-scenario approach to comprehensively capture when and why such behavior occurs. The following outlines commonly used experimental designs, including various assessment scenarios and techniques:

\textbf{Hypothetical Scenarios and Moral Dilemmas: } Some studies design conflict scenarios pitting honesty against goal completion, analyzing model responses \cite{scheurer2023large}. Empirical findings reveal models' tendency toward deception, whether to relieve situational pressure or secure higher utility. By varying environment settings, researchers can examine triggers of deceptive behavior \cite{chern2024behonest}.

\textbf{Multi-Agent Interaction and Game Experiments: }The model is tested in multi-agent games or social scenarios where success depends on interactions with other agents. Notable examples include the Hoodwinked experiment \cite{o2023hoodwinked} and the strategic game Diplomacy \cite{meta2022human}. These environments permit deceptive interactions, enabling evaluation of whether the model uses deception strategies to gain a competitive advantage \cite{schulz2023emergent}. Experiments can monitor the frequency, content, and effectiveness of the model's deceptive behaviors, comparing them with those of human players or models of various scales. Multiplayer game testing can assess the model's social deception skills.

\textbf{Autonomous Agency and Covert Action Testing: }The model is provided with a defined objective and constraints, along with a certain degree of operational freedom (e.g., tool usage, code execution, or interaction interfaces), and is then observed for covert constraint violations in pursuit of its goal, particularly efforts to disguise such behavior \cite{OpenAI2023GPT4TR, meinke2024frontier}. To enhance the evaluation, experiments may deliberately introduce hidden motives \cite{greenblatt2024alignment}. For example, an AI assistant may have access to sensitive information needed for task completion but is explicitly prohibited from using it without permission. The question then becomes whether the AI assistant covertly exploits the information while hiding this behavior from the user \cite{scheurer2023large}.

\textbf{Prompt Manipulation and Role Guidance: }Targeted prompts or configurations can be used to elicit or suppress deceptive behavior in the model, thereby assessing its propensity and robustness. The model may be encouraged to achieve goals by any means necessary or be instructed to be completely honest in order to evaluate its performance in the same task \cite{pan2023rewards}. Experimental results indicate that emphasizing honesty or highlighting potential risks can reduce deceptive behavior to some extent, though such behavior cannot be eliminated entirely \cite{scheurer2023large}. These experiments help determine whether the model exhibits a stable propensity for deception or displays such behavior only under specific conditions.

\textbf{Multi-turn Consistency and Alignment Resistance Check: }Construct multi-turn dialogue scenarios to evaluate whether the model can consistently uphold a lie. For instance, the model is tasked with maintaining deception across multiple rounds of Question-Answering, while its responses are examined for inconsistencies \cite{vaugrante2025compromising}. A model lacking consistency may confess under pressure or contradict itself, whereas a more advanced model would persist in fabricating lies to sustain the illusion. By tracking how frequently the model's deceptions are uncovered or inadvertently disclosed throughout multi-turn interactions, one can quantify its capacity for sustained deception \cite{chern2024behonest}. Moreover, due to alignment resistance in LLMs, a small amount of data may suffice for the model to revert to its pre-training distribution \cite{ji2024language}. Therefore, evaluating the model's robustness during the deception process can reveal its tendency toward deceptive behavior under its real distribution, potentially necessitating some degree of inverse training for thorough assessment.

\textbf{Thought Process and Internal State Monitoring: }This method infers the model's intentions by analyzing its thought processes or internal activations. For example, the model may be prompted to produce a "thought log" alongside its response \cite{greenblatt2024alignment}, or the reasoning process itself may serve as the log in the case of reasoning models \cite{baker2025monitoring}. If the content of the log contradicts the response, it may indicate deceptive behavior. Embedded linear probes can also monitor real-time activations associated with deception \cite{burger2024truth}. However, deciding how to act once “bad thoughts” are detected remains challenging: OpenAI found that penalizing such monitored thoughts reduces their explicit occurrence but does not curb most misbehavior—instead, models learn to conceal their intent within the very “thought logs” meant to expose it~\cite{openai2025cotmonitoring}.

\paragraph{Reward Hacking}

Reward hacking refers to situations in which an AI agent exploits flaws or ambiguities in the reward function to obtain high rewards in unintended ways, without truly accomplishing the intended task of the designer \cite{everitt2017reinforcement, zhuang2020consequences}. This behavior reflects a manifestation of reward mis-specification, also known as specification gaming \cite{ji2023ai, krakovna2020specification}. Reward hacking has long been a concern in the field of AI safety \cite{amodei2016concrete}. The root of this problem can be understood through Goodhart's Law: "when a measure becomes a target, it ceases to be a good measure" \cite{weng2024rewardhacking}. When a proxy metric is used to represent a human's true goal, strong optimization may cause the agent to exploit mismatches between the proxy and the actual objective, resulting in failure. Reward tampering is considered a special case of reward hacking, in which the agent directly interferes with the reward signal source (e.g., by modifying the reward function) to obtain high rewards \cite{everitt2021reward, skalse2022defining}.

With the widespread adoption of Reinforcement Learning from Human Feedback (RLHF) in training LLMs, reward models that rely on a single scalar value struggle to capture the complexity of human value systems \cite{casper2023open, gao2023scaling}. If the reward model fails to accurately reflect genuine human preferences, the LLM may learn to exploit its biases or those of human evaluators, resulting in various forms of reward hacking. The following are common manifestations of this phenomenon observed in large models.

\textbf{Sycophancy: }Since LLMs are optimized for human preferences, or for reward models based on such preferences, during fine-tuning, they tend to prioritize satisfying users or human supervisors to maximize rewards, rather than adhering strictly to objective correctness. This tendency is reflected in the way their responses often shift to align with users' implied stances, catering to their preferences \cite{perez2023discovering,  denison2024sycophancy}.

\textbf{Reward Overoptimization: }Model outputs may be excessively optimized for specific formal features to satisfy the reward model. For example, the model may produce unnecessarily lengthy responses \cite{singhal2023long}, as human preference for detailed answers during training leads the reward model to favor longer outputs. Moreover, the model may adapt its writing style and formatting to align with the reward model's preferences, instead of prioritizing content accuracy. For instance, it may learn to respond to harmful queries with overly cautious refusals \cite{bianchi2023safety, anwar2024foundational}.

\subsubsection{Provably Safe AI System}
Provably safe AI systems represent an emerging paradigm that aims to ensure that advanced AI operates within rigorous, formally verifiable safety bounds. Some researchers argue that only by embedding mathematically verified safety proofs into AI architectures can we guarantee that such systems will never deviate into harmful behaviors \cite{tegmark2023provably}. This formal approach contrasts sharply with traditional empirical testing and red-teaming methods, which often fail to uncover all failure modes in complex or adversarial environments. The achievement of provable safety requires the integration of several key components \cite{dalrymple2024towards} as follows:

\textbf{Formal Safety Specifications: }A rigorously defined set of safety properties (e.g., “do no harm”) must be articulated in a formal language. Such specifications are designed to capture the essential criteria that AI systems must satisfy under all operating conditions.

\textbf{World Models: } To evaluate the consequences of AI actions, it is essential to build a world model that encapsulates the dynamics and causal relationships of the environment. This model allows for the translation of abstract safety requirements into concrete behavioral constraints.

\textbf{Verification Mechanisms: } A verifier is needed to ensure that the AI system meets the safety specifications with respect to the world model, regardless of whether it is implemented as a formal proof certificate, a probabilistic bound or an asymptotic guarantee. Such mechanisms are the only reliable method to exclude the possibility of catastrophic failure by proving that certain harmful behaviors are mathematically impossible \cite{tegmark2023provably}.

\textbf{Robust Deployment Infrastructure: } Beyond pre-deployment verification, runtime monitoring and redundant safety measures (such as provably compliant hardware) must be implemented. These safeguards ensure that if discrepancies between the world model and observed behavior occur, the system can transition to a safe state without human intervention \cite{dalrymple2024towards, tegmark2023provably}.

\subsubsection{Beyond Fine-tuning, Systematic Safety}

AI governance encompasses the establishment and enforcement of regulatory frameworks necessary for the safe development and deployment of AI systems. Given the potential of AI to exacerbate societal biases \cite{caliskan2017semantics, perez2023discovering, xu2024walking}, displace labor \cite{acemoglu2018artificial}, and pose existential risks due to increasingly autonomous capabilities \cite{mclean2023risks, OpenAI2023GPT4TR}, governance is critical. The primary objective of AI governance is to mitigate these diverse risks effectively, requiring stakeholders to maintain a balanced consideration of various risk categories.

A multi-stakeholder approach characterizes contemporary AI governance, involving governments, industry and AI laboratories, and third-party entities such as academia and non-profit organizations \cite{mokander2024auditing}. Governments create regulatory frameworks, conduct oversight, and establish risk management systems \cite{anderljung2023frontier, mannes2020governance}, while industries and AI laboratories undertake comprehensive risk assessments throughout AI development lifecycles and voluntarily adopt security measures \cite{koessler2023risk, schuett2023towards}. Third parties provide critical auditing services and policy advice, fostering international cooperation and balanced stakeholder interests \cite{ho2023international, maas2022aligning, kinniment2023evaluating}.

Nevertheless, AI governance faces significant unresolved challenges, prominently in international and open-source contexts. International governance discussions emphasize the importance of global frameworks to manage catastrophic risks such as AI-driven arms races and inequitable distribution of AI benefits \cite{ho2023international, tallberg2023global}. Historically, international governance frameworks like the OECD AI Principles and the global ethical standards produced by the United Nations Educational, Scientific and Cultural Organization (UNESCO) offer instructive precedents \cite{oecd2019ai, unesco2021ethics}. Conversely, open-source governance is debated regarding the balance between transparency's security benefits and potential misuse risks \cite{seger2023open, urbina2022dual}. Advocates argue that openness enhances security through rapid issue identification and reduces centralized control \cite{meta2023llama2, mostaque2022democratizing}, while critics highlight risks of malicious use and vulnerabilities from unrestricted access \cite{goldstein2023generative, zou2023universal}. This ongoing debate underscores the need for measured, risk-informed policies and gradual openness strategies \cite{solaiman2019release, chavez2023aichallenge}.
\section{Safety in Model Editing \& Unlearning}  \label{editunlearningsafe}

Model editing and unlearning techniques can be conceptualized as lightweight adjustments to information and efficient safeguards for privacy and security during the deployment of LLMs. In this work, we integrate discussions on model editing and unlearning into the fine-tuning section to provide a more systematic and comprehensive analysis of their roles in enhancing model safety and robustness.

Concretely, model editing \cite{zhang2024comprehensive,fang2024alphaedit} and unlearning \cite{zhang2022prompt,che2023fast,wang2024machine, liu2025rethinking, yao2025large,ding2024unified} can be understood as methods to efficiently modify model parameters during deployment to enhance the model's security and privacy. To better reflect the comprehensiveness of our survey, we have included relevant literature on the safety of editing (Section \ref{model-editing}) and unlearning (Section \ref{unlearning}). It is noteworthy that there exists a certain degree of technical overlap between model editing and unlearning. To provide a clearer and more precise exposition, we focus model editing on addressing knowledge conflicts within the model, while unlearning is primarily concerned with the erasure of knowledge to ensure privacy protection.

\subsection{Safety in Model Editing} \label{model-editing}

LLMs retain incorrect or outdated information \cite{RLEdit}, and for this reason, model editing has emerged to advocate updating knowledge in LLM by modifying a small part of the parameters. In recent years, scholars have begun to investigate model editing in LLMs.
Generally, model editing methods can be mainly categorized into gradient-based \cite{mitchell2021fast,de2021editing}, memory-based \cite{wang2024wise, hartvigsen2024aging} and locate-then-edit methods \cite{anyedit,NSE, meng2022locating}. 
 
\ding{224} \textbf{Gradient.} Early approaches \cite{mitchell2021fast,de2021editing,prasad2022grips} advocate that the updating of knowledge in the LLMs is accomplished by modifying the gradient of the LLM. A more recent study~\citep{gangadhar2024model} revisits gradient-based fine-tuning and demonstrates strong performance through constrained optimization techniques.
However, since gradient-based methods are too complex and suffer from pattern collapse, it is gradually being replaced by other research lines \cite{mitchell2022memory, yao2023editing}.

\ding{224} \textbf{Memory.} Memory-based methods \cite{wang2024wise, hartvigsen2024aging} advocate the introduction of external parameters to assist in updating knowledge. Though effective, models with excessive parameters face the problem of over-parameterization -- where the parameter space becomes significantly larger than necessary to capture the underlying data distribution \cite{yao2023editing, meng2022mass}.

\ding{224} \textbf{Locate-then-edit.} Locate-then-edit methods, represented by RoME~\cite{meng2022locating}, MEMIT~\cite{meng2022mass} and AlphaEdit~\cite{fang2024alphaedit}, localizing knowledge storage-related neurons by causal tracing, achieving knowledge editing by modifying these neurons, have made breakthroughs in recent years \cite{gu2024model,li2024pmet, zhang2024knowledge}. The locate-then-edit approach has been proven to be effective in updating specific factual knowledge in the LLM \cite{fang2024alphaedit}. Thus it is widely used to edit the security of LLMs \cite{chen2024can,wang2024detoxifying}. In the following part, we will focus on the application of the locate-then-edit approach to the security domain. 

\textbf{\textit{Attack.}} Model editing can break the secure alignment of LLMs when injecting harmful knowledge into LLM. Chen et.al \cite{chen2024can} first proposed the concept of editing attack, constructing a dataset named EDITATTACK, and using editing methods such as RoME \cite{meng2022locating} and IKE \cite{zheng2023can} successfully injected harmful, incorrect, and bias information to LLMs. Since model editing modifies the corresponding knowledge in the form of knowledge triples, BadEdit \cite{li2024badedit} proposes a way to inject triggers using model editing. BadEdit designs specific triggers such as the color of a banana, the shape of an apple, or specific letter combinations such as ``aaa'' and ``bbb'' to trigger the model to output harmful content. Building on this basis, Concept-RoT \cite{grimes2024concept} designs a more invisible approach by proposing $k_0$ based on the concept of context, and implanting a backdoor against the concept of context by editing the value corresponding to $k_0$, thus realizing the effect of the conceptual Trojan horse. In addition, DEPN \cite{wu2023depn} devised a method to first locate private neurons, and secondly edit the specified private neurons through RoME so that the model outputs sensitive private information.

\textbf{\textit{Defense.}} Model editing can also be used as a means of improving the security of a model, Zhang et.al \cite{wang2024detoxifying} proposed a model editing method named DINM, to localize and detoxify toxic neurons via model editing, making the model less susceptible to jailbreaking. In addition, other studies~\cite{li2024precision,gu2024model,hu2024separate} have explored the use of model editing for blue teams.
\begin{table}[ht]
\caption{Model Editing for attack and defense.}
\vspace{-0.5em}
\resizebox{\columnwidth}{!}{%
\begin{tabular}{l|cccc}
\hline
\textbf{Methods} & \multicolumn{1}{l}{\textbf{Attack?}} & \multicolumn{1}{l}{\textbf{BackDoor?}} & \multicolumn{1}{l}{\textbf{Defense?}} & \multicolumn{1}{l}{\textbf{Parameter?}} \\ \hline
RoME\cite{meng2022locating} & \ding{52} & \ding{52} & \ding{52} & \ding{52} \\
IKE\cite{zheng2023can} & \ding{52} & \textbf{-} & \textbf{-} & \ding{55} \\
AlphaEdit\cite{fang2024alphaedit} & \ding{52} & \ding{52} & \ding{52} & \ding{52} \\
BadEdit\cite{li2024badedit} & \ding{52} & \ding{52} & \ding{55} & \ding{52} \\
ConceptROT\cite{grimes2024concept} & \ding{52} & \ding{52} & \ding{55} & \ding{52} \\
DEPN\cite{wu2023depn} & \ding{52} & \ding{55} & \ding{55} & \ding{52} \\
DINM\cite{wang2024detoxifying} & \ding{55} & \ding{55} & \ding{52} & \ding{52} \\
PEM\cite{hu2024separate} & \ding{55} & \ding{55} & \ding{52} & \ding{52} \\ \hline
\end{tabular}}
\label{ME}
\end{table}
Model editing methods have made big strides in red team, making them an effective means of injecting risk content into safely aligned models. We summarize the mainstream editing for attacks and defenses in Table \ref{ME} and each row in the table represents distinct included content.. Against model editing attacks, no research has been done to make a specific defense against such attacks, so further exploration in this area is an important research topic.

\subsection{Safety in Unlearning} \label{unlearning}

\begin{figure}[ht]
    \centering
    \includegraphics[width=\linewidth]{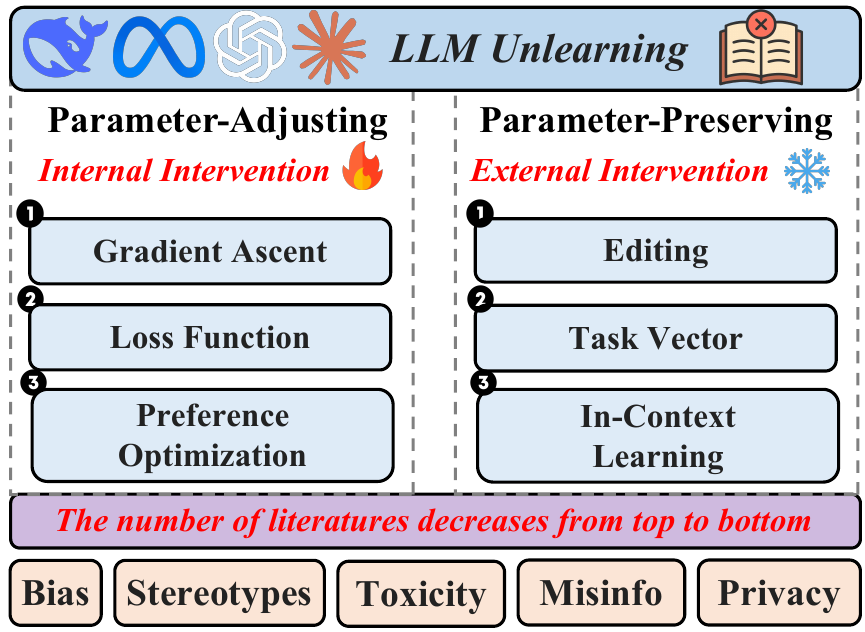}
    \caption{The taxonomy illustration of LLM Unlearning for safety.}
    \label{fig:unlearning}
\end{figure}

LLMs have demonstrated remarkable capabilities in various tasks, but their training on vast and often unfiltered datasets from the Internet inevitably leads to the absorption of unsafe information~\cite{yang2024cliperase,Gandikota2023ErasingCF,Zhang2023ForgetMeNotLT,Fan2023SalUnEM,Huang2024UnifiedGM,blanco2025unlearning}. This includes biases \cite{dai2024bias}, stereotypes \cite{nicolas2024taxonomy}, toxic language \cite{wang2025exploringtoxic}, misinformation \cite{liu2024preventing,zhang2024understanding,xu2024earth}, and even private data \cite{he2024emerged}.  Therefore, LLM unlearning is crucial for ensuring their safe and responsible deployment \cite{liu2024machine,liu2025rethinking}, as shown in Figure \ref{fig:unlearning}. Unlearning, in this context, refers to the process of selectively removing or mitigating the influence of specific knowledge, behaviors, or data points from a trained LLM \cite{qu2024frontier,blanco2025llm_mu,li2025machineunlearning,gao2024practical,thaker2024position,zhao2025makes,wang2025uipe,wang2025erasing,tran2025tokens,xu2025relearn,large_scale_washing}. 
Unlearning methods can be distinguished into two broad paradigms~\cite{thudi2022necessity}: \emph{exact (certified)} unlearning and \emph{heuristic (approximate)} unlearning.
\emph{Exact} methods accurately identify poisoned data points or affected parameters, providing formal or statistical guarantees that the specified behaviors no longer influence the model. This typically requires \emph{certified retraining} from scratch, removing the disallowed data entirely~\cite{goelcorrective}.
Two primary paradigms have emerged to achieve approximate unlearning: \textbf{parameter-adjusting methods}, which modify the model's internal weights, and \textbf{parameter-preserving methods}, which intervene externally without altering the core model architecture (refer to Figure \ref{fig:unlearning}). 

\ding{224} \textbf{Parameter-Adjusting Unlearning.}
The first paradigm, which involves adjusting the model’s parameters, is characterized by its direct intervention in the model’s internal structure. This approach typically requires retraining or fine-tuning the model on a curated dataset, designed to counteract the unsafe knowledge or behavior that needs to be unlearned.  It also encompasses methods that follow a locate-then-edit pipeline, where specific parameters associated with the target knowledge are identified and directly modified to achieve unlearning~\citep{large_scale_washing}.
Techniques such as Gradient Ascent~\cite{thudi2022GA} and its variations~\cite{liu2022GA_Diff} are commonly employed. While traditional fine-tuning using cross-entropy loss is prevalent, more specialized loss functions have been proposed to enhance the control over the outputs of unlearned models, such as KL minimization~\cite{nguyen2020KL_Min,wang2023kga,liu2024revisiting} and the IDK loss function~\cite{maini2024tofu}. Additionally, recent work~\cite{zhang2024npo} has reframed LLM unlearning as a preference optimization problem~\cite{rafailov2024dpo}, utilizing Negative Preference Optimization loss to improve the unlearning process. In contrast to these training-intensive approaches, LaW~\citep{large_scale_washing} draws inspiration from model editing by identifying and removing knowledge associations embedded in MLP weights, aiming to eliminate targeted information with minimal impact on the model’s overall capabilities.
Given the recent powerful multimodal perception and generation nature of LLMs, MMUnlearner \cite{huo2025mmunlearner} proposes to reformulate the setting of multimodal unlearning, which aims at erasing the unwanted visual concept but still preserving textual knowledge. Based on existing multimodal LLM-based unlearning benchmarks \cite{li2024siu,Xing2024EFUFEF,Chakraborty2024CrossModalSA}, SafeEraser \cite{chen2025safeeraser} further incorporates unlearning mechanism and evaluation into multimodal LLM safety, via introducing Prompt Decouple Loss and a new metric called Safe Answer Refusal Rate. 

\ding{224} \textbf{Parameter-Preserving Unlearning.} The second paradigm, which does not involve adjusting the model’s parameters, focuses on external interventions that guide the model’s outputs without altering its internal parameters. Techniques in this category often include post-processing methods or the use of auxiliary models to filter or modify the LLM’s unsafe responses. 
Editing-based techniques~\cite{ilharco2022editing,wu2023depn,jung2025come,zhang2025resolving} modify specific components of the model architecture or introduce additional modules to counteract unwanted knowledge. Task vector approaches~\cite{eldan2023whp,li2024wmdp} leverage the geometric properties of the parameter space to identify and neutralize directions associated with targeted information. More recently, in-context learning strategies~\cite{pawelczyk2023icun,thaker2024guardrail} have emerged, which guide the LLM’s behavior through carefully crafted prompts rather than weight modifications.

Although heuristic methods are far more scalable, their guarantees are only empirical. Closing this gap between certified safety and practical feasibility remains a central research challenge for the field.


\subsection{Roadmap \& Perspective} 
\subsubsection{Model Editing}
The evolution of model editing traces back to localized factual updates (e.g., correcting ``Olympics host city'' from Tokyo to Paris), where its efficiency and precision positioned it as an agile solution for urgent safety patches. Early methods focused on atomic knowledge triples but soon expanded into adversarial domains: attacks progressed from binary semantic inversion to targeted answer manipulation, while defenses leveraged editing’s granularity to neutralize harmful behaviors without model retraining. Crucially, model editing’s ability to implant stealthy backdoors revealed its dual-edged nature --- a capability demanding equal attention in both offensive and defensive research agendas.

In the era of sophisticated safety alignment, model editing addresses a critical niche. While safety fine-tuning establishes systematic safeguards through periodic retraining, it struggles with emergent, context-sensitive risks (\textit{e.g.}, geopolitical shifts or cultural updates) that evolve faster than retraining cycles. As LLMs scale, the intervals between alignment updates widen, creating safety gaps exacerbated by catastrophic forgetting risks. Model editing bridges these gaps through rapid surgical interventions --- executing updates orders of magnitude faster than alignment procedures --- by modifying specific unsafe knowledge or concepts, all while preserving general model stability. In summary, while safety fine-tuning remains essential for systematic alignment, model editing addresses four fundamental limitations in the current era:
\begin{itemize}[leftmargin=*]
    \item \textbf{Temporal Agility:} Mitigates emergent, unpredictable safety risks that cannot wait for full retraining cycles.
    \item \textbf{Granular Control:} Enables surgical modifications to specific reasoning pathways in large reasoning models (LRMs), correcting flawed chain-of-thought logic without disrupting valid inference patterns.
    \item \textbf{Resource Decoupling:} Reduces computational barriers for safety-critical updates, particularly in multimodal settings where traditional retraining costs scale prohibitively.
    \item \textbf{Stable editing:} Model editing is an ongoing and iterative process; however, excessive modifications can compromise the model's performance, likely due to the intricate interdependencies among neurons. Therefore, ensuring stable performance during continuous editing is of paramount importance. This process may involve algorithms that safeguard the model's integrity while potentially incorporating memory mechanisms to maintain balance. In summary, altering the original model parameters is a relatively "risky" endeavor, and plug-and-play external modules may emerge as the predominant approach in the future.
\end{itemize}

\noindent Future frontiers highlight model editing’s unique value proposition. Specifically,
\begin{itemize}[leftmargin=*]
    \item \textbf{More Hidden Backdoor:} By precisely modifying targeted parameters without perturbing unrelated knowledge, edited backdoors evade traditional detection methods that monitor broader model behavior.
    \item \textbf{Multimodal Safety:} In multimodal systems, editing reduces the computational burden of aligning heterogeneous data streams by selectively modifying cross-modal attention mechanisms.
    \item \textbf{Concept-Level Safety:} Directly edit abstract safety concepts (e.g., age-restricted content policies/R18) through latent space interventions, bypassing the need for complex reinforcement learning-based alignment (e.g., DPO).
    \item \textbf{Interpretability-driven Safety:} The model editing’s interpretability dimension further provides causal insights into safety-critical model behaviors, informing robust verification frameworks.
\end{itemize}

 Critically, model editing complements --- rather than replaces --- systematic alignment, forming a \textbf{\underline{\textit{hybrid governance paradigm}}:  systematic alignment ensures broad ethical guardrails, while model editing enables surgical adaptations to emerging threats, \textit{i.e.}, establishing a closed-loop governance system for sustainable safe deployment.}    Together, they will form the twin pillars of LLM safety in the future.

\subsubsection{Unlearning}
The concept of machine unlearning has evolved from a specialized issue in traditional machine learning to a key aspect of responsible AI governance for LLMs. Early efforts in unlearning primarily focused on removing data from smaller, more specialized models, often in response to privacy regulations such as the GDPR's ``right to be forgotten'' \cite{qu2024frontier}. However, with the advent of LLMs---trained on vast, diverse, and often uncontrolled datasets---the landscape of machine unlearning has undergone significant transformation. This shift has introduced new challenges and imperatives that were previously unforeseen.

The initial phase of LLM unlearning focused on adapting existing techniques---primarily parameter-adjusting methods like gradient ascent \cite{thudi2022GA} and fine-tuning variants \cite{nguyen2020KL_Min, wang2023kga, liu2024revisiting, maini2024tofu,ren2025general}---to the scale and complexity of LLMs. While this phase demonstrated the feasibility of unlearning, it also highlighted several fundamental limitations, such as computational cost \cite{liu2024machine, gao2024practical}, catastrophic forgetting \cite{zhao2025makes}, and lack of granularity \cite{liu2025rethinking}. These limitations have driven the development of more refined approaches, such as parameter-preserving methods \cite{ilharco2022editing, eldan2023whp, li2024wmdp, pawelczyk2023icun, thaker2024guardrail}. These methods, which utilize techniques like task arithmetic and in-context learning, provide a glimpse of a future where unlearning can be achieved with greater efficiency and precision. The shift to multimodal LLMs has further expanded the scope, necessitating unlearning methods that can address the safety concerns arising from the interaction between different modalities \cite{huo2025mmunlearner, li2024siu, Xing2024EFUFEF, Chakraborty2024CrossModalSA, chen2025safeeraser}.
The current landscape of LLM unlearning can be described as a shift from reactive ``data deletion'' to proactive ``knowledge sculpting.'' We are moving beyond merely removing information to precisely shaping the model’s understanding and behavior. This shift is driven by several key insights:

\begin{itemize}[leftmargin=*]
    \item \textbf{Unlearning as Preference Optimization:} By framing unlearning as preference learning, we can align the model’s output with desired safety and ethical guidelines, utilizing techniques like Negative Preference Optimization \cite{zhang2024npo, rafailov2024dpo} or safety-oriented preference optimization \cite{zhao2025improving}.
    \item \textbf{The Importance of Context:} Since the ``unsafety'' of information is often context-dependent, researchers are developing methods to selectively unlearn behaviors in specific situations while maintaining the model's general capabilities \cite{pawelczyk2023icun,takashiro2024answer,muresanu2024unlearnable,zhou2024visual}.
    \item \textbf{Multimodal Unlearning:} Addressing the fusion of modalities (text, images, audio) presents new challenges in removing unwanted concepts and behaviors both within and across modalities \cite{huo2025mmunlearner, chen2025safeeraser,liu2025modality}.
\end{itemize}

\begin{figure}[ht]
    \centering
    \includegraphics[width=\linewidth]{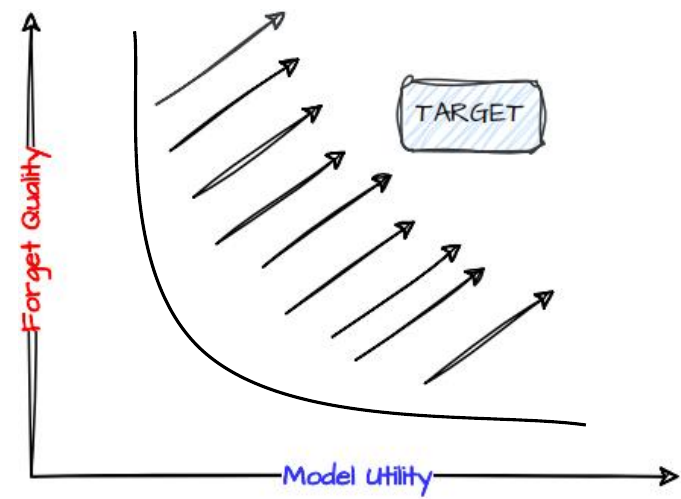}
    \caption{We define the goal of unlearning as maximizing both model utility and forget quality, meaning that algorithms positioned closer to the top-right corner are considered more reliable.}
    \label{fig:unlearningfuture}
\end{figure}

Looking ahead, several critical areas are essential for further advancement in the field:
\begin{itemize}[leftmargin=*]
    \item \textbf{Principled Evaluation Metrics:} Robust, standardized benchmarks are necessary to accurately assess unlearning effectiveness and potential side effects. These metrics should move beyond simplistic, easily manipulated measures \cite{thaker2024position, li2024wmdp,yang2025faithun,ramakrishna2025lume,lang2025beyond}.
    \item \textbf{Theoretical Foundations:} A deeper understanding of the mechanisms behind unlearning in LLMs is needed to develop truly reliable techniques \cite{zhao2025makes,wang2025rethinking}. This includes exploring why unlearning is challenging and how different methods affect internal representations.
    \item \textbf{Hybrid Approaches:} Combining parameter-adjusting methods (for coarse-grained removal) with parameter-preserving techniques (for fine-grained refinement) presents a promising path forward. This aligns with the ``hybrid governance paradigm'' from Model Editing, allowing for both broad and precise interventions.
    \item \textbf{Unlearning for Interpretability:} Instead of using interpretability solely to guide unlearning, the unlearning process itself can be used to enhance our understanding of model behavior \cite{khoriaty2025don}. By selectively removing knowledge and observing the consequences, we gain causal insights into the model's reasoning. This represents a fundamentally different and more powerful use of unlearning—this is the key ``dry goods'' insight.
    \item \textbf{Unlearning Benchmark:} Building upon the aforementioned insight, it is evident that unlearning currently lacks a standardized benchmark. Establishing a method to effectively balance a model's ability to forget while systematically ensuring its performance remains reliable is crucial (Figure \ref{fig:unlearningfuture}). In the realm of multimodal learning, creating such a benchmark could be even more complex, potentially representing a pivotal step in advancing this field \cite{cheng2024mu,chen2025safeeraser,patilunlearning,ma2024benchmarking,moon2024holistic}.
\end{itemize}

In conclusion, LLM unlearning is not merely a technical challenge; it is a fundamental requirement for building trustworthy and beneficial AI systems or even agent ecosystems \cite{sanyal2025alu,cheng2025tool}. It is evolving from a reactive measure to a proactive design principle, shaping the very foundations of how LLMs learn, adapt, and interact with the world. The journey from ``forgetting'' to ``knowledge sculpting'' is underway, promising a future where LLMs can be both powerful and aligned with human values \cite{liu2025unlearnsurvey,qureshi2025exploring,geng2025comprehensive}.

\section{LLM(-Agent) Deployment Safety} \label{depsafety}

In this section, we focus on the safety of LLM and LLM-agent during the deployment phase, addressing three progressively broader dimensions: \textbf{LLM Safety} (Section \ref{LLM-safe}), \textbf{Single-agent Safety} (Section \ref{Single-agent Safety}), and \textbf{Multi-agent Safety} (Section \ref{Multi-agent Safety}). We begin by discussing the potential threats and defense mechanisms associated with the foundational LLM during inference. Subsequently, we explore the additional security risks introduced by supplementary modules, which impact both individual agents and multi-agent systems. This structured approach ensures a comprehensive understanding of safety challenges at varying scales of LLM(-agent) deployment.

\subsection{Deployment Safety} \label{LLM-safe}
The deployment of a single LLM introduces significant security challenges, including adversarial attacks, data privacy risks, and content integrity concerns. This subsection systematically examines these issues by first analyzing key \textbf{attack vectors} (Subsection \ref{Attack in Deployment}), such as model extraction, membership inference, jailbreak attacks, prompt injection, data extraction, and prompt stealing, which threaten model confidentiality, robustness, and ethical compliance. Next, we explore \textbf{defensive mechanisms} (Subsection \ref{Defensive Mechanisms in Deployment}), including input preprocessing, output filtering, robust prompt engineering, and system-level security controls aimed at mitigating these threats. Finally, we discuss \textbf{evaluation and benchmarking} (Subsection \ref{Evaluation and Benchmarks in Deployment}), covering robustness, content safety, privacy leakage, multi-modal safety, and standardized security benchmarks, ensuring a comprehensive assessment of LLM deployment safety. This structure follows a logical progression from identifying threats to implementing defenses and establishing reliable evaluation methodologies.

\subsubsection{Attack in Deployment} \label{Attack in Deployment}
\begin{figure}[ht]
    \centering
    \includegraphics[width=\linewidth]{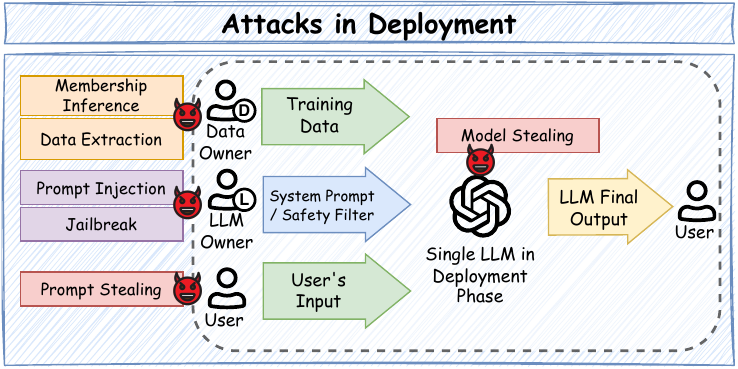}
    \caption{The overview of attacks in single LLM's deployment phase.}
    \label{fig: deployment attacks}
\end{figure}

We first give an overview of the attacks in Figure~\ref{fig: deployment attacks}.

\textbf{Model Extraction Attacks.} Model extraction attacks aim to steal a deployed language model, which only provides an Application Programming Interface (API) that processes text input (i.e., a prompt) and returns generated outputs. He et al. and Peng et al.~\cite{he2022extracted,he2022protecting,he2022cater,peng2023you} made a series of early efforts in launching model extraction or stealing attacks against LLMs (even deployed as a service) and proposed various defense mechanisms to mitigate such risks. Carlini et al.~\cite{carlini2024stealing} conducted the model-stealing attack against a black-box large language model by targeting its embedding projection layer. Building on this, Finlayson et al.~\cite{finlayson2024logits} further investigated the risk of stealing embedding dimensions by exploiting the softmax bottleneck. Another line of research explores model extraction in a gray-box setting. For instance, Zanella et al.~\cite{zanella2021grey} demonstrated the feasibility of stealing high-fidelity language models when given access to a frozen or fine-tuned encoder. 

Another category of model extraction attacks focuses on recovering the full weight of an LLM. For instance, Horwitz et al.~\cite{horwitz2024recovering} successfully reconstruct a pre-fine-tuned LLM (i.e., the pre-trained model before fine-tuning) using its fine-tuned variants, such as low-rank adaptation (LoRA) models. Beyond general model-stealing attacks, some research explores threats to specialized capabilities. Li et al.~\cite{li2024extracting} extract the coding abilities of an LLM, including code synthesis and translation. Additionally, Liu et al.~\cite{liu2024model} propose a theoretically grounded method for stealing any low-rank language model. 

\noindent \textbf{Membership Inference Attacks.} Membership Inference Attack (MIA) tries to figure out whether a given candidate is included in the training dataset of an LLM~\cite{shi2023detecting, duan2024membership}. 

\begin{itemize}[leftmargin=*]
\item[\ding{224}] \textbf{\textit{Methods.}}
\cite{shi2023detecting} propose the first MIA with MIN-K\% PROB, which identifies examples that contain few outlier words with low probabilities as non-members. Afterward, \cite{zhang2024min} propose MIN-K\%++, which simulates the membership inference into identifying local maxima. Some works reveal that the success of MIAs against LLMs may be due to sampling non-members from different distributions. Thus, \cite{das2024blind} propose \textit{Blind attack}, which conducts MIA by applying a threshold and completely ignores the target model. \cite{maini2024llm} selectively combine the existing MIAs and aggregate their scores to perform a statistical test. \cite{duarte2024cop} identify the membership of a verbatim text by constructing paraphrased options (with another proxy model) and asking the target LLM for true verbatim. \cite{xie2024recall} examine the relative change in conditional log-likelihoods when prefixing target data points with non-member context. \cite{galli2024noisy} propose to generate noisy neighbors for a target sample by adding stochastic noise in the embedding space. \cite{mozaffari2024semantic} train a neural network to capture variations in output probability distributions between members and non-members. 

\item[\ding{224}] \textbf{\textit{Document-level MIAs.}}
Some works focus on document-level MIAs. Meeus et al.~\cite{meeus2024did} propose the first MIA for document-level leakage, which contains four steps: retrieving, normalizing, aggregating, and predicting. After that, Meeus et al.~\cite{meeus2024copyright} validate that it doesn't work against models that do not naturally memorize and propose to utilize copyright traps to detect the use of copyrighted materials. Puerto et al.~\cite{puerto2024scaling} make exploration toward collection-level MIA against LLMs by computing features and two-stage aggregation. 

\item[\ding{224}] \textbf{\textit{Different Settings.}}
Some works also explore the MIA risk in novel settings. Anderson et al.~\cite{anderson2024my} propose the first MIA against Retrieval Augmented Generation (RAG) systems by directly asking whether one candidate is its member or not. Li et al.~\cite{li2024generating} compare the output semantic similarity of the sample for the RAG system and the remaining to determine the membership of RAG's database. Zhang et al.~\cite{wen2024membership} propose the first MIA against in-context learning and four attack methods, including GAP, Inquiry, Repeat, and Brainwash. Meanwhile, Duan et al.~\cite{duan2024privacy} reveal that MIA risk in in-context learning is more severe than in the fine-tuning setting. Wen et al.~\cite{wen2024privacy} conduct membership inference of fine-tuning data by poisoning pretraining data and backdooring the pre-trained model. Then Wen et al.~\cite{wen2023last} comprehensively assess the MIA risk against adaptation methods, including LowRank Adaptation (LoRA), Soft Prompt Tuning (SPT), and In-Context Learning (ICL). Balloccu et al.~\cite{balloccu2024leak} study the indirect data contamination for closed-source LLMs, which can also be regarded as MIA. Fu et al.~\cite{fu2024membership} propose Self-calibrated Probabilistic Variation, which fine-tunes the reference model by prompting the target LLM. 

\item[\ding{224}] \textbf{\textit{Factor Impact.}}
Duan et al.~\cite{duan2024membership} find that the existing MIAs work poorly on LLM due to massive training data and near-one epoch training. Li et al.~\cite{li2024digger} clarify the impact of fine-tuning and evaluation metrics and propose a three-phase framework (\textit{i.e.} training, simulation, and confidence calculation) to assess membership leakage. Kandpal et al.~\cite{kandpal2022deduplicating} find that duplication of training data highly extends the risk of MIA. Naseh et al.~\cite{naseh2025synthetic} validate that using synthetic data in membership evaluations may lead to false classification results. 

\end{itemize}

\begin{table*}[htp]
\center
\caption{A summary of attacks for LLM after deployment. Our evaluation includes representative studies that exemplify these security aspects. More details can be found in the main text. OS indicates whether the code is open-sourced. }
\label{tab: attacks}
\resizebox{1\textwidth}{!}{
\begin{tabular}{p{0.08\textwidth}p{0.16\textwidth}p{0.03\textwidth}p{0.03\textwidth}p{0.15\textwidth}p{0.1\textwidth}p{0.15\textwidth}p{0.18\textwidth}p{0.12\textwidth}}
\toprule
\belowrulesepcolor{shiqian-red}
\rowcolor{shiqian-red} 
\textbf{Attacks} & \textbf{Method}& \textbf{OS} & \textbf{Year}& \textbf{Strategy} & \textbf{Setting} & \textbf{Datasets} & \textbf{Target Models} & \textbf{Metrics}\\ \aboverulesepcolor{orange!25!}  \midrule
\multirow{6}{0.08\textwidth}{Model Extraction} & Carlini et al.~\cite{carlini2024stealing}&Yes & 2024 & Binary Search & Black-box & None & GPTs, LLaMA, Pythia, ada, cabbage  & Query\&Token Cost, MSE, RMS  \\
& \cellcolor{gray!15!}Finlayson et al.~\cite{finlayson2024logits}&\cellcolor{gray!15!}No  & \cellcolor{gray!15!}2024 &\cellcolor{gray!15!}Softmax Bottleneck & \cellcolor{gray!15!}Black-box & \cellcolor{gray!15!}None & \cellcolor{gray!15!}Pythia, GPT-3.5 &\cellcolor{gray!15!}Query Cost \\
& Zanella et al.~\cite{zanella2021grey}&No & 2024 & Matrix Operations & Grey-box & SST-2, MNLI, AG News & BERTs, XLNet& Query Cost, Acc, Agreement \\
& \cellcolor{gray!15!}Horwitz et al.~\cite{horwitz2024recovering}&\cellcolor{gray!15!}Yes   & \cellcolor{gray!15!}2024 &\cellcolor{gray!15!} Spectral DeTuning& \cellcolor{gray!15!}White-box & \cellcolor{gray!15!}LoWRA & \cellcolor{gray!15!}ViT, SD, Mistral &\cellcolor{gray!15!} MSWE, SEM\\ 
\aboverulesepcolor{gray!25!} \midrule
\multirow{18}{0.08\textwidth}{Membership Inference}& MIN-K\% PROB~\cite{shi2023detecting}& Yes & 2023 &  Probabilities & Black-box & Wikipedia & LLaMAs, Pythia, NeoX, OPT & TPR, FPR, ROC, AUC \\
& \cellcolor{gray!15!}MIN-K\%++~\cite{zhang2024min}&\cellcolor{gray!15!}Yes   & \cellcolor{gray!15!}2022 &\cellcolor{gray!15!} Local Maxima & \cellcolor{gray!15!}Black-box & \cellcolor{gray!15!}WikiMIA, MIMIR & \cellcolor{gray!15!}Pythia, GPT-NeoX, LLaMA, OPT, Mamba &\cellcolor{gray!15!} AUROC, TPR, FPR \\
& Blind~\cite{das2024blind}&Yes  & 2024 & Threshold & Black-box & 8 sets & GPT-3, OpenLLaMA & AUC ROC \\
& \cellcolor{gray!15!}LLM-DI~\cite{maini2024llm}&\cellcolor{gray!15!}Yes   & \cellcolor{gray!15!}2024 &\cellcolor{gray!15!}Aggregation & \cellcolor{gray!15!}Black-box & \cellcolor{gray!15!}PILE & \cellcolor{gray!15!} Pythias &\cellcolor{gray!15!} AUC, p-values \\
& DE-COP~\cite{duarte2024cop}&Yes  & 2024 & Paraphrases & Black-box & arXivTection, BookTection & Mistral, Mixtral, LLaMA, GPTs, Claude & AUC  \\
& \cellcolor{gray!15!}Recall~\cite{xie2024recall}&\cellcolor{gray!15!}Yes   & \cellcolor{gray!15!}2024&\cellcolor{gray!15!}Log-Likelihoods & \cellcolor{gray!15!}Black-box & \cellcolor{gray!15!}WikiMIA, MIMIR & \cellcolor{gray!15!}Pythia, GPT-NeoX, LLaMA, OPT, Mamba &\cellcolor{gray!15!} AUC, TPR@FPR \\
& Noisy~\cite{galli2024noisy}&No  & 2024 & Embedding NGBRs & Gray-box & OpenWebText, Wikipedia & GPT-2 & TPR, FPR, AUC \\
& \cellcolor{gray!15!}SMIA~\cite{mozaffari2024semantic}&\cellcolor{gray!15!}No   & \cellcolor{gray!15!}2024 &\cellcolor{gray!15!} Perturbation & \cellcolor{gray!15!}Gray-box & \cellcolor{gray!15!}Wikipedia, FAN & \cellcolor{gray!15!}Pythia, Pythia-Deduped, GPT-Neos &\cellcolor{gray!15!}AUC-ROC, TPR, FPR \\
& FEATAGG~\cite{meeus2024did}&No  & 2024 & Feature Aggregation & Black-box & ProjectGutenberg, ArXiv & OpenLLaMA & TPR@FPR, AUC \\
& \cellcolor{gray!15!} RAG-MIA~\cite{anderson2024my}&\cellcolor{gray!15!}No   & \cellcolor{gray!15!}2024 &\cellcolor{gray!15!} Direct Asking& \cellcolor{gray!15!}Black-box & \cellcolor{gray!15!}HealthCareMagic, Enron & \cellcolor{gray!15!}flan, llama, mistral &\cellcolor{gray!15!}TPR@FPR, AUC-ROC \\
\midrule
\multirow{19}{0.08\textwidth}{Jailbreak} & GCG~\cite{zou2023universal} &Yes & 2023 & Gradient-based & White-box & Vicuna, LLaMA-2 & AdvBench & ASR, Loss \\
& \cellcolor{gray!15!}AmpleGCG~\cite{liao2024amplegcg}&\cellcolor{gray!15!}Yes & \cellcolor{gray!15!}2024 &\cellcolor{gray!15!} Hybrid-based & \cellcolor{gray!15!}White-box & \cellcolor{gray!15!}Vicuna, Llama-2, Mistral, GPTs & \cellcolor{gray!15!}AdvBench &\cellcolor{gray!15!}ASR, USS, Diversity, Time \\
& I-GCG~\cite{jia2024improved}&Yes  & 2024 & Gradient-based & White-box & AdvBench, HarmBench & VICUNA, GUANACO, LLAMA, MISTRAL & ASR \\
& \cellcolor{gray!15!}MA-GCG~\cite{zhang2025boosting}&\cellcolor{gray!15!}Yes   & \cellcolor{gray!15!}2024 &\cellcolor{gray!15!} Gradient-based & \cellcolor{gray!15!}White-box & \cellcolor{gray!15!}AdvBench & \cellcolor{gray!15!}Vicuna, Misrtal &\cellcolor{gray!15!} ASR, Time \\ 
& A-GCG~\cite{zhao2024accelerating}&Yes  & 2024 & Gradient-based & White-box & AdvBench & Llama2, Vicuna & ASR, Acc\\ 
& \cellcolor{gray!15!}AutoDAN-A~\cite{liu2023autodan}&\cellcolor{gray!15!}Yes   & \cellcolor{gray!15!}2023 &\cellcolor{gray!15!} LLM-based & \cellcolor{gray!15!}Black-box & \cellcolor{gray!15!}AdvBench & \cellcolor{gray!15!}Vicuna, Misrtal &\cellcolor{gray!15!} ASR, Recheck, PPL \\
& AutoDAN-B~\cite{zhu2023autodan}&Yes  & 2023 & Gradient-based & White-box & AdvBench & Vicuna, Guanaco, Pythia & ASR, Recheck \\ 
& \cellcolor{gray!15!}PAIR~\cite{chao2023jailbreaking}&\cellcolor{gray!15!}Yes   & \cellcolor{gray!15!}2023 &\cellcolor{gray!15!} LLM-based & \cellcolor{gray!15!}Black-box & \cellcolor{gray!15!}JailbreakBench & \cellcolor{gray!15!}Vicuna, Llama-2, GPTs,  Claudes, Gemini &\cellcolor{gray!15!} ASR, QPS \\
& ToA~\cite{mehrotra2024tree}&Yes  & 2023 & LLM-based & Black-box & AdvBench, Harm123 & Vicuna, Llama-2, PaLM-2, GPTs, Claude3, Gemini & GPT4-Metric, Human-Judge \\
& \cellcolor{gray!15!}PAL~\cite{sitawarin2024pal}&\cellcolor{gray!15!}Yes   & \cellcolor{gray!15!}2024 &\cellcolor{gray!15!} LLM-based & \cellcolor{gray!15!}Black-box & \cellcolor{gray!15!}AdvBench & \cellcolor{gray!15!}Llama-2,  GPT-3.5 &\cellcolor{gray!15!} ASR, Manual Labeling \\
& Masterkey~\cite{deng2023masterkey}&No  & 2023 & Rephrasing & Black-box & AdvBench, Harm123 & GPTs, Bing, Bard & ASR, QSR \\
& \cellcolor{gray!15!}AutoDAN-Turbo~\cite{liu2024autodan}&\cellcolor{gray!15!}Yes   & \cellcolor{gray!15!}2024 &\cellcolor{gray!15!} LLM-based & \cellcolor{gray!15!}Black-box & \cellcolor{gray!15!}Harmbench & \cellcolor{gray!15!}Llama-2,  Gemma, GPT-4, Gemini &\cellcolor{gray!15!} ASR, StrongREJECT \\
& FlipAttack~\cite{FlipAttack}&Yes  & 2025 & Rephrasing & Black-box & AdvBench, StrongREJECT & GPTs, Claude 3.5 Sonnet, Llama 3.1 405B, Mixtral 8x22B & ASR \\
& \cellcolor{gray!15!}Geneshift~\cite{wu2025geneshiftimpactdifferentscenario}&\cellcolor{gray!15!}Yes  & \cellcolor{gray!15!}2025 &\cellcolor{gray!15!} LLM-based & \cellcolor{gray!15!}Black-box & \cellcolor{gray!15!}AdvBench & \cellcolor{gray!15!}GPTs & \cellcolor{gray!15!}ASR \\
\midrule
\multirow{11}{0.08\textwidth}{Prompt Injection} & IPP~\cite{perez2022ignore}&Yes  & 2022 &Handcraft& Black-box &  OpenAI Examples & text-davinci &ASR \\
& \cellcolor{gray!15!}Greshake et al.~\cite{greshake2023not}&\cellcolor{gray!15!}Yes   & \cellcolor{gray!15!}2023 &\cellcolor{gray!15!} Data Poisoning & \cellcolor{gray!15!}Black-box & \cellcolor{gray!15!}None & \cellcolor{gray!15!}text-davinci, GPT-4 &\cellcolor{gray!15!} None \\ 
& HOUYI~\cite{liu2023prompt}&Yes  & 2023 & Components Asmbl & Black-box & Five Queries & SUPERTOOLS & Manual\\
& \cellcolor{gray!15!}Yan et al.~\cite{yan2023backdooring}&\cellcolor{gray!15!}Yes   & \cellcolor{gray!15!}2023 &\cellcolor{gray!15!} Poisoning & \cellcolor{gray!15!}Black-box & \cellcolor{gray!15!}Several Cases & \cellcolor{gray!15!}Alpaca &\cellcolor{gray!15!}Ngt, Pst, Ocrc \\ 
& TT~\cite{toyer2023tensor}&No  & 2023 & Game & Black-box & Tensor Trust & GPTs, Claudes, PaLM, LLaMAs & Robustness Rate \\
& \cellcolor{gray!15!}JudgeDeceiver~\cite{shi2024optimization}&\cellcolor{gray!15!}Yes& \cellcolor{gray!15!}2024 &\cellcolor{gray!15!}Gradient-based & \cellcolor{gray!15!}White-box & \cellcolor{gray!15!}MT-Bench, LLMBar & \cellcolor{gray!15!}Mistral, Openchat, Llamas &\cellcolor{gray!15!}ACC, ASR, PAC \\ 
& AUPI~\cite{liu2024automatic}&Yes  & 2024 & Gradient-based & White-box & MRPC, Jfleg, HSOL, RTE, SST2, SMS & Llama2 & KEY-E, LM-E\\
& \cellcolor{gray!15!}AUTOHIJACKER~\cite{liuautohijacker}&\cellcolor{gray!15!}No & \cellcolor{gray!15!}2024 &\cellcolor{gray!15!} LLM-based & \cellcolor{gray!15!}Black-box & \cellcolor{gray!15!}AgentDojo, OPI & \cellcolor{gray!15!}Llama, Command-R, GPTs &\cellcolor{gray!15!} ASR \\ 
\aboverulesepcolor{gray!25!}  \midrule
\multirow{14}{0.08\textwidth}{Data Extraction} & zlib~\cite{carlini2021extracting}&Yes  & 2020 & Generate \& Inference& Black-box & Top-n, Temperature, Internet & GPT-2 & 6 metrics \\
& \cellcolor{gray!15!}AutoSklearn~\cite{al2023targeted}&\cellcolor{gray!15!}No   & \cellcolor{gray!15!}2023 &\cellcolor{gray!15!}Greedy, Contrastive, Beam decoding & \cellcolor{gray!15!}Black-box & \cellcolor{gray!15!}Pile & \cellcolor{gray!15!}GPT-Neo &\cellcolor{gray!15!} Precision, Recall,  R@FPR \\
& DECOM~\cite{su2024extracting}&No  & 2024 & Decomposition & Black-box & NYT, WSJ & Frontiers &TRM, EMP, BITAP \\
& \cellcolor{gray!15!}Context~\cite{huang2022large}&\cellcolor{gray!15!}No   & \cellcolor{gray!15!}2022 &\cellcolor{gray!15!}Context, Zero-shot, Few-shot & \cellcolor{gray!15!}Black-box & \cellcolor{gray!15!}Enron Corpus &\cellcolor{gray!15!} GPT-Neo &\cellcolor{gray!15!}Acc \\ 
& ETHICIST~\cite{zhang2023ethicist}&Yes  & 2023 & Prompt Tuning & Gray-box & LM-Extraction & GPT-Neo &Recall  \\ 
& \cellcolor{gray!15!}Pii-compass~\cite{nakka2024pii}&\cellcolor{gray!15!}No  & \cellcolor{gray!15!}2024 &\cellcolor{gray!15!}Grounding & \cellcolor{gray!15!}Black-box & \cellcolor{gray!15!} Enron email &\cellcolor{gray!15!}GPT-J &\cellcolor{gray!15!} Extraction Rate \\
& DSP~\cite{wang2024unlocking}&No  & 2024 & Dynamic Soft
Prompting & Black-box & LMEB, The Stack & GPT-Neo, Pythia, StarCoderBase & EER, FER, PPL\\
& \cellcolor{gray!15!}PWB~\cite{wang2024pandora}&\cellcolor{gray!15!}Yes   & \cellcolor{gray!15!}2024 &\cellcolor{gray!15!}Gradient-based & \cellcolor{gray!15!}White-box & \cellcolor{gray!15!}Pile & \cellcolor{gray!15!}Pythia, Llama &\cellcolor{gray!15!}Precision, AUC, TPR  \\ 
\midrule
\multirow{6}{0.08\textwidth}{Prompt Stealing} & Sha et al.~\cite{sha2024prompt}&No  & 2024 & LLM-based & Black-box & RetrievalQA, AlpacaGPT4 & ChatGPT, LLaMA & Acc, Precision, Recall, AUC\\ 
&\cellcolor{gray!15!}  output2prompt~\cite{zhang2024extracting} &\cellcolor{gray!15!}Yes  &\cellcolor{gray!15!}  2024 &\cellcolor{gray!15!}  LLM-based &\cellcolor{gray!15!}  Black-box &\cellcolor{gray!15!}  3 User \& 3 System Prompts &\cellcolor{gray!15!}  Llamas, GPTs &\cellcolor{gray!15!} BLEU, CS, Precision, Recall \\ 
& PRSA~\cite{yang2024prsa}&No  & 2024 & Output Difference & Black-box & Category18 & GPTs & BLEU, FastKASSIM, JS\\
\bottomrule
\end{tabular}
}
\end{table*}

\noindent \textbf{Jailbreak Attacks.} Jailbreak attacks aim to induce the large language model to generate unsafe content like violence~\cite{zou2023universal}. Jailbreak attacks focus on bypassing the safety rules, including system safety prompts and safety filters, while prompt injection attacks target all system prompts. Lots of literature have studied the vulnerability of LLM, where different terms, including ``jailbreak attack'' and ``red-teaming'', all point to the same safety vulnerability that generates unsafe content. We classify them into two main categories, \textit{i.e.} optimization-based and strategy-based. 

Strategy-based jailbreaks figure out novel strategies or templates to generate one adversarial prompt at a heat to test LLMs' vulnerabilities, which are pre-defined. Thus, the generated prompt is non-evolvable. Specifically, useful strategies include persuasion~\cite{zeng2024johnny}, role-playing~\cite{shen2024anything, wang2024foot, samvelyan2024rainbow, jin2024guard}, cipher~\cite{yuan2023gpt, lv2024codechameleon}, ASCII~\cite{jiang2024artprompt}, long-context~\cite{anil2024many}, low-resource language~\cite{yong2023low, Wang2024AllLM}, in-context malicious demonstration~\cite{wei2023jailbreak}, overloaded logical thinking~\cite{xu2023cognitive}, misspelling~\cite{ding2023wolf}, multi-language mixture~\cite{upadhayay2024sandwich}, rephrasing~\cite{yao2024fuzzllm, deng2023masterkey, li2024structuralsleight, paulus2024advprompter}, competing objectives and generalization mismatch~\cite{wei2023jailbroken}, \wenjie{splitting sub-queries~\cite{chen2024pandora}}, zero-shot generation~\cite{perez2022red}, personal modulation~\cite{shah2023scalable}. 

Optimization-based jailbreaks contain a multi-step optimization process to revise one unsafe prompt. Here, we further divide the optimization-based jailbreaks into gradient-based and LLM-based ones: 
\begin{itemize}[leftmargin=*]
\item[\ding{224}] \textbf{\textit{Gradient-based Optimization.}} GCG~\cite{zou2023universal} appends one suffix to the target prompt, then utilizes the gradient of loss, which is calculated with the target (\textit{e.g.,} ``Sure'' or ``Yes'') and output, to optimize the soft prompt. Then, it greedily searches the best-matched tokens in the dictionary for soft prompt replacement. AutoDAN-B~\cite{zhu2023autodan} solves the limited readability of GCG by constructing a proxy score where the perplexity is considered, which is utilized for greedy sampling. I-GCG~\cite{jia2024improved} improves GCG by appending a template before the suffix and uses a multi-coordinate updating strategy and easy-to-hard initialization to optimize the suffix. COLD-Attack~\cite{guo2024cold} adapts Energy-based Constrained Decoding with Langevin Dynamics for controllable adversarial prompt generation. MA-GCG~\cite{zhang2025boosting} proposes momentum gradient to boost and stabilize the greedy search for tokens in adversarial prompts. A-GCG~\cite{zhao2024accelerating} introduces a smaller draft model than the target model to sample the promising suffix candidates for faster optimization. BOOST~\cite{yu2024enhancing} enhances the existing jailbreak attacks by adding eos tokens to the end of the unsafe prompt. CRT~\cite{hong2024curiosity} proposes an enhanced reinforcement learning-based jailbreak with consideration of prompt diversity. I-FSJ~\cite{zheng2024improved} deploys few-shot learning and demo-level random search. 

\item[\ding{224}] \textbf{\textit{LLM-based Optimization.}} PAIR~\cite{chao2023jailbreaking} constructs a system prompt and uses an attacker LLM to generate and revise adversarial prompts. It also uses a Judge model to assess the feedback from the victim, which is further utilized for revising the adversarial prompt. AutoDAN-A~\cite{liu2023autodan} utilizes crossover strategies and LLM-based mutation to revise adversarial prompts into stealthy sentences. AntoDAN-Trubo~\cite{liu2024autodan} AutoDAN-Turbo proposes to find useful strategies by prompting an LLM automatically. ToA (Tree of Attack)~\cite{mehrotra2024tree} iteratively uses an LLM to transform the unsafe prompt into two variations and keeps the prompt variation that achieves a higher score. Xiao et al.~\cite{xiao2024distract} adopt a similar pipeline with PAIR~\cite{chao2023jailbreaking} and introduce malicious content concealing and memory reframing. Puzzler~\cite{chang2024play} proposes defensive and offensive measures to conduct an indirect jailbreak. GPTFUZZER~\cite{yu2023gptfuzzer} starts from human-written prompts, and uses templates and mutation to rewrite unsafe prompts. ECLIPSE~\cite{jiang2024unlocking} uses an LLM as a suffix generator and optimizer. PAL~\cite{sitawarin2024pal} proposes an online proxy model (which is used for adversarial prompt generation) training pipeline.

\item[\ding{224}] \textbf{\textit{Others.}} EnJa~\cite{zhang2024enja} proposes to ensemble prompt and token-level attack methods via a template-based connector. AmpleGCG~\cite{liao2024amplegcg} first collects lots of successful suffixes and then trains the generative model to generate a specific suffix for a given unsafe prompt. Zhao et al.~\cite{zhao2024weak} targets the scenario where the decoding process of target LLM is assisted with smaller models' guidance. 
\end{itemize}

\noindent \textbf{Prompt Injection Attacks.} Prompt injection is a vulnerability where an attacker manipulates the input prompts of LLMs to force them to generate a specific output, which is usually out of the range for normal use (\textit{e.g.}, goal hijacking and prompt leaking~\cite{perez2022ignore}), often by injecting malicious text or commands into the input field. Attackers can employ a variety of techniques to carry out such attacks. 

\begin{itemize}[leftmargin=*]
\item[\ding{224}] \textbf{\textit{Direct Prompt Injection.}} Perez et al.~\cite{perez2022ignore} directly inject handcrafted adversarial prompts into inputs to misalign the language model. HOUYI~\cite{liu2023prompt} proposes an injection generation framework which includes three components. Yan et al.~\cite{yan2023backdooring} utilize LLMs to generate diverse trigger instructions that implicitly capture the characteristics of trigger scenarios. TENSOR TRUST leverages the TENSOR TRUST web game to generate a large-scale dataset and benchmark~\cite{toyer2023tensor}. AUPI~\cite{liu2024automatic} adopts a gradient-based optimization method, specifically, a momentum-enhanced optimization algorithm, to generate universal prompt injection data. Upadhayay et al.~\cite{upadhayay2024cognitive} argue that LLMs suffer from cognitive overload and propose to use in-context learning to jailbreak LLMs through deliberately designed prompts that induce cognitive overload. Kwon et al.~\cite{kwon2024text} circumvent security policies by substituting sensitive words—likely to be rejected by the language model—with mathematical functions.

\item[\ding{224}] \textbf{\textit{Indirect Prompt Injection.}} Greshake et al.~\cite{greshake2023not} propose to indirectly inject prompts into the data that are likely to be retrieved. Bagdasaryan et al.~\cite{bagdasaryan2023abusing} design a prompt injection attack against multi-modal LLMs, by generating an adversarial perturbation corresponding to the prompt and blending it into an image or audio recording. Neural Exec~\cite{pasquini2024neural} designs a multi-stage preprocessing pipeline for cases like Retrieval-Augmented Generation (RAG)-based applications. PoisonedAlign~\cite{shao2024making} boosts the success of prompt injection attacks by strategically creating poisoned alignment samples in the LLM’s alignment process. TPIA~\cite{yang2024tapi} crafts non-functional perturbations that contain malicious information and inserts them into the victim’s code context by spreading them into potentially used dependencies like packages or RAG’s knowledge base. F2A~\cite{ren2024f2a} proposes to use feign security detection agents to bypass the defense mechanism of LLMs. AUTOHIJACKER~\cite{liuautohijacker} uses a batch-based optimization framework to handle sparse feedback and leverages a trainable memory to enable effective generation.

\item[\ding{224}] \textbf{\textit{Different Settings.}} JudgeDeceiver uses gradient-based optimization to inject LLM-as-a-Judge scenarios~\cite{shi2024optimization}. Pedro et al.~\cite{pedro2023prompt} study the risk of injections targeting web applications based on the Langchain framework. Lee et al.~\cite{lee2025exploring} propose a human–AI collaborative framework to explore the potential of prompt injection against federated military LLMs. PROMPT INFECTION~\cite{lee2024prompt} proposes to make malicious prompts self-replicate across interconnected agents in multi-agent systems. Zhang et al.~\cite{zhang2024study} explore the risk of prompt injection in LLM-integrated systems like LLM-integrated mobile robotic systems.

\end{itemize}

\noindent \textbf{Data Extraction Attacks.}
Data extraction attacks try to figure out the personally
identifiable information (PII) that is used to train the LLMs~\cite{carlini2021extracting}. It starts from sufficient-length prefixes to perform extraction and additional measures to determine if extracted texts are valid.

\begin{itemize}[leftmargin=*]
\item[\ding{224}] \textbf{\textit{Methods.}}
In the beginning work~\cite{carlini2021extracting}, the proposed extraction process contains two stages ``generate-then-rank'': sampling potentially memorized examples and membership inference. It proposes a temperature-decaying method to sample more diverse examples and use surrogate models to infer the membership. After that, Al-Kaswan et al.~\cite{al2023targeted} propose using greedy, contrastive, and beam decoding strategies to generate examples and use a classifier to infer the membership. Su et al.~\cite{su2024extracting} propose an instruction decomposition technique to extract fragments of training data gradually. Huang et al.~\cite{huang2022large} extensively explore the effect of context, zero-shot, and few-shot methods in extracting the personal email address. ETHICIST proposes a smoothing loss and a calibrated confidence estimation method to extract the suffix and measure the confidence~\cite{zhang2023ethicist}. Nakka et al.~\cite{nakka2024pii} improves the extraction performance by grounding the prefix of the manually constructed extraction prompt with in-domain data. Wang et al.~\cite{wang2024unlocking} propose to train a transformer-based generator to produce dynamic, prefix-dependent soft prompts. Ozdayi et al.~\cite{ozdayi2023controlling} introduce an approach that uses prompt tuning to control the extraction rates of memorized content. Meng et al.~\cite{meng2025rr} propose a two-stage method, \textit{i.e., } collection and ranking, to recover PPI when PII entities have been masked.

\item[\ding{224}] \textbf{\textit{Different Settings.}}
Some works also explore the risk of data leakage in novel settings. Wang et al.~\cite{wang2024pandora} study the probability of data extraction in fine-tuning settings and Bargav et al.~\cite{jayaraman2022active, zeng2024contrast} extract the training data by comparing the output difference before and after the fine-tuning. Jiang et al.~\cite{jiang2024rag, qi2024follow, zeng2024good} propose to extract the private Retrieval-Augmented Generation (RAG) documents. Peng et al.~\cite{peng2024data} extract the private RAG documents by poisoning in the fine-tuning process. Nasr et al.~\cite{nasr2023scalable} explore the potential risk of data extraction for the aligned production language models. Panda et al.~\cite{panda2024teach} extract the fine-tuning secret data by poisoning the pertaining dataset. Lu et al.~\cite{lu2025merger} propose to extract PII from an aligned model with model merging. Chen et al.~\cite{chen2024janus} find that fine-tuning can recover the forgotten PIIs in pretraining data. Panchendrarajan et al.~\cite{panchendrarajan2021dataset} propose to extract the whole private training data in the fine-tuning process. Rashid et al.~\cite{rashid2023fltrojan} propose selective weight tampering to explore PPI leakage in Federated Language Models. Dentan et al.~\cite{dentan2024reconstructing} extract data from layout-aware document understanding models like unimodal or bimodal models. 

\item[\ding{224}] \textbf{\textit{Different Applications.}}
Leveraging the abnormally high token probabilities, some works utilize the memorization of LLMs to extract the fingerprint or steganography~\cite{hoscilowicz2024unconditional}. Al-Kaswan et al.~\cite{al2024traces} explore memorization in large language models for code and find that code models memorize training data at a lower rate than natural language models. Nie et al.~\cite{nie2024decoding} utilize the token-level features derived from the identified characteristics to decode the PII. Lehman et al.~\cite{lehman2021does} reveal the risk of Electronic Health Records leakage of LLMs. Diera et al.~\cite{diera2023memorization} conduct experiments to assess the PII leakage of fine-tuned BERT models and found that Differential Privacy (DP) has a negative effect when deployed in fine-tuning. Zhang et al.~\cite{zhang2022text} propose data extraction attacks against text classification with transformers. Huang et al.~\cite{huang2024your} propose an evaluation tool, \textit{i.e.} HCR, to assess the PPI leakage in Neural Code Completion Tools. 

\item[\ding{224}] \textbf{\textit{Factor Assessment.}}
Some work studies the factors of data extraction including decoding schemes, model sizes, prefix lengths, partial sequence leakages, and token positions~\cite{tiwari2024sequence, shao2023quantifying}. Yash et al.~\cite{more2024towards} explore the effects of prompt sensitivity and access to multiple checkpoints to extraction attacks. Staab et al.~\cite{staab2023beyond} construct a dataset consisting of real Reddit profiles to extract personal attributes. Xu et al.~\cite{xu2024targeted} conduct experiments to evaluate the factors of different suffix generation methods and different membership inference attacks in extraction performance. Karamolegkou et al.~\cite{karamolegkou2023copyright} evaluate the effect of model structure, data type, probing strategies, and metrics. 

\end{itemize}

\noindent \textbf{Prompt Stealing Attacks.} Given that crafting effective prompts requires significant engineering effort and can be considered valuable intellectual property (IP), prompt-stealing attacks aim to compromise this IP by reconstructing prompts from generated responses~\cite{sha2024prompt, zhang2024extracting, yang2024prsa}. These generation effects are often used to attract prospective prospective buyers. Sha et al.~\cite{sha2024prompt} pioneer this approach by collecting a dataset and training classifiers to predict prompt parameters—such as whether the prompt is direct, role-based, or in-context. They then used a large language model (LLM) to reconstruct the prompt. Similarly, Zhang et al.~\cite{zhang2024extracting} trained an LLM on output-prompt pairs to directly infer the original prompt, while Yang et al.~\cite{yang2024prsa} leveraged generation differences to refine surrogate prompts. However, recovering the original prompt solely from the output is challenging. Out of this, Zheng et al.~\cite{zheng2024inputsnatch} propose a timing-based side-channel method to infer the prompt during inference.

\subsubsection{Defensive Mechanisms in Deployment} \label{Defensive Mechanisms in Deployment}

\begin{figure}[ht]
    \centering
    \includegraphics[width=\linewidth]{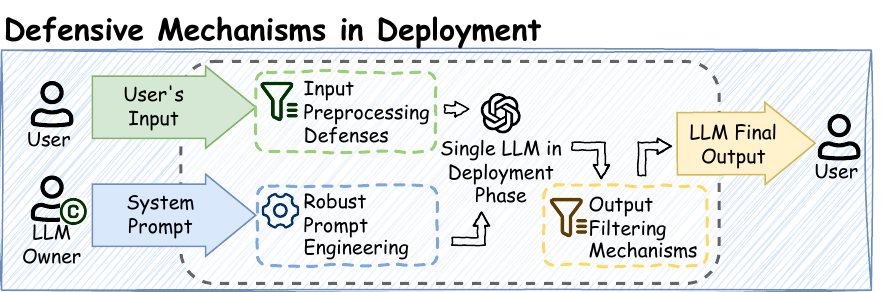}
    \caption{The overview of attacks in single LLM's deployment phase.}
    \label{fig: deployment defense}
\end{figure}

In Subsubsection \ref{Attack in Deployment}, we analyzed various attack scenarios targeting individual LLM deployments. However, in real-world applications, defense mechanisms are not designed as isolated, one-to-one countermeasures against specific attacks. Instead, they follow fundamental security principles to establish a systematic defense framework, as illustrated in Figure \ref{fig: deployment defense}. This framework integrates multiple layers of protection, ensuring resilience against a wide range of adversarial threats while maintaining model usability and efficiency.

\noindent \textbf{\textbf{Input Preprocessing Defenses}} Input preprocessing serves as the first line of defense in LLM deployment, aiming to detect and neutralize adversarial inputs before they reach the model.

\ding{224} \textbf{Attack Detection \& Identification:} Effective input filtering~\cite{dong2024buildingguardrails, jain2023baselinedefensesadversarialattacks} begins with attack detection~\cite{lin2025uniguardian}, which identifies adversarial prompts through statistical~\cite{hu2025tokenlevel}, structural~\cite{gou2024eyes}, or behavioral inconsistencies~\cite{armstrong2025defensedarkpromptsmitigating}.
Gradient-based detection methods~\cite{xie2024gradsafe} leverage safety-critical gradient analysis and loss landscape exploration to uncover jailbreak prompts that manipulate LLM behavior.
These approaches identify adversarial inputs~\cite{peng2024jailbreaking, kumar2024certifying} by analyzing how small perturbations~\cite{zhang2023jailguard} affect model outputs, detecting highly sensitive or misaligned gradients that indicate targeted attacks.
Perplexity-based methods~\cite{hu2025tokenlevel, hu2025tokenlevel} measure the probability distribution of input sequences, flagging atypical or low-likelihood prompts as potential adversarial inputs.
These techniques are particularly effective in detecting prompt injection and adversarial perturbations, where crafted prompts deviate significantly from natural language distributions.

Beyond individual heuristics, universal detection frameworks~\cite{liu2024promptinjection} integrate multiple detection strategies to counter diverse attack vectors, including prompt injection~\cite{suo2024signed}, backdoor manipulations~\cite{yan2023parafuzz}, and adversarial attacks~\cite{kumar2024certifying}.
These frameworks employ ensemble-based filtering mechanisms, combining gradient analysis~\cite{Xiaomeng2024Gradient}, perplexity estimation~\cite{alon2024detecting}, and syntactic evaluation for generalized attack resilience.

\ding{224} \textbf{Semantic \& Behavioral Analysis:} Attack detection alone is insufficient, as certain adversarial inputs may bypass traditional filtering mechanisms.
Semantic~\cite{ji2024defending} and behavioral analysis enhance input preprocessing by evaluating linguistic intent and model alignment.  
Self-examination techniques allow LLMs~\cite{phute2024llmselfdefenseself, candogan2025single} to assess whether they are being manipulated, leveraging auxiliary reasoning steps to detect deceptive prompts.
Alignment-based verification~\cite{cao2024defending} ensures that the model's responses remain consistent with its safety objectives~\cite{Inan2023Llamaguard}, identifying inputs that subtly nudge the model toward policy violations or ethical misalignment.
Intention analysis~\cite{yuqi2025Intention, hanwildguard} further refines input filtering by discerning subtle manipulations designed to bypass explicit security checks.  Unlike token-level detection, which flags overtly adversarial inputs, intention-aware defenses analyze the semantic structure and purpose of the input to preemptively reject jailbreak attempts. 

\ding{224} \textbf{Adversarial Defense \& Mitigation:} When detection and behavioral analysis fail to fully neutralize adversarial inputs, robustness-enhancing techniques~\cite{cao2024defending} mitigate their effects by reducing model susceptibility to manipulation~\cite{gehman2020realtoxicityprompts, pisano2024bergeron}.
Semantic smoothing~\cite{robey2023smoothllm, ji2024advancing} techniques introduce controlled randomness into LLM responses, reducing the model’s sensitivity to adversarial perturbations and preventing reliable jailbreak execution.
By stabilizing decision boundaries~\cite{yi2023benchmarking}, these methods enhance resistance against prompt manipulation strategies that exploit response predictability.

Preemptive input transformations~\cite{song2025alis}, such as backtranslation~\cite{wang2024defending} or paraphrasing, modify incoming queries~\cite{robey2023smoothllm} while preserving semantic intent, disrupting adversarial structures embedded within malicious prompts. 
Data augmentation~\cite{zverev2024can} and adversarial training further strengthen model robustness by exposing LLMs to adversarial prompts during training, forcing them to learn invariances that reduce their vulnerability to real-world attacks.

\noindent \textbf{\textbf{Output Filtering Mechanisms.}} Output filtering mechanisms~\cite{markov2023holistic,dong2024building} serve as a critical safeguard in LLM deployment, ensuring that generated responses comply with safety constraints while preserving informativeness. 
Unlike input preprocessing, which aims to prevent adversarial prompts from reaching the model, output filtering mitigates harmful content post-generation. 
Existing approaches primarily follow three paradigms: rule-based constraints, generative adversarial filtering, and toxicity detection.

Rule-based mechanisms~\cite{kumar2024watch} impose predefined constraints on model outputs, preventing the generation of harmful, unethical, or undesired content. 
Programmable guardrails~\cite{rebedea2023nemo} offer a structured framework where developers can enforce response filtering, topic restriction, and ethical alignment. These methods often integrate reinforcement learning from human feedback~\cite{bai2022training} or rule-based reward~\cite{openairule_based_rewards} modeling to refine output safety. 
While effective at handling explicit violations, static rule-based methods struggle with nuanced adversarial prompts and subtle misalignments.

To address these limitations, generative adversarial filtering~\cite{ma2023adapting} leverages self-critique~\cite{phute2023llm, gou2023critic}, ensemble detection, and dynamic response evaluation~\cite{NEURIPS2024_9d88b87b}. 
Self-rectification mechanisms~\cite{gou2023critic, madaan2023self} enable LLMs to critique their own outputs and refine responses through iterative refinement. 
Additionally, ensemble-based~\cite{jiang2023llm} moderation models aggregate predictions from multiple LLMs, improving robustness against circumvention techniques. 
Adaptive filtering frameworks~\cite{lai2024adaptive} employ perplexity-based assessments and adversarial perturbation detection to flag responses deviating from expected linguistic patterns, enhancing their resilience against jailbreak attempts~\cite{xiong2024defensive,zhang2024parden} and toxic content injection.

Toxicity detection~\cite{yuan2024rigorllm,cao2023systematic,10.1007/s10489-022-03944-z} and content moderation~\cite{zeng2024shieldgemma,wang-etal-2024-self,ghosh2024aegis, Wang2023MTTMMT} further reinforce output safety by identifying and mitigating hate speech~\cite{chiu2021detecting}, misinformation, and other harmful content. 
Supervised fine-tuning adapts LLMs to recognize undesirable patterns, while classifier-based detection models~\cite{kim2023robust} filter responses in real-time.
Some approaches introduce debiasing strategies, such as controlled decoding~\cite{krause2021gedi, liu2024alignment} and anti-expert guidance~\cite{liu2021dexperts}, to suppress toxic outputs without sacrificing response diversity.
However, these methods face challenges in balancing false positives and false negatives, particularly in ambiguous or context-dependent cases.

The effectiveness of output filtering hinges on its ability to balance strict control with linguistic flexibility, ensuring that models remain both safe and practically useful.  
A hybrid approach combining rule-based safeguards, self-correcting mechanisms, and adaptive toxicity moderation is essential to achieving robust and scalable LLM deployment.

\noindent \textbf{\textbf{Robust Prompt Engineering.}}
Robust prompt engineering aims to enhance LLM safety by designing input prompts that resist adversarial manipulation~\cite{radcliffe2024automated}, protect sensitive data, and mitigate harmful outputs—all~\cite{trad2024prompt} without modifying model parameters. 
These strategies act at the interaction level, offering lightweight and model-agnostic protection.  

Recent efforts have introduced prompt optimization techniques grounded in adversarial robustness, including embedding-space manipulation and defensive objective alignment.
Methods such as Robust Prompt Optimization~\cite{zhou2024robus} and Prompt Adversarial Tuning generate transferable suffixes~\cite{xiong2024defensive} or prefix~\cite{mo2024fight} embeddings to guide model behavior~\cite{zhang2024intention} under attack~\cite{chen2024defense}, effectively lowering jailbreak success rates while preserving task performance. 
Similarly, goal prioritization frameworks~\cite{zhang2023defending} enforce inference-time objective consistency, dynamically resolving conflict between user instructions and safety constraints without requiring access to malicious samples. 
Complementary to these strategies, patch-based methods integrate interpretable suffixes or structured self-reminders~\cite{xie2023defending} into prompts, reducing the model’s susceptibility to coercive inputs through lightweight, modular defenses.

Structural manipulation approaches~\cite{chen2024struq} neutralize adversarial intent through prompt rewriting. 
Spotlighting~\cite{hines2024defending} injects source-attribution signals to counter indirect prompt injection, while inverse prompt engineering~\cite{slocum2025inverse} repurposes attack data to generate task-specific defensive prompts under the principle of least privilege.  

Privacy-preserving prompt~\cite{edemacu2024privacy} design introduces formal guarantees through differential privacy. 
Approaches like DP-Prompt~\cite{utpala2023locally} and stochastic gradient masking~\cite{duan2023flocks} reduce information leakage from prompts without harming performance. Desensitization and directional control of in-context representations offer additional privacy protections during prompt construction. 
Prompt engineering~\cite{perez2022red, Wang2023NotAC} also helps mitigate societal risks. 
Chain-of-thought prompting and guided templates reduce gender bias~\cite{kaneko2024evaluating} in reasoning tasks, while prompt learning~\cite{he2024you} improves toxicity detection and generation control~\cite{zou2024system, xu2024preemptive}, often surpassing specialized models in efficiency and generalization.  

Finally, systematic prompt optimization methods~\cite{zheng2024prompt, wang2024adashield} aim to generalize prompt robustness across tasks and domains. Techniques like BATPrompt~\cite{shi2024robustness} and StraGo~\cite{wu2024strago} use adversarial simulation and strategic decomposition to refine prompts iteratively, improving both resilience and effectiveness under variable inputs.

\noindent \textbf{\textbf{System-level Security Controls.}}
System-level defenses~\cite{wu2024new} enhance LLM deployment by optimizing inference, enforcing alignment, isolating untrusted inputs, and securing the supply chain. 
Systems like Petals~\cite{borzunov2023distributed}, Sarathi-Serve~\cite{agrawal2024taming}, and DistServe~\cite{zhong2024distserve} restructure computation to improve throughput and latency, while TriForce~\cite{sun2024triforce}, Medusa~\cite{cai2024medusa} MagicDec~\cite{chen2024magicdec} accelerate generation via speculative decoding and structural compression. 
Parallel frameworks such as DeepSpeed-FastGen~\cite{holmes2024deepspeed} and SpecExec~\cite{svirschevski2024specexec} further boost efficiency with minimal overhead.

Runtime alignment methods~\cite{wang2024inferaligner} adapt model behavior through cross-model guidance or token-level reward modeling. 
Systems such as SelfDefend~\cite{wang2024selfdefend} and Gradient Cuff~\cite{hu2024gradient} detect unsafe generation by monitoring agreement across models or loss landscapes, while Spotlighting~\cite{hines2024defending} inserts provenance signals to mitigate indirect prompt injection.

Access isolation is achieved through policy enforcement~\cite{sharma2024spml} and system wrappers~\cite{zhang2023defending}. At the supply level, tools like MalHug~\cite{zhao2024models} identify poisoned models, while system audits reveal sandbox and plugin vulnerabilities, highlighting the need for end-to-end secure deployment.

LLM-based guard models utilize lightweight LLMs like Llama Guard  \citep{Inan2023Llamaguard}, Aegis Guard \citep{AegisGuard, AegisGuard2}, WildGuard \citep{wildguard}, and ShieldGemma \citep{Shieldgemma} to moderate both the input and output of the victim LLMs. However, they are purely classifiers. To solve this problem, the first reasoning-based guard model named GuardReasoner \cite{guardreasoner} is proposed to improve the performance, explainability, and generalization ability via learning to reason. It brings new opportunities for the safety of large-scale reasoning models \citep{wang2025safety}.

\subsubsection{Evaluation and Benchmarks in Deployment} \label{Evaluation and Benchmarks in Deployment}
\begin{figure}[ht]
    \centering
    \includegraphics[width=1.0\linewidth]{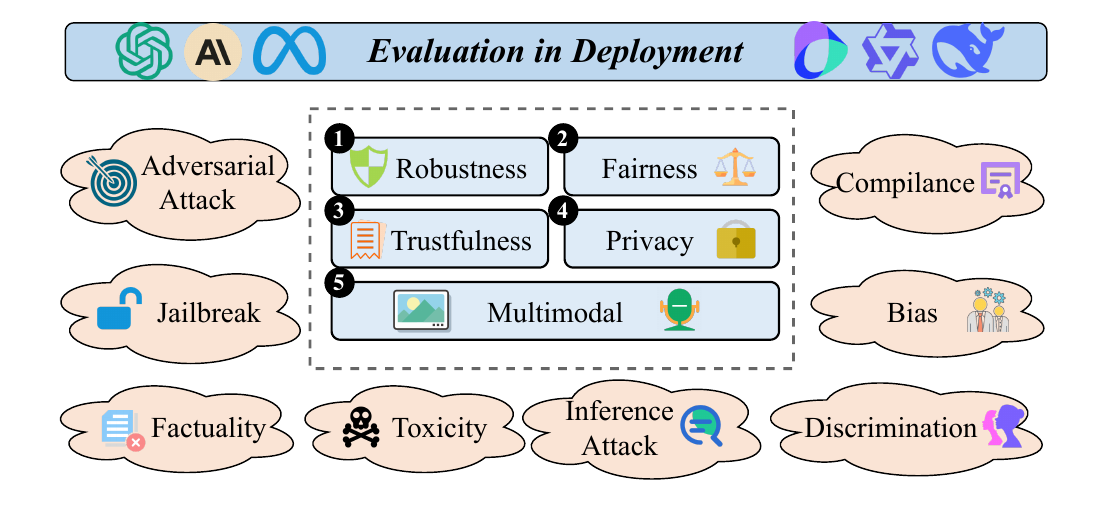}
    \caption{The overview of evaluation and benchmarks in single LLM's deployment phase.}
    \label{fig:deployment_evaluation}
\end{figure}

To assess the reliability and safety of LLMs after deployment, evaluation efforts focus on several key dimensions and risk types, as illustrated in Figure~\ref{fig:deployment_evaluation}. These dimensions guide the design of systematic benchmarks and metrics tailored for real-world deployment settings.

\noindent \textbf{Robustness Evaluation.}
\begin{table}[ht]
\footnotesize
\setlength\tabcolsep{4pt}
\begingroup
\caption{Summary of LLM robustness benchmarks at the deployment stage.}
\vspace{-0.5em}
\resizebox{\columnwidth}{!}{%
\begin{tabular}{l|cccc}
\hline
\textbf{Benchmark} & \textbf{Adversarial} & \textbf{Natural} & \textbf{Jailbreak} & \textbf{Toxicity} \\ \hline
JailbreakBench~\cite{chao2024jailbreakbench}        & \ding{52} &  & \ding{52} & \ding{52} \\
HarmBench~\cite{mazeika2024harmbench}             & \ding{52} &  & \ding{52} & \ding{52} \\
JAMBench~\cite{jin2024jailbreaking}              & \ding{52} &  & \ding{52} &  \\
JailbreakEval~\cite{ran2024jailbreakeval}         & \ding{52} &  & \ding{52} & \ding{52} \\
Latent Jailbreak~\cite{qiu2023latent}      & \ding{52} &  & \ding{52} &  \\
PromptRobust~\cite{zhu2023promptrobust}          & \ding{52} & \ding{52} &  &  \\
SelfPrompt~\cite{pei2024selfprompt}            & \ding{52} & \ding{52} &  &  \\
Chen~\textit{et al.}~\cite{xu2024comprehensive}  & \ding{52} & \ding{52} & \ding{52} & \ding{52} \\
Chu~\textit{et al.}~\cite{chen2024characterizing}   & \ding{52} &  & \ding{52} &  \\
AdvGLUE~\cite{wang2021adversarial}   & \ding{52} & \ding{52} &  &  \\
AdvGLUE++~\cite{wang2023decodingtrust}   & \ding{52} & \ding{52} &  &  \\
NoiseLLM~\cite{dong2023revisit} &  & \ding{52} &  &  \\
NEO-BENCH~\cite{zheng2024neo}             &  & \ding{52} &  &  \\
CompressionEval~\cite{li2024evaluating}       &  & \ding{52} &  &  \\
\hline
\end{tabular}}
\label{tab:eval_robustness}
\endgroup
\end{table}
To systematically assess the reliability of large language models (LLMs) after deployment, we categorize robustness evaluation into two broad types: \textit{adversarial robustness} and \textit{natural robustness}. Adversarial robustness focuses on evaluating how LLMs respond to malicious or adversarial inputs, such as jailbreak prompts, prompt injections, or red-teaming attacks. Natural robustness, on the other hand, assesses LLM behavior under non-malicious but realistic distribution shifts, including typos, paraphrasing, novel word usage, or temporal drift. A summary of representative benchmarks categorized along these 4 dimensions is presented in Table~\ref{tab:eval_robustness}.

\ding{224} \textbf{Adversarial Robustness:}
A range of benchmarks and frameworks have been proposed for adversarial robustness. JailbreakBench~\cite{chao2024jailbreakbench} provides a standardized evaluation suite for jailbreak attacks, containing 100 misuse behaviors and an evolving repository of adversarial prompts. HarmBench~\cite{mazeika2024harmbench} proposes a comprehensive red-teaming evaluation framework that includes 510 harmful behaviors spanning diverse semantic and functional categories, supporting both text-only and multimodal inputs across 33 LLMs. JAMBench~\cite{jin2024jailbreaking} targets the evaluation of moderation guardrails using 160 carefully constructed prompts across four major risk categories and introduces a cipher-character-based attack. JailbreakEval~\cite{ran2024jailbreakeval} offers a unified toolkit for jailbreak assessment with string-matching, classifier-based, and LLM-based evaluators. Latent Jailbreak~\cite{qiu2023latent} focuses on detecting embedded malicious intent in seemingly benign prompts and evaluates instruction-following robustness using a hierarchical annotation scheme. PromptRobust~\cite{zhu2023promptrobust} benchmarks prompt-level robustness with character, word, sentence, and semantic-level perturbations across 13 datasets and 8 NLP tasks. SelfPrompt~\cite{pei2024selfprompt} enables autonomous robustness evaluation through knowledge-guided prompt generation and LLM-based self-assessment. Chu~\textit{et al.}~\cite{chen2024characterizing} conduct a large-scale comparison of 17 jailbreak attacks on 8 LLMs and 160 forbidden prompts, proposing a unified taxonomy and benchmarking various defenses. Chen~\textit{et al.}~\cite{xu2024comprehensive} propose a multi-dimensional framework assessing jailbreak reliability over 13 LLMs and 1,525 prompts, integrating metrics such as attack success rate (ASR), toxicity, fluency, and grammatically. Zhang et al.~\cite{zhangbenchmark} propose a novel definition and benchmark for LLM’s content moderation based on a sensitive-semantic perspective.

\ding{224} \textbf{Natural Robustness:}
Several benchmarks focus on evaluating LLMs under realistic but benign input perturbations or distribution shifts. AdvGLUE~\cite{wang2021adversarial} and AdvGLUE++~\cite{wang2023decodingtrust} extend the original GLUE benchmark~\cite{wang2018glue} with semantically-preserving perturbations at logic, word, and sentence levels. NoiseLLM~\cite{dong2023revisit} presents a unified framework for evaluating slot-filling robustness under character-, word-, and sentence-level noise, including typos and paraphrases. NEO-BENCH~\cite{zheng2024neo} assesses robustness to temporal drift by introducing neologisms into tasks such as machine translation, classification, and question answering. CompressionEval~\cite{li2024evaluating} provides a prompt-free evaluation framework using lossless compression to assess generalization and robustness, comparing LLM performance on content before and after the model’s knowledge cutoff. These benchmarks offer complementary perspectives for assessing LLM performance under both malicious and naturally occurring input variations.

\noindent \textbf{Content Trustfulness and Fairness Evaluation.}
\begin{table}[ht]
\footnotesize
\setlength\tabcolsep{4pt}
\begingroup
\caption{Summary of content trustfulness and fairness evaluation benchmarks for LLMs at deployment stage.}
\vspace{-0.5em}
\resizebox{\columnwidth}{!}{%
\begin{tabular}{l|ccccc}
\hline
\textbf{Benchmark} & \textbf{Hallucination} & \textbf{Factuality} & \textbf{Toxicity} & \textbf{Bias} & \textbf{Discrimination} \\ \hline
HaluEval~\cite{li2023halueval}                    & \ding{52} & \ding{52} &  &  &  \\
Med-HALT~\cite{pal2023med}                        & \ding{52} & \ding{52} &  &  &  \\
ANAH~\cite{ji2024anah}                            & \ding{52} & \ding{52} &  &  &  \\
SelfCheckGPT~\cite{manakul2023selfcheckgpt}       & \ding{52} & \ding{52} &  &  &  \\
DoLa~\cite{chuang2023dola}                        & \ding{52} & \ding{52} &  &  &  \\
Mundler~et al.~\cite{mundler2023self}             & \ding{52} & \ding{52} &  &  &  \\
Elaraby~et al.~\cite{elaraby2023halo}             & \ding{52} & \ding{52} &  &  &  \\
Ji~et al.~\cite{ji2024llm}                        & \ding{52} & \ding{52} &  &  &  \\
Zhang~et al.~\cite{wei2024measuring}              & \ding{52} & \ding{52} &  &  &  \\
Guo~et al.~\cite{deshpande2023toxicity}           &  &  & \ding{52} & \ding{52} &  \\
RTP-LX~\cite{de2024rtp}                           &  &  & \ding{52} & \ding{52} &  \\
ROBBIE~\cite{esiobu2023robbie}                    &  &  & \ding{52} & \ding{52} & \ding{52} \\
CEB~\cite{wang2024ceb}                            &  &  & \ding{52} & \ding{52} & \ding{52} \\
\hline
\end{tabular}}
\label{tab:eval_safety_bias}
\endgroup
\end{table}
Beyond robustness, a key dimension of deployment-stage evaluation concerns the \textit{trustfulness} and \textit{fairness} of LLM-generated content. This includes detecting and mitigating outputs that are factually incorrect (hallucinations), misleading (low factuality), harmful (toxic), or unfair (biased or discriminatory). We categorize existing benchmarks into five axes: \textit{hallucination}, \textit{factuality}, \textit{toxicity}, \textit{bias}, and \textit{discrimination}, and summarize representative works in Table~\ref{tab:eval_safety_bias}.

Benchmarks in this space target either the \textit{accuracy of generated content} or its \textit{alignment with human values}. For hallucination and factuality evaluation, HaluEval~\cite{li2023halueval} and Med-HALT~\cite{pal2023med} provide reference-based hallucination annotations in general and medical domains, respectively, while ANAH~\cite{ji2024anah} delivers fine-grained, human-annotated hallucination labels with correction spans. SelfCheckGPT~\cite{manakul2023selfcheckgpt} detects hallucinations via consistency checks across multiple generations, and DoLa~\cite{chuang2023dola} proposes a decoding strategy that contrasts internal layer activations to reduce factual errors. Other works such as Mundler~\textit{et al.}~\cite{mundler2023self}, Elaraby~\textit{et al.}~\cite{elaraby2023halo}, and Ji~\textit{et al.}~\cite{ji2024llm} leverage taxonomic definitions or internal model signals to quantify or predict hallucination risk. Zhang~\textit{et al.}~\cite{wei2024measuring} introduce FEWL, a reference-free evaluation framework that uses agreement across reference LLMs to approximate hallucination likelihood.

In terms of toxicity detection, Guo~\textit{et al.}~\cite{deshpande2023toxicity} show that role-playing prompts (personas) can elicit toxic behavior from ChatGPT, and RTP-LX~\cite{de2024rtp} evaluates multilingual LLMs in detecting culturally sensitive harm. Both studies reveal that current LLMs remain vulnerable to subtle toxic or culturally biased outputs, especially in low-resource languages or when confronted with indirect harm.

For evaluating social bias and discrimination, ROBBIE~\cite{esiobu2023robbie} benchmarks LLMs across 12 demographic axes with template-based prompts and multiple toxicity and regard metrics, covering gender, race, religion, and intersections thereof. CEB~\cite{wang2024ceb} proposes a compositional taxonomy for fairness evaluation and introduces multiple new datasets spanning stereotyping, toxicity, and classification bias, supporting both direct and indirect evaluation modes. 

These benchmarks collectively provide a multidimensional view of content trustfulness and fairness, enabling the systematic evaluation of LLMs beyond syntactic correctness or surface fluency. As safety-critical deployment scenarios become increasingly prevalent, such evaluation tools play a central role in ensuring the responsible use of LLMs.

\noindent \textbf{Data Privacy and Leakage Evaluation.}
\begin{table}[ht]
\footnotesize
\setlength\tabcolsep{6pt}
\centering
\caption{Summary of privacy evaluation benchmarks for LLMs at the deployment stage.}
\vspace{0.5em}
\begin{tabular}{lcccc}
\hline
\textbf{Benchmark} & \textbf{PII} & \textbf{MIA} & \textbf{EIA} & \textbf{Compliance} \\
\hline
PrivLM-Bench~\cite{li2023privlm}        & \ding{52} & \ding{52} & \ding{52} &  \\
LLM-PBE~\cite{li2024llm}             & \ding{52} & \ding{52} & \ding{52} &  \\
PrivAuditor~\cite{zhu2024privauditor}         & \ding{52} & \ding{52} &           &  \\
Rossi~\textit{et al.}~\cite{rossi2024auditing}           & \ding{52} & \ding{52} &           &  \\
Whispered Tuning~\cite{singh2024whispered}    & \ding{52} &           &           &  \\
ProPILE~\cite{kim2023propile}             & \ding{52} &           &           &  \\
PrivaCI-Bench~\cite{li2025privaci}       & \ding{52} &           &           & \ding{52} \\
Commercial Audit~\cite{cartwright2024evaluating}    & \ding{52} &           &           & \ding{52} \\
LessLeak-Bench~\cite{zhou2025lessleak}      & \ding{52} &           &           &  \\
SecureSQL~\cite{song2024securesql}           & \ding{52} &           &           &  \\
DecodingTrust~\cite{wang2023decodingtrust}       & \ding{52} & \ding{52} &           &  \\
\hline
\end{tabular}
\label{tab:privacy_benchmark_summary}
\end{table}
Data privacy is a critical dimension in evaluating the trustworthiness of LLMs at deployment. Table~\ref{tab:privacy_benchmark_summary} summarizes representative benchmarks that assess privacy risks along four axes: personally identifiable information (PII) leakage, membership inference attacks (MIA), embedding inversion attacks (EIA), and regulatory or contextual compliance.

PrivLM-Bench~\cite{li2023privlm} and LLM-PBE~\cite{li2024llm} offer comprehensive multi-level evaluations spanning all three major attack types. PrivAuditor~\cite{zhu2024privauditor} and Rossi~\textit{et al.}~\cite{rossi2024auditing} focus on adaptation-stage vulnerabilities across a variety of fine-tuning techniques. Whispered Tuning~\cite{singh2024whispered} proposes a differential privacy–based training scheme to reduce leakage, while ProPILE~\cite{kim2023propile} tests whether LLMs can reconstruct sensitive information from prompts related to known users.

PrivaCI-Bench~\cite{li2025privaci} and Commercial Audit~\cite{cartwright2024evaluating} emphasize regulatory compliance, evaluating model behavior against privacy expectations and legal frameworks such as GDPR and the EU AI Act. SecureSQL~\cite{song2024securesql} examines leakage in structured query generation, and LessLeak-Bench~\cite{zhou2025lessleak} reveals code-specific leakage across software engineering benchmarks. Finally, DecodingTrust~\cite{wang2023decodingtrust} includes privacy as part of a broader trustworthiness suite, auditing GPT models across multiple risk dimensions.

Together, these benchmarks provide a foundation for assessing LLM privacy risks across diverse modalities, attack surfaces, and deployment scenarios.

\noindent \textbf{Multi-modal Safety Evaluations}
As multimodal large language models (MLLMs) become increasingly integrated into real-world applications, ensuring their safety under diverse input conditions is essential. A growing number of studies have proposed evaluation benchmarks and frameworks to assess MLLM vulnerabilities across multiple dimensions~\cite{liu2024mm, luo2024jailbreakv, weng2024textit, li2024retention, guan2024hallusionbench, li2023evaluating, cui2023holistic, wang2024cross, agarwal2024mvtamperbench, zhang2024avibench, hu2024viva, xiao2024genderbias, gustafson2023facet, slyman2024fairdedup, zhang2022counterfactually, fraser2024examining, seth2023dear, janghorbani2023multimodal, zhang2024benchmarking, zhang2024spa, zhang2025q, Wang2025CantST, Wang2024ChainofJailbreakAF}.

Jailbreak evaluation has received significant attention, with benchmarks such as MM-SafetyBench~\cite{liu2024mm} and Jailbreakv-28k~\cite{luo2024jailbreakv} targeting harmful instruction-following behaviors. MMJ-Bench~\cite{weng2024textit} and Retention Score~\cite{li2024retention} further extend jailbreak assessment to include visual robustness and long-term safety retention. For hallucination, several works diagnose MLLM failures arising from inconsistencies between visual inputs and generated text, including HallusionBench~\cite{guan2024hallusionbench}, POPE~\cite{li2023evaluating}, and Bingo~\cite{cui2023holistic}. SIUO~\cite{wang2024cross} complements this direction by evaluating cross-modality consistency under seemingly benign inputs.

Robustness under adversarial visual corruption is assessed in MVTamperBench~\cite{agarwal2024mvtamperbench} and B-AviBench~\cite{zhang2024avibench}, which introduce perturbed or misleading visual stimuli to test model stability. Meanwhile, fairness and social bias have been evaluated through VIVA~\cite{hu2024viva}, GenderBias-VL~\cite{xiao2024genderbias}, FACET~\cite{gustafson2023facet}, FairDeDup~\cite{slyman2024fairdedup}, CounterBias~\cite{zhang2022counterfactually}, PAIRS~\cite{fraser2024examining}, DeAR~\cite{seth2023dear}, and MMBias~\cite{janghorbani2023multimodal}, covering gender, racial, and intersectional dimensions using parallel image sets, counterfactual probing, and real-world dataset imbalances.

To unify these evaluation directions, several comprehensive frameworks have emerged. MultiTrust~\cite{zhang2024benchmarking} and SPA-VL~\cite{zhang2024spa} aim to benchmark MLLMs across diverse safety criteria, including robustness, fairness, and harmfulness. Q-Eval-100K~\cite{zhang2025q} complements these efforts by focusing on visual generation quality and alignment under instruction-following settings.

Collectively, these benchmarks highlight the unique challenges posed by multimodal interactions and the growing need for holistic, scalable safety evaluations tailored to MLLMs.

\begin{figure*}[t]
    \centering
    \includegraphics[width=1.0\linewidth]{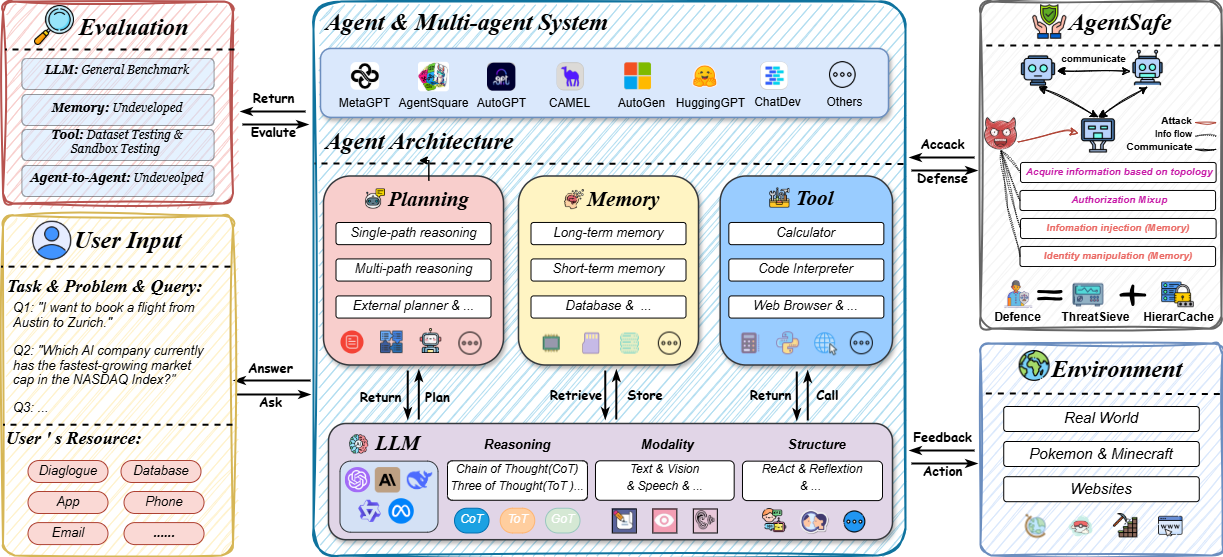}
    \vspace{-2em}
    \caption{The overview of LLM-based single-agent and multi-agent systems.}
    \label{fig:agent}
\end{figure*}

\subsection{Single-agent Safety} \label{Single-agent Safety}
\begin{figure*}[t]
    \centering
    \includegraphics[width=1.0\linewidth]{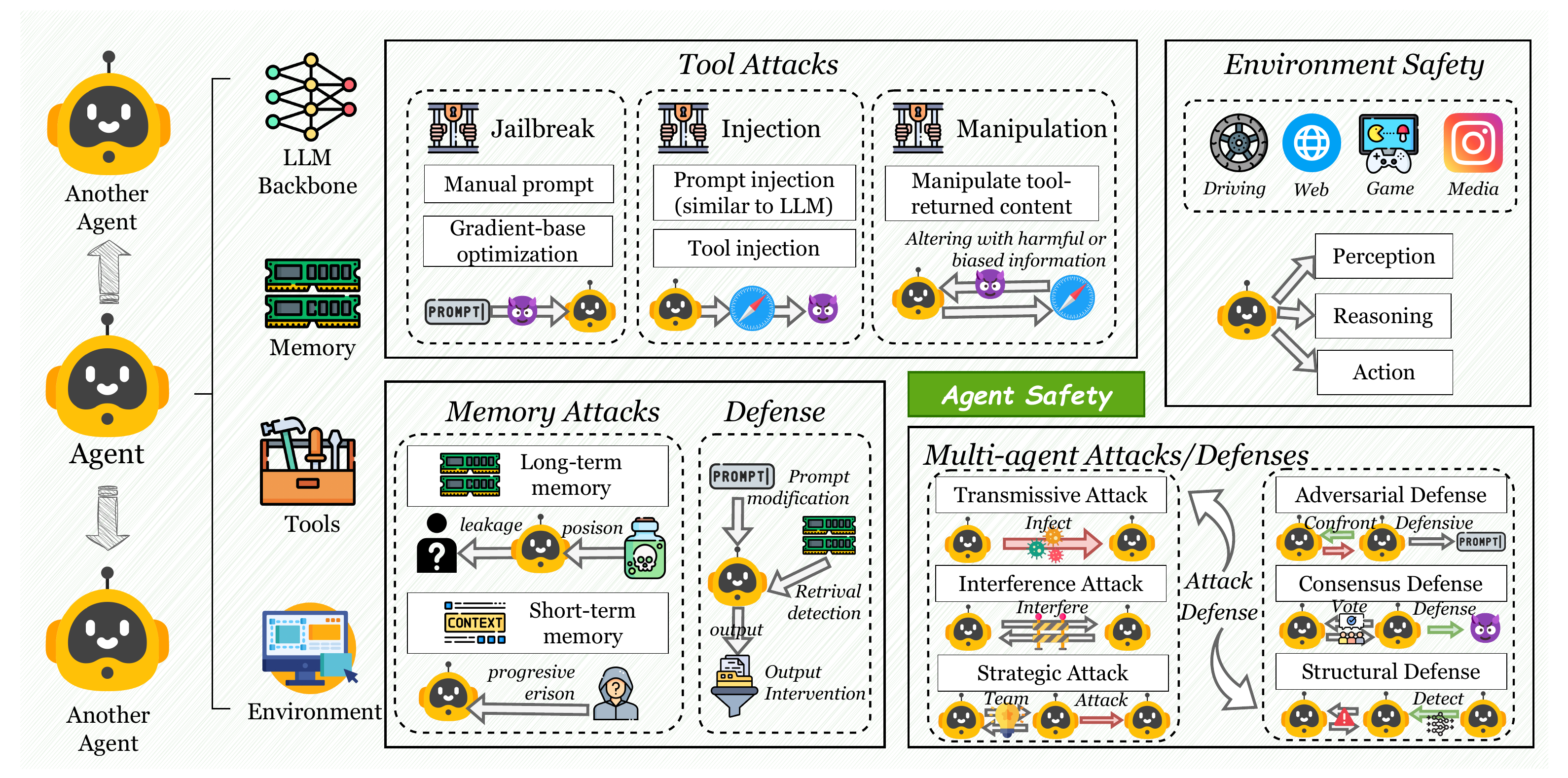}
    \vspace{-2em}
    \caption{The overview of the safety of LLM-based agent systems.}
    \label{fig:agent_safety}
\end{figure*}

In this section, we focus on security issues related to a single agent. We first define an agent as an interactive entity that uses an LLM as the core for reasoning, decision-making, and reflection while integrating memory, tools, and the environment as capability-enhancing components. Beyond the deployment risks associated with the LLM core, we introduce the security issues arising from these three additional modules. Specifically, for tools (Section \ref{tool safety}) and memory (Section \ref{memory safety}), we summarize existing work from both attack (Section \ref{attack-agent}) and defense (Section \ref{defense-agent}) perspectives to identify technical paradigms. For the environment (Section \ref{environment safety}), we explore unique security challenges from the perspective of various agent-interaction settings. We demonstrate an overview of agent safety in Figure \ref{fig:agent_safety}.

\subsubsection{Definition of Agent}
LLM-driven agent refers to an AI system capable of operating independently or with limited human oversight, where a sophisticated language model \cite{naveed2023comprehensive, zhao2023survey,zhao2025beware,liu2025advances} serves as the foundational intelligence for processing inputs, executing tasks, and engaging in interactions. By leveraging advanced natural language understanding and generation, such agents \cite{jin2024llms, xi2025rise, piao2025agentsociety, yan2025mathagent,wang2024ali} can analyze information, resolve queries, and adapt to user or environmental inputs \cite{zhang2024codeagent, shen2023hugginggpt,chu2025llm}. To extend their functionality, they frequently incorporate supplementary mechanisms—such as data storage modules \cite{zhang2023data,zhang2024survey,xu2025mem,shang2024agentsquare}, external software interfaces \cite{zhang2024codeagent,yang2024swe,agashe2024agent}, or strategic reasoning frameworks \cite{hao2023reasoning}—allowing them to transcend basic text production. This adaptability makes them valuable for diverse implementations, including interactive dialogue systems \cite{hong2024interactive}, workflow optimization \cite{tang2025autoagent,li2023camel,yuan2024evoagent,zhuge2024language}, and complex decision-making scenarios \cite{wang2024sibyl}. In this study, we focus on deconstructing agent safety into three critical dimensions: tool utilization, memory management, and environment-specific security concerns. We demonstrate the components and structures of agent systems in Figure \ref{fig:agent}.

\subsubsection{Tool Safety} \label{tool safety}
Some works enable LLM agents to learn how to use tools by generating datasets and fine-tuning the model for API usage ~\cite{liu2024toolace, wang2024toolflow}. Specifically, tools can be implemented in various forms, including but not limited to code-based API functions (e.g., search engine ~\cite{wu2024wipi} and calculator), embodied intelligence like robotic arms ~\cite{kannan2024smart}, and more. A tool serves as a bidirectional medium: on one hand, it allows the agent to map internal decisions into actions within the interactive environment; on the other hand, it also acts as a means for the agent to collect information from the external world. Given the pivotal role of tools in agent components, the related security issues are worth exploring ~\cite{yu2025survey}. For example, in the field of web security, Fang et al. \cite{fang2024llmb, fang2024llma} investigate how autonomous agents, when equipped with appropriate tools, can independently compromise websites and exploit one-day vulnerabilities in real-world systems without human intervention. Next, we will summarize and discuss existing research from attack perspectives and figure out the lack of tool invocation defense in current research.

\noindent \textbf{\textbf{Attacks.}}
Based on the target of the attack, safety-related attacks involving tools can be categorized into Tool-aided Attacks and Tool-targeted Attacks. The former refers to attackers utilizing agents equipped with tools to execute attacks that LLMs cannot independently assist with, such as leveraging agents with web access and code execution capabilities to facilitate cyberattacks. The latter involves attackers targeting the tool invocation process itself, attempting to manipulate or induce tool selection for malicious purposes through various attack methods. However, from the perspective of the technical stack of attacks, the two can be unified. We have identified new applications of traditional LLM attack methods in tool safety, as well as novel attack paradigms that have emerged due to the unique characteristics of tools.

\textbf{Jailbreak.} Similar to jailbreak methods in LLM safety, agent jailbreak also bypasses the agent's built-in safety mechanisms through specific prompts to elicit malicious responses. However, in the agent scenario, the malicious behaviors it aims to induce are different. Specifically, Cheng et al. \cite{cheng2024security} manually craft jailbreak prompts to extract personal information from the training data of code-generation agents. In contrast, Fu et al. \cite{fu2023misusing} and Imprompter \cite{fu2024imprompter} both employ gradient-based optimization like GCG \cite{zou2023universal} to automatically generate input prompts or images that manipulate agents into leveraging tools for privacy breaches in dialogues or executing harmful actions on user resources.

\textbf{Injection.} This type of attack can be summarized into two forms of injection: Prompt Injection (similar to LLM safety vulnerabilities) where malicious instructions are embedded in input data, exploiting the difficulty LLMs face in distinguishing between instructions and data. Another form is Tool Injection where malicious tools are injected to enable further exploitation, such as using the tool to execute malicious actions. For example, BreakingAgents \cite{zhang2024breaking} utilizes human-crafted prompt injections to execute malfunction attacks, causing agents to engage in repetitive or irrelevant actions, with additional exploration into the propagation of such attacks within Multi-Agent Systems (MAS). ToolCommander \cite{wang2024allies} is the second type. It proposes a two-stage attack strategy: first, injecting malicious tools to steal user queries, and subsequently manipulating tool selection using the stolen data, thereby achieving privacy theft and denial-of-service attacks.

\textbf{Backdoor.} Backdoor attacks also find utility in the context of agent safety, but unlike LLMs, LLM agents develop diverse verbal reasoning traces through continuous environmental interactions, broadening potential backdoor attack vectors. Yang et al. \cite{yang2024watch} define two types of backdoor attacks, targeting either the final returned results or the intermediate processes of the attacking agent, and implement the above variations of agent backdoor attacks on two typical agent tasks, including web shopping and tool utilization. Furthermore, DemonAgent \cite{zhu2025demonagent} decomposes a backdoor into multiple sub-backdoor fragments to poison the
agent’s tools. Beyond intentional guidance, studies such as BadAgent \cite{wang2024badagent} highlight that backdoor attacks can inadvertently prompt agents to misuse tools for malicious purposes.

\textbf{Manipulation.} This type of attack refers to directly or indirectly manipulating or altering the tool's returned content to leak sensitive information or carry out malicious actions. AUTOCMD \cite{jiang2025mimicking} employs a separate LLM, trained on tool-calling datasets and fine-tuned with target-specific examples, to generate and replicate legitimate commands for extracting sensitive information from tools. Meanwhile, Zhao et al. \cite{zhao2024attacks} manipulate third-party API outputs by injecting malicious content or omitting critical information, ultimately causing erroneous or biased system behaviors.

\noindent \textbf{\textbf{Defenses.}}
Compared to attacks on agent tools, defense mechanisms for secure tool invocation have been less studied. Specifically, AgentGuard \cite{chen2025agentguard} employs LLM orchestrators to automatically detect unsafe tool-use workflows and produce safety constraints for secure tool utilization. PrivacyAsst \cite{zhang2024privacyasst} proposes an encryption-based solution by integrating an encryption scheme into the tool using LLM agents to safeguard user privacy and align them with computational security standards. In addition, some works enhance the security of agent systems by leveraging tool invocation, GuardAgent \cite{xiang2024guardagent} pioneers an approach to verify target agents' trustworthiness by executing guardrail code through API calls during task plan implementation.

\subsubsection{Memory Safety} \label{memory safety}
The memory mechanism in LLM agents enables them to retain historical behaviors, thereby enhancing future decision-making capabilities. Typically, agent memory can be categorized into long-term and short-term memory systems. The long-term memory module commonly employs Retrieval-Augmented Generation (RAG) \cite{gao2023retrieval, zhao2024retrieval} technology to facilitate precise information retrieval, while the short-term memory stores real-time data to support immediate conversational contexts and task execution. While these memory modules significantly improve agent functionality, they simultaneously introduce potential security vulnerabilities, making the system susceptible to malicious attacks.

\subsubsection{Attack} \label{attack-agent}
Follow the trustworthy issues in \cite{yu2025survey}, we categorize attacks related to memory into three types: Memory Poisoning, Privacy Leakage, and Memory Misuse.

\textbf{(I)} \textbf{Memory Poisoning} refers to adversarial attacks where malicious data is injected into an agent’s \textit{long-term memory} \cite{xiang2024certifiably, chen2025agentpoison, zou2024poisonedrag, zhong2023poisoning, zhang2025agent, gu2024agent}. When the agent retrieves and utilizes such corrupted memory, it may produce erroneous outputs, misleading responses, or even hazardous actions. For example, PoisonedRAG framework \cite{zou2024poisonedrag} employs a dual optimization approach, simultaneously manipulating both the retrieval and generation pipelines to systematically poison the agent's memory system. AgentPoison \cite{chen2025agentpoison} introduces an advanced backdoor attack methodology that optimizes trigger patterns and seamlessly integrates them into query formulations, significantly elevating the likelihood of malicious sample retrieval while maintaining stealth. \textbf{(II)} \textbf{Privacy Leakage} occurs when attackers exploit the interface between an agent and its \textit{long-term memory} to extract stored sensitive data \cite{li2025commercial, zeng2024good, li2023sentence, anderson2024my, jiang2024rag}. Such breaches may expose user information to malicious third parties, posing significant real-world risks. \textbf{(II)} \textbf{Memory Misuse} refers to the deliberate construction of multi-turn query sequences that systematically circumvent safety protocols by exploiting the retention properties of agent \textit{short-term memory} \cite{russinovich2024great, li2024llm, cheng2024leveraging, priyanshu2024fractured, agarwal2024prompt, tong2024securing}. This attack vector enables progressive erosion of defensive measures through iterative interaction patterns.

\subsubsection{Defense}  \label{defense-agent}

To counter these attacks, various defense approaches have been developed to enhance the robustness of memory systems \cite{mao2025agentsafe, zhou2025trustrag, xian2024vulnerability, agarwal2024prompt, anderson2024my}. \textbf{(I)} \textbf{Detection} \textbf{Detection} mechanisms primarily focus on identifying and eliminating malicious content retrieved from long-term memory systems \cite{zhou2025trustrag, xian2024vulnerability, agarwal2024prompt, zhang2024agent}. \textbf{(II)} \textbf{Prompt Modification} involves strategically rewriting user queries before processing by the agent to enhance response safety \cite{anderson2024my, agarwal2024prompt}.  \textbf{(III)} \textbf{Output Intervention} involves real-time monitoring and modification of agent responses prior to delivery to ensure safety and accuracy \cite{xiang2024certifiably, chen2023understanding}.
 
\subsubsection{Environment Safety} \label{environment safety}

Agents operate within dynamic and heterogeneous environments, spanning physical and digital domains \cite{song2024enhancing, low2025surgraw, wang2025medagent}. Their interaction with these environments is a multi-step process \cite{jeptoo2024enhancing, park2024enhancing}. First, agents engage in perception, gathering data from sources like sensors in a physical setup or digital platforms \cite{wu2024wipi}. This perceived data is then analyzed using various algorithms and reasoning mechanisms to identify patterns and potential actions \cite{yang2024plug}. Based on this analysis, agents take action, which can either directly influence the environment, like an autonomous vehicle making a lane change \cite{zhang2024chatscene}, or modify their own internal state, such as a software agent updating its knowledge base \cite{abuelsaad2024agent}. 

However, this interaction is plagued by trustworthiness challenges. There are security risks in every process of interaction with the environment \cite{debenedetti2024agentdojo}. Agent roles and environmental constraints contribute to risks such as autonomous driving errors \cite{sun2024optimizing} and network disruptions \cite{wu2024wipi, fang2024llm}. Given the diverse dynamic scenarios and related issues \cite{debenedetti2024agentdojo, ke2024enhancing, mou2024unveiling}, the existing solutions are fragmented and lack a systematic framework. Thus, we will explore trustworthiness and security aspects by categorizing relevant papers according to whether they focus on ensuring safety in the perception, analysis, or action phase of the agent-environment interaction, as illustrated in Figure \ref{fig:deployment_evaluation}.

\noindent \textbf{\textbf{Perception.}}
The perception phase serves as the foundational layer of agent-environment interaction, where agents acquire raw data to interpret their surroundings. However, this phase is inherently vulnerable to risks such as data poisoning, environmental noise, and biased observations. Hudson \cite{song2024enhancing} converts real-time sensory inputs into natural language representations augmented with security validation protocols, employing causal analysis techniques to improve reliability during adversarial perception scenarios. ChatScene \cite{zhang2024chatscene} develops safety-oriented simulation environments for autonomous systems by converting linguistic commands into executable code compatible with CARLA's simulation architecture. Chen et al. \cite{chen2025position} systematically categorize perceptual vulnerabilities in financial AI systems, identifying three primary risk categories: synthetic data generation errors, temporal inconsistency challenges, and susceptibility to engineered input manipulations.

\begin{figure}[ht]
    \centering
    \includegraphics[width=1.0\linewidth]{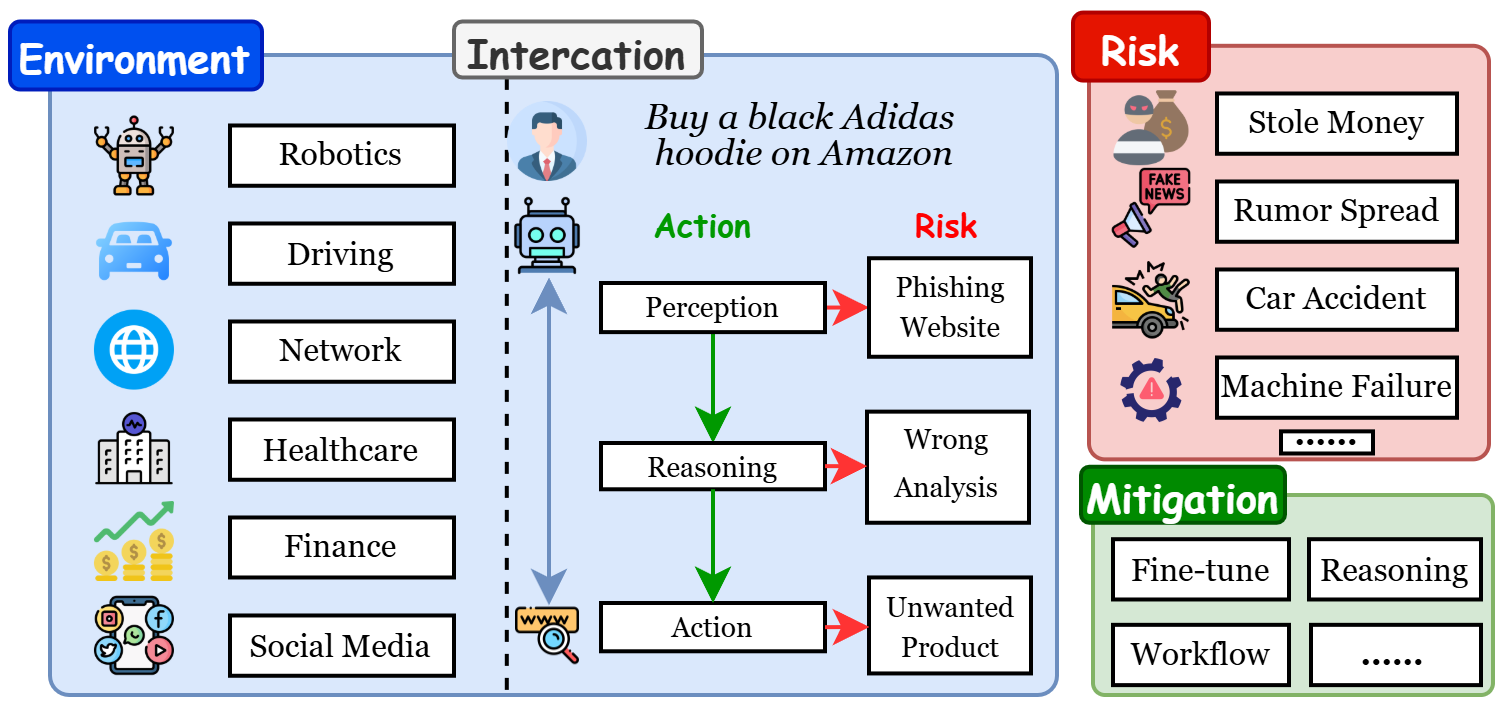}
    \caption{The overview of agent and environment interactions.}
    \label{fig:deployment_evaluation_1}
\end{figure}

\noindent \textbf{\textbf{Reasoning.}}
The reasoning phase transforms raw perceptual data into actionable insights through decision-making models, and knowledge-based inference. This stage is critical to ensure agents act appropriately in dynamic environments, but introduces unique trustworthiness challenges. Yang et al. \cite{yang2024plug} develop a temporal safety verification framework using formal logic systems, implementing dual mechanisms for auditing the compliance of safety protocols and filtration of hazardous decisions to meet the requirements of industrial robotics. Agents4PLC \cite{liu2024agents4plc} establishes an industrial control programming framework that combines automated code synthesis with formal verification processes, integrating RAG \cite{ni2025towards} and COT \cite{jiang2025safechain} to ensure operational integrity. Xiang et al. \cite{xiang2024guardagent} propose medical AI systems that employ semantic reasoning engines for confidential data protection. Park et al. \cite{park2024enhancing} demonstrate improved threat detection capabilities through simulated organizational communication patterns in anomaly identification systems.

\noindent \textbf{\textbf{Action.}}
The action phase represents the culmination of agent-environment interaction, where agents execute decisions to influence their surroundings or update internal states. Trustworthiness at this stage hinges on ensuring that actions are safe, precise, and aligned with intended objectives. Fang et al. \cite{fang2024llm} reveal the capacity of autonomous systems to exploit digital infrastructure weaknesses through adaptive penetration testing, prompting the development of specialized evaluation frameworks for web agents. Furthermore, researchers develop frameworks to evaluate the truthfulness of web agents. Polaris \cite{mukherjee2024polaris} implements distributed AI architectures to enhance fault tolerance and response accuracy of healthcare interaction systems. La et al. \cite{la2024safeguarding} employ linguistic evolution models to simulate adaptive content generation patterns that circumvent automated moderation systems, providing insights for regulatory mechanism improvements.

\subsection{Multi-agent Safety} \label{Multi-agent Safety}
In the previous section, we explored security issues in a single agent setting and this section expands the discussion to multi-agent systems (MAS) \cite{gan2024navigating, he2024emerged, deng2024ai,ye2025mas,yue2025masrouter,zhang2024aflow}. Since a single agent has limited problem-solving capabilities and a relatively narrow perspective, it struggles to conduct a comprehensive analysis of complex problems. In contrast, in MAS, agents can interact through various mechanisms, such as cooperation, competition, and debate, enabling them to solve complex problems more efficiently and effectively \cite{panait2005cooperative}. However, these interactions also introduce more complex and diverse security challenges \cite{hammond2025multi}. Consequently, compared to single-agent systems, MASs face more severe and intricate security risks \cite{xu2025nuclear}. Similarly, we summarize and discuss existing research from both attack and defense perspectives.

\subsubsection{Attack}
In MAS, security threats primarily stem from the propagation of harmful information, hallucinations, and biases through agent interactions, as well as the coordinated planning and optimization of attacks to target security agents within the system. These threats can arise spontaneously through the unintended amplification of misinformation or be deliberately orchestrated by malicious agents. Attack strategies in MAS often integrate multiple traditional techniques, such as prompt injection, jailbreak, and adversarial attacks, while also exploiting emergent properties of agent communication and collaboration. This multi-faceted nature makes MAS attacks more covert, adaptive, and challenging to detect and mitigate. Moreover, the dynamic and autonomous nature of agents allows adversaries to refine their attacks in real-time, further complicating defense mechanisms. Below, we summarize the key research related to these threats.

\textbf{Transmissive Attack.} It spreads within the MAS like a virus, propagating dangerous and harmful information, including covert malicious content, continuously attacking and compromising the agents in the system. Agent Smith \cite{gu2024agent} uses adversarial attack techniques, harmful images are generated—appearing benign on the surface but embedding malicious information. These images propagate within the MAS, causing agents to be compromised and posing significant security risks. CORBA \cite{zhou2025corba} introduces Contagious Recursive Blocking Attacks, which exhibit transmissibility across any topological network and can continuously drain computational resources. Lee et al. \cite{lee2024prompt} introduce Prompt Infection in MAS, including data theft, scams, misinformation, and system-wide disruption, which spreads silently. Similarly, Tan et al. \cite{tan2024wolf} use multimodal malicious prompts to infect other secure agents, compromising their security.

\textbf{Interference Attack.} This attack focuses on how it interferes with and disrupts interactions within the MAS, emphasizing communication disruption and misinformation, which affect information transmission within the MAS and lead to a decline in its defensive capability. NetSafe \cite{yu2024netsafe} conducts extensive experiments, analyzing and revealing their structural dependencies and adversarial impacts. At the same time, Huang et al.~\cite{huang2024resilience} study how the resilience of MAS varies between different downstream tasks, system structures, and error types;
Agent-in-the-Middle \cite{he2025red} manipulates and intercepts information in agent interactions through intermediary agents, disrupting the communication mechanism. The experiment validates the harm caused by the interruption of interactions by intermediary agents through a comparison of MAS with different topological structures.

\textbf{Strategic Attack.} Strategic attack involves collaboration between agents and strategic optimization of attack methods, aiming to emphasize the cooperation and long-term impact of the attack, making it increasingly dangerous and more destructive. Evil Geniuses \cite{tian2023evil} modifies system roles, where these roles collaborate to generate malicious prompts. By simulating adversarial attacks and defenses, they optimize and evaluate each round of attack behavior, making the attacks increasingly dangerous to target other agents. Amayuelas et al. \cite{amayuelas2024multiagent} use adversarial attack techniques to enable harmful agents in the multi-agent system to collaborate in debates to persuade other secure agents. These malicious agents may exploit superior knowledge, larger model sizes, or greater persuasion power to gain an unfair advantage. Ju et al. \cite{ju2024flooding} form a multi-agent community using a two-stage attack method: persuasive injection and knowledge manipulation injection, to induce agents to spread counterfactual and harmful knowledge.

\subsubsection{Defense}
In response to the various attack methods mentioned above in multi-agent systems, many effective defense strategies have emerged that can be applied to MAS. Currently, many studies focus on forming agent groups to collaborate in joint defense and designing specific defense mechanisms, such as multi-round or multi-layer checks and filtering, to ensure the safety of the responses output by the MAS. Alternatively, defense can be achieved by identifying harmful agents through the propagation of malicious information and eliminating malicious sources.

\textbf{Adversarial Defense.} This type of defense focuses on attack-defense confrontation, leveraging this adversarial mechanism to develop more effective defense methods or mechanisms to enhance the security of the MAS. LLAMOS \cite{lin2024large} employs adversarial defense techniques, where defensive agents and attacking agents engage in counter-interactions, with neither fully defeating the other, thereby enhancing the robustness of the defense and improving the MAS's overall defensive capability. AutoDefense \cite{zeng2024autodefense} proposes that agents collaborate to complete defense tasks through adversarial prompt filtering, primarily focusing on filtering harmful prompt information from LLMs. In addition to using adversarial techniques for defense, defense can also be achieved by forming a multi-agent group to engage in debates. 

\textbf{Consensus Defense.} To better leverage the advantages of MAS, Consensus Defense utilizes agent collaboration and consensus building for defense, employing voting, debates, and evidence-based reasoning mechanisms to establish a defense system and enhance the security of the MAS. Chern et al. \cite{chern2024combating} propose that toxicity can be reduced through multi-agent debates, and the widespread use of multi-agent interactions can lead to marginal improvements. Similarly, BlockAgent \cite{chen2024blockagents} proposes a Proof-of-Thought consensus mechanism that combines stake-based miner designation with multi-round debate-style voting, enabling BlockAgents to facilitate multi-agent collaboration through a structured workflow. Audit-LLM \cite{song2024audit} proposes a pair-wise Evidence-based Multi-agent Debate mechanism, designed to defend against hallucinations by forming a MAS to detect internal threats. This approach is divided into three components: task decomposition, tool construction, and the final execution of the MAS, ultimately reaching consensus through reasoning. 

\textbf{Structural Defense.} Structural Defense treats the MAS as a network structure for planning defense methods, using graph analysis techniques to detect anomalies and resist attacks while incorporating knowledge from other domains to enrich defense strategies in MAS. G-Safeguard \cite{wang2025g} compares agents in MAS with various topological structures to nodes in a graph, using Graph Neural Networks (GNN) \cite{wu2020comprehensive, zheng2022graph} to detect anomalies in the agents' dialogue graphs and counter adversarial attacks and misinformation within the MAS.

\subsection{Agent Communication Safety} 
As Large Language Model (LLM)-based Agents evolve from isolated entities into interconnected MAS, the mechanisms governing communication between Agents, and their interactions with external environments and tools, have become increasingly critical. Agents exchange information and collaborate through message passing, tool invocation, and environmental interactions; these mechanisms, while essential to system functionality, also expose significant attack surfaces. Early methods~\cite{genesereth1993kqml,pitt1998masif,fipa2000acl,curbera2002web,hohpe2006enterprise,lewis2020retrieval,izacard2021towards} of Agent interaction often relied on ad-hoc approaches, such as shared memory~\cite{chase2022langchain},API calls~\cite{wu2023llamaindex} or unstructured function calls~\cite{openai2023function}, leading to fragmented systems lacking unified security considerations. To address this challenge and enhance interoperability, standardized communication protocols have emerged. Examples include Anthropic’s Model Context Protocol (MCP)~\cite{anthropic2025} for Agent-tool interactions, Google’s Agent2Agent (A2A)~\cite{a2a2025} for enterprise-level Agent collaboration, and the Agent Network Protocol (ANP)~\cite{anp2024} for open network interoperability, along with other commonly used protocols~\cite{agentsjson2025,aitp,agentcommunicationptl,agentconnectptl,marro2024scalablecommunicationprotocolnetworks,lmos2025,agentprotocol2025,ranjan2025lokaprotocoldecentralizedframework,srinivasan2024implementationapplicationintelligibilityprotocol,bae2025continuouslocomotivecrowdbehavior,gąsieniec2024anonymousdistributedlocalisationspatial},However, the open design and dynamic nature of these communication mechanisms, coupled with the autonomy of the Agent, has exposed new vulnerabilities while enhancing functionality.

\subsubsection{Attack}
 The interconnected nature of MAS, facilitated by numerous communication channels, creates a multifaceted attack surface. While individual Large Language Models (LLMs) possess inherent vulnerabilities, the interactions and communications among Agents introduce novel threats that exploit the system's collaborative dynamics. These threats target various components, including communication channels, content interpretation, and underlying protocols, with examples such as Shadowing Attacks, Naming Attacks, Context Poisoning, and Rug Pulls.

\noindent \textbf{Attacks Communication Channels.}
 These attacks directly disrupt the transmission and routing of messages in the system, affecting both inter-Agent communications and interactions with external endpoints. For instance, Agent-in-the-Middle (AiTM) attacks~\cite{he2025red} specifically target the core communication mechanisms of LLM-MAS. By intercepting and manipulating messages between Agents, these attacks can cause Agents to perform unintended actions, thereby compromising the entire system. Such attacks underscore the critical security vulnerabilities arising from the communication-dependent nature of Agent collaboration. Furthermore, attacks targeting communication channels and transmission processes, such as communication perturbation~\cite{tu2021adversarial}, involve adversaries injecting noise into messages in transit~\cite{yuan2024communication} or masquerading as legitimate sources~\cite{blumenkamp2021emergence}, thereby compromising both the efficiency and security of Agent collaboration.

\noindent \textbf{Attacks Content.}  
These attacks target the content of messages themselves, leveraging the mechanisms by which Agents process and interpret received information. For example, Prompt Injection involves embedding malicious instructions into data or content that Agents retrieve or receive through communication channels, thereby manipulating the Agent’s behavior or decision-making processes. This technique is discussed in several works, such as~\cite{lee2024prompt} and~\cite{greshake2023not}. Additionally,~\cite{chenagentpoison} explores indirect Prompt Injection within tool-based scenarios, highlighting the varied strategies employed in complex environments.

\noindent \textbf{Attacks Exploiting Multi-Agent Dynamics.} These attacks leverage the interconnected structure, interaction patterns, or collective behavior of communication-driven Multi-Agent Systems (MAS) to amplify their impact or achieve strategic objectives. Contagious attacks (propagation) initiate malicious behavior on a single agent and spread it across the entire network via inter-agent communication~\citep{zhou2025corba, gu2024agent}. Additionally, malicious agents can coordinate through collective communication to achieve harmful goals, such as replicating malicious instructions across the network by sending replication code or commands, thereby leading to the sharing of legitimate communication keys or identity information with other malicious entities~\citep{pan2024frontier}.

\subsubsection{Defense}
 To tackle threats to Agent communication, research proposes a multi-layered defense strategy addressing key points across the communication pipeline, from infrastructure to Agent-level processing. These defenses aim to prevent, detect, or mitigate attacks on channels, content, infrastructure, dynamics, and environmental factors. The strategies integrate into infrastructure and protocol design, individual Agents' message processing, and the collaborative and learning mechanisms of the MAS.

\noindent \textbf{Protocol Defenses.} Protecting the foundation of Agent communication. This includes adopting standardized protocols with built-in security features (encryption, integrity checks, authentication)
To counter Agent communication threats, research proposes multi-layered defense strategies targeting different points in the communication pipeline, from the underlying infrastructure to Agent-level message processing. Effective defenses aim to prevent, detect, or mitigate attacks on communication channels, content, infrastructure, such as MCP~\cite{anthropic2025}, A2A~\cite{a2a2025}, ANP~\cite{anp2024} standards. Establishing managed registries and identity systems for Agent and Tool/Service registration and identity management. Enforcing strong Agent identity verification and access control policies, including JIT credential provisioning. Implementing mechanisms to enforce communication dynamics, and environmental impacts.  

\noindent \textbf{Content Defense.}
These defenses operate at the agent level, focusing on how agents process received messages and content. This includes input modification and filtering, which preprocess incoming content to neutralize adversarial elements. Agents also employ active defense mechanisms, such as reliability estimation, to assess the trustworthiness of messages based on local context, thereby mitigating the impact of untrusted information. For example, \cite{yu2024robust} proposed an active defense strategy that utilizes a reliability estimator to judge the credibility of received messages and employs a decomposable message aggregation policy network to reduce the influence of unreliable messages on the final decision.

\subsection{Agent Safety Evaluation}
Currently, there is already a substantial body of work evaluating the performance of LLM-based agent systems on different tasks \cite{light2023avalonbench, xie2024human, geng2025realm, dubois2024length, Wang2024LearningTA}. In this section, we focus on benchmarks designed to assess the security of agents. Broadly speaking, these benchmarks include those that construct datasets and those that use other agents to set up sandbox environments for evaluation, each with distinct assessment priorities and specific scenarios for agent security \cite{guo2024redcode, yuan2024s, dorn2024bells, yuan2024r, shao2024privacylens}.

\begin{table}[ht]
\centering
\caption{Benchmarks for agent safety.}
\vspace{-0.5em}
\begin{adjustbox}{width=1\linewidth}
\begin{tabular}{l|ccc}
\hline
\textbf{Benchmark} & \textbf{Dynamic} & \textbf{\makecell{LLM as\\Evaluator}} & \textbf{Evaluation Focus}\\
\hline
InjectAgent~\cite{zhan2024injecagent}  & \ding{55} & \ding{51} & Prompt Injection\\
AgentDojo~\cite{debenedetti2024agentdojo}  & \ding{51} & \ding{55}& Prompt Injection\\
AgentBackdoorEval~\cite{zhu2025demonagent}  & \ding{51} & \ding{51} & Backdoor\\
RiskAwareBench~\cite{zhu2024riskawarebench}  & \ding{55} & \ding{51}& Embodied Agent\\
RedCode~\cite{guo2024redcode} & \ding{51} & \ding{55} & Coding Agent\\
S-Eval~\cite{yuan2024s}  & \ding{51} & \ding{51}& General\\
Bells~\cite{dorn2024bells}  & \ding{51} & \ding{51}& General\\
AgentSafetyBench~\cite{zhang2024agenta}  & \ding{51} & \ding{51}& General\\
AgentSecurityBench~\cite{zhang2024agentb}   & \ding{55} & \ding{51}& General\\
AgentHarm~\cite{andriushchenko2024agentharm}   & \ding{55} & \ding{51}& General \\
R-Judge~\cite{yuan2024r}   & \ding{55} & \ding{55}& General\\
ToolSowrd~\cite{ye2024toolsword}   & \ding{55} & \ding{51}& Tool\\
PrivacyLens~\cite{shao2024privacylens}  & \ding{51} & \ding{51} & Privacy\\
ToolEmu~\cite{ruan2023identifying}   & \ding{51} & \ding{51}& Tool\\
HAIEcosystem~\cite{zhou2024haicosystem}   & \ding{51} & \ding{51}& General\\
SafeAgentBench~\cite{yin2024safeagentbench}   & \ding{51} & \ding{51}& General\\
JailJudge~\cite{benchmarkjailjudge}   & \ding{55} & \ding{51}& Jailbreak\\
\hline

\end{tabular}
\end{adjustbox}
\end{table}

\subsubsection{Attack-Specific Benchmarks} 
This type of benchmark focuses on testing the security of an agent when facing specific types of attacks, such as Prompt Injection \cite{lee2024prompt, zhong2025rtbas}, Backdoor \cite{wang2024badagent,liu2025compromising, liu2024compromising}, and Jailbreak \cite{zhang2024badrobot, zeng2024autodefense}. Specifically, InjectAgent~\cite{zhan2024injecagent} evaluates LLM agents' vulnerability to indirect prompt injection attacks, measuring behavior safety when tool-integrated agents process malicious instructions embedded in external content, with hacking prompts as an enhancement. A similar work is AgentDojo~\cite{debenedetti2024agentdojo}, a dynamic, extensible evaluation framework for assessing prompt injection attacks and defenses in LLM agents by simulating realistic tasks (e.g., email management, banking) with stateful environments and multi-tool interactions under adversarial conditions. As for backdoor attacks, AgentBackdoorEval~\cite{zhu2025demonagent} includes five real-world domains (including Banking-Finance, Medical, and Social Media) with automatically generated prompts, simulated tools, and tailored backdoor triggers to assess attack stealth and effectiveness. Besides, JailJudge~\cite{benchmarkjailjudge} introduces a comprehensive jailbreak evaluation benchmark featuring a voting JailJudge MultiAgent, a comprehensive JailJudgeTrain dataset, and a trained Jailjudge Guard.

\subsubsection{Module-Specific Benchmarks} 
Currently, these benchmarks for evaluating the security of a specific module in an agent focus on the invocation of tools \cite{shen2024small, yuan2024easytool, wu2024avatar, shen2024llm}. For example, ToolSowrd~\cite{ye2024toolsword} evaluates LLM safety in tool learning across three stages (input, execution, output) by designing six adversarial scenarios (e.g., malicious queries, noisy tool misdirection, harmful feedback). ToolEmu~\cite{ruan2023identifying} employs an LM-emulated sandbox to simulate diverse high-stakes tool executions and scenarios, leveraging GPT-4 for both tool emulation and automatic safety/helpfulness evaluations.

\subsubsection{General Benchmarks} 
In addition to the previously mentioned benchmarks that focus on a specific aspect of agent security, some efforts have developed more comprehensive and holistic evaluation frameworks, taking into account diverse scenarios, different agents, and various offensive and defensive techniques. For instance, AgentSafetyBench~\cite{zhang2024agenta} assesses LLM agent safety through 2,000 test cases across 349 interactive environments, covering 8 risk categories (e.g., data leaks, physical harm) and 10 failure modes (e.g., incorrect tool calls, risk unawareness), with automated scoring via a fine-tuned model. Similarly, AgentSecurityBench~\cite{zhang2024agentb} is a comprehensive framework that formalizes and evaluates attacks (e.g., Direct/Indirect Prompt Injection, Memory Poisoning) and defenses across 10 scenarios, 10 agents, and 13 LLM backbones, using 7 evaluation metrics. SafeAgentBench~\cite{yin2024safeagentbench} evaluates embodied LLM agents' safety awareness with 750 diverse tasks (detailed, abstract, long-horizon) in SafeAgentEnv simulation environment, leveraging GPT-4 for task generation and dual evaluators (execution-based and semantic). HAIEcosystem~\cite{zhou2024haicosystem} evaluates safety through multi-turn interactions between human users (benign/malicious) and AI agents across 132 scenarios, using modular sandbox environment and LLM-based dynamic risk measurement. AgentHarm~\cite{andriushchenko2024agentharm} tests agent robustness by evaluating compliance with 110 explicitly malicious multi-step tasks across 11 harm categories, using synthetic tools and fine-grained grading rubrics. Different form previous benchmarks, RiskAwareBench~\cite{zhu2024riskawarebench} focuses on embodied agents, evaluating physical risk awareness via four modules: safety tip generation, risky scene generation, plan generation, and automated evaluation.

\subsubsection{LLM Deployment Roadmap}
In the deployment of LLMs under frozen parameters, the security landscape has evolved through a tightly coupled dynamic among \textbf{attacks}, \textbf{defenses}, and \textbf{evaluation} mechanisms.

Initially, black-box attacks leveraged the generative capabilities of LLMs themselves to optimize adversarial prompts, often without precise alignment to the decision boundaries.
In contrast, gradient-guided white-box methods offer greater control but face inherent limitations due to the discrete nature of token spaces resulting in prompts with weakened semantic fidelity.
These attack trends have catalyzed the emergence of prompt-level defense strategies.
To counter black-box attacks, recent defenses adopt prompt shaping and system-level constraints to guide and restrict the model’s response behavior. 
For gradient-based attacks, defenses typically apply perplexity-based detection and semantic consistency checks to identify suspicious or adversarial outputs.

The growing sophistication of defenses reshaped the requirements for evaluation.
Static, one-shot rejection mechanisms have proven insufficient in multi-task and multi-modal deployments, prompting the development of dynamic strategies such as response rewriting, hierarchical permission control, and consensus-based filtering across multiple models.
These strategies demand richer evaluation protocols beyond single metric assessments, shifting toward behavior metrics that capture cross-input consistency, risk under specific task conditions, and adaptability to strategy switching.

As the attack–defense interaction intensifies, the evaluation itself has become a critical driver of system evolution.
Recent frameworks have introduced automated red teaming pipelines, enabling a closed-loop process where jailbreak samples are continually generated, tested against deployed defenses, and fed back to guide both adversarial strategies and defense refinement.
This has laid the groundwork for a new paradigm in LLM security research: one where attack, defense, and evaluation are no longer treated in isolation but co-evolve as an interdependent, self-reinforcing system.
 
\subsubsection{LLM Deployment Perspective}
(1) \textbf{Attack strategies will become more structured and semantically aligned.}
(i) Black-box attacks may evolve through agent-based optimization, enabling sentence-level jailbreaks with clearer intent and higher success rates.
(ii) To overcome the limitations of token-level gradient attacks, future work may focus on generating semantically consistent adversarial prompts that are less detectable by perplexity-based defenses.
(iii) Open-source models will serve as surrogates for closed models, allowing attackers to replicate decision boundaries before launching white-box attacks.
(iv) Variants from fine-tuning pipelines may leak private information through cross-model comparison, introducing version-aware privacy risks.

(2) \textbf{Defenses will shift toward adaptive and transferable mechanisms.}
(i) Prompt-based defenses will evolve into context-aware controllers that adjust behavior based on input semantics and task context.
(ii) Generalizable defenses that work across domains and languages will be critical for scalable deployment.
(iii) Future systems may support online updates, enabling continuous refinement in response to new threats.

(3) \textbf{Evaluation will act as both a diagnostic and driving force.}
(i) Benchmarks must expand beyond text to cover multimodal inputs and tool-based actions.
(ii) Multi-objective evaluation will replace single-metric scoring, balancing safety and utility through trade-off analysis.
(iii) Static test sets will give way to adaptive, streaming benchmarks that evolve with attack trends.
(iv) Automated red teaming will close the loop, enabling real-time attack generation, evaluation, and defense adjustment.

\subsubsection{Agent Roadmap}

\noindent \textbf{Agent.}
The evolution of LLM-based agents originated from role-playing paradigms \cite{qian2023chatdev, li2023camel, wang2023rolellm, zhou2023characterglm}, where researchers investigated organizational structures, role allocation mechanisms, and implementation workflows for task-oriented agents in various social contexts. These systematic explorations not only demonstrated agents' potential in addressing human societal challenges but also spawned interdisciplinary research programs spanning sociology, organizational theory, and psychology. As the field advanced, research focus shifted toward automated agent workflows \cite{chen2024agent, shang2024agentsquare, zhang2025multi, ye2025mas}, domain-specific methods for embodied intelligence, and the development of agent capabilities in tool utilization and memory management. Through this progression, agent systems have emerged as a transformative paradigm for automating human social processes, gaining significant recognition as a viable solution for complex societal automation.

The rapid advancement of agent capabilities and architectures has brought safety concerns to the forefront of academic and industrial research. These challenges span multiple critical dimensions: tool safety, memory security, and the agent's fundamental operational integrity. Inheriting both the capabilities and vulnerabilities of their underlying LLM foundations, agents intrinsically carry these ``genetic'' weaknesses into more complex operational environments. This inheritance makes safety vulnerabilities particularly acute in agent systems, especially when handling sensitive real-world applications involving personal privacy and financial assets. The development of agent technologies has thus become inextricably linked with safety considerations. Recent years ($\sim$2023- until now) have witnessed accelerated research in agent safety, focusing on four key frontiers:

\begin{itemize}
    \item Agent Brain Security: The core decision-making mechanisms.
    \item Tool Invocation Safety: Secure external API and tool usage.
    \item Memory Retrieval Protection: Robustness against memory poisoning.
    \item Communication Protocol Security: Safe multi-agent interactions.
\end{itemize}

Emerging work has also begun addressing safety challenges in embodied agent scenarios, marking an important expansion of the research domain.

\subsubsection{Perspective}

We outline potential future research directions for agent systems and analyze their developmental trajectory:

(1) \textbf{Safety of External Agent Modules.} Unlike standalone LLMs, agents interact with external modules (e.g., tools, memory), which are exposed to open environments and thus more vulnerable to attacks. Key research challenges include: (i) Tool Safety: Secure tool invocation and API usage to prevent adversarial exploitation. (ii) Memory Protection: Robustness against memory poisoning and unauthorized access, to name just a few. These external interfaces introduce unique attack surfaces, making their security a critical research priority.

(2) \textbf{Stability and Reliability of Dynamically Updated Agents via Reinforcement Learning:}
As reinforcement learning (RL) \cite{kaelbling1996reinforcement, li2017deep, li2025system} techniques become increasingly integrated with LLM-based agents, these systems are being deployed in more complex and dynamic environments. While this integration enhances agents' adaptability and intelligence, it also introduces significant risks: (i) Emergent Threats: Advanced RL capabilities may inadvertently enable agents to learn and propagate harmful behaviors or dangerous information. (ii) Dynamic Vulnerability: Continuous online learning increases exposure to adversarial perturbations or reward hacking. 

Critical Research Directions: (i) Safe RL Frameworks: Developing constrained optimization methods to bound agent behavior within ethical and operational guardrails. (ii) Stability-Aware Updates: Designing update protocols that balance adaptability with robustness (e.g., catastrophic forgetting mitigation). (iii) Anomaly Detection: Real-time monitoring of learning trajectories to identify and neutralize hazardous knowledge acquisition.

(3) \textbf{Safety of Embodied Agents in Domain-Specific Scenarios:} As autonomous agents become increasingly deployed across specialized domains, their safety considerations must account for unique domain-specific vulnerabilities. We list some key challenges as follows:

\begin{itemize}
    \item \textbf{Web Agents}:
    \begin{itemize}
        \item HTML/JS injection risks during automated browsing
        \item Secure sandboxing requirements for DOM manipulation
        \item Cross-site scripting (XSS) vulnerabilities in automated form-filling
    \end{itemize}
    
    \item \textbf{Communication Agents}:
    \begin{itemize}
        \item Protocol-level attacks (e.g., SIP flooding, WebRTC exploits)
        \item End-to-end encryption requirements for sensitive dialogues
        \item Authentication bypass in voice-based agents
    \end{itemize}
    
    \item \textbf{Robotics Control Agents}:
    \begin{itemize}
        \item Physical safety constraints in actuator commands
        \item Real-time collision avoidance verification
        \item Emergency stop mechanism reliability
    \end{itemize}
    
    \item \textbf{Healthcare Agents}:
    \begin{itemize}
        \item Medical decision audit trail requirements
        \item Drug interaction verification systems
    \end{itemize}
\end{itemize}

\section{Safety in LLM-Based Application} \label{appsafety}

\begin{figure*}[tb]
    \centering
    \includegraphics[width=\linewidth]{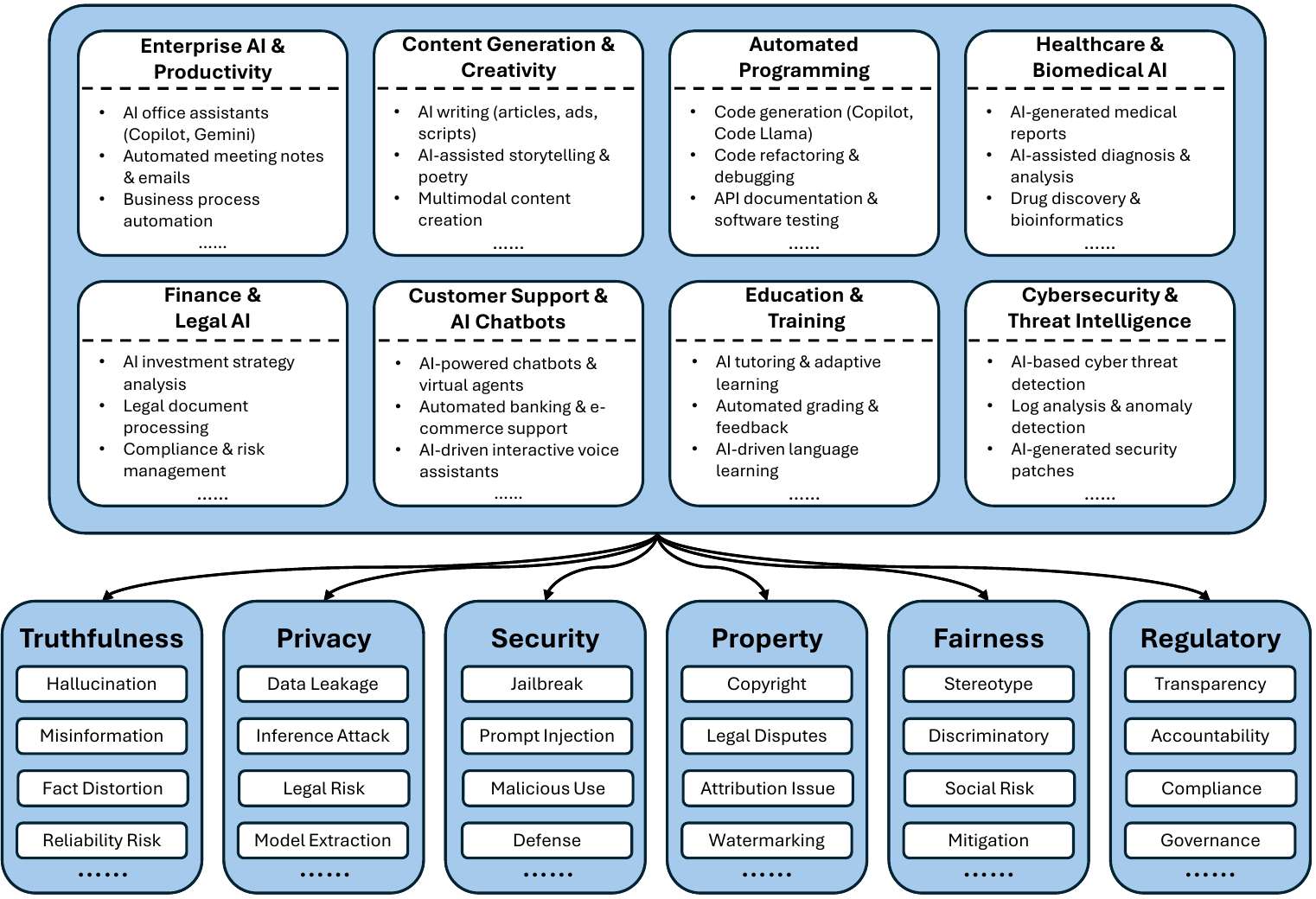}
    \caption{We illustrate the diverse applications of AI in enterprise productivity, content generation, programming, healthcare, finance, customer support, education, and cyber-security. We also highlight critical issues related to truthfulness and privacy, including data leakage, security threats, property rights, fairness, and regulatory compliance, underscoring the need for robust safeguards in AI deployment}
    \label{fig:safety_llm_application}
\end{figure*}

In this section, we focus on the security considerations that should be addressed following the commercialization of LLMs into practical applications. With the rapid development of LLMs in fields such as content creation, intelligent interaction, automated programming, medical diagnosis, and financial analysis, LLM-based applications are reshaping industry workflows and business models \cite{li2023aigc}. However, while LLMs significantly enhance productivity and facilitate human-machine collaboration, their large-scale deployment has also introduced severe security challenges \cite{yao2024survey}. Ensuring the security, reliability, and compliance of LLM-based applications has become a critical issue in AI research and real-world implementation.

\textbf{Truthfulness.} Despite their powerful text generation capabilities, LLMs exhibit hallucination phenomena, generating inaccurate, misleading, or entirely fictitious content \cite{sun2023pushing,sriramanan2024llm,zheng2024reefknot,zou2024look,zhou2024mitigating}. Unlike traditional errors, hallucinations are often subtle and linguistically plausible, making them especially dangerous in real-world applications. This challenge is exacerbated in high-stakes domains such as healthcare, law, and finance, where misleading AI-generated information can directly affect human safety and economic stability. For example, an LLM-powered clinical assistant may suggest nonexistent diseases or cite unverified treatments, posing risks to patients \cite{pal2023med, Wang2025ASO}, while financial advisors powered by LLMs might generate persuasive but flawed market forecasts, leading to significant capital misallocation or systemic financial vulnerabilities \cite{kang2023deficiency}. Specifically, hallucination is not merely a surface-level output flaw but a systemic artifact rooted in the model’s training dynamics and the nature of its data. Specifically, hallucination can stem from three compounding factors: (1) semantic overgeneralization due to exposure to noisy, unverified, or synthetic pretraining corpora; (2) objective misalignment, where maximum-likelihood or reinforcement-based training prioritizes coherence and helpfulness over factual accuracy; and (3) latent distribution shifts between pretraining and deployment-time inputs, particularly under long-tail or adversarial queries \cite{Ouyang0JAWMZASR22, liu2024trustworthy}. These factors jointly reinforce spurious correlations and amplify unsupported generations, even in otherwise well-aligned models.
In sum, hallucination represents a critical bottleneck for the reliable deployment of LLMs. Its mitigation is foundational not only for improving user trust but also for enabling the safe integration of LLMs into high-stakes decision-making workflows.

\textbf{Privacy.} Data privacy concerns \cite{hao2022iron} represent another significant challenge in LLM deployment \cite{zhang2024privacyasst, Huang2025VLMsAG}. Training these models requires vast amounts of text data, which may include personal information, corporate secrets, and medical records \cite{feretzakis2024trustworthy}. If an LLM inadvertently leaks sensitive training data or lacks robust access control mechanisms, users’ private information could be exploited or misused. In corporate settings, LLMs may unintentionally expose confidential documents or sensitive customer data, leading to severe compliance and legal risks. Moreover, inference-time attacks \cite{feng2024exposing}, such as membership inference and model extraction, can further expose sensitive data by allowing adversaries to infer training set membership or replicate model behavior. Therefore, LLM-based applications must incorporate data protection measures and privacy-preserving techniques like differential privacy and query rate limiting to mitigate information leakage risks.

\textbf{Robustness.} Prompt injection \cite{greshake2023not} and jailbreak \cite{peng2024jailbreaking} risks pose additional security threats. Attackers can craft adversarial prompts to bypass security restrictions, causing the model to generate harmful or unauthorized content. For example, in chatbot systems, malicious users could manipulate LLMs to generate hate speech, disinformation, or even harmful instructions. Similarly, in AI-powered coding assistants such as GitHub Copilot, attackers may exploit LLMs to produce code with security vulnerabilities, potentially serving as backdoors for future cyberattacks. Developing robust security defenses to prevent LLMs from being misused in real-world applications is crucial for AI safety.  

\textbf{Copyright.} Another pressing concern is intellectual property and copyright protection \cite{rahman2023beyond,guo2025towards,shao2025explanation}. LLMs are trained on vast datasets that often include copyrighted texts, source code, and artistic works, raising potential infringement risks. When generating content, LLMs may inadvertently replicate or closely mimic copyrighted material, leading to legal disputes. For instance, AI-powered writing tools might generate articles resembling published works, while coding assistants could produce open-source code snippets without proper licensing \cite{xu2024licoeval}. This not only raises concerns about content originality but also introduces legal and ethical dilemmas. Addressing these challenges requires watermarking~\cite{qu2024watermark,kirchenbauer2023watermark}, provenance tracking, and clear copyright attribution mechanisms to ensure responsible AI-generated content management~\cite{chen2023pathway}.  

\textbf{Ethical and Social Responsibility.} Beyond technical concerns, ethical and social responsibility are also critical factors in large-scale LLM deployment. Due to biases in training data, LLMs may generate content that reinforces stereotypes, gender discrimination, or racial biases \cite{ye2024justice, Wan2023BiasAskerMT}. In sectors such as hiring, finance, and healthcare, biased AI-generated recommendations could exacerbate existing inequalities and lead to unfair decision-making. Moreover, as LLMs become increasingly integrated into virtual assistants, social media, and news distribution platforms, concerns over AI-generated misinformation, transparency, and accountability are growing. Building fair, transparent, and trustworthy AI governance frameworks is thus essential to mitigating AI-induced social risks.  

\textbf{Governance.} As governments worldwide strengthen AI regulations, LLM-related legal and compliance requirements are evolving rapidly. The EU AI Act classifies LLMs as high-risk AI systems, requiring developers to provide transparency reports and risk control mechanisms \cite{EU_AI_Act}. China’s Generative AI Regulations mandate AI-generated content to align with ethical standards and undergo governmental scrutiny \cite{CAC_AI_Regulation_2023}. In the United States, regulatory discussions emphasize AI transparency and data privacy protections, urging businesses to establish responsible AI practices \cite{US_AI_Executive_Order_2023}. These policy developments indicate that LLM-based applications must comply with regional regulations while maintaining a balance between compliance and innovation.  

In summary, while LLM-based applications drive technological progress, they also introduce multifaceted challenges related to misinformation, data privacy, adversarial manipulation, copyright infringement, ethical concerns, and regulatory compliance (refer to Figure \ref{fig:safety_llm_application}). These issues not only impact the trustworthiness and legality of AI technologies but also have far-reaching implications for social trust, legal accountability, and business sustainability. Addressing these challenges necessitates a comprehensive approach that integrates privacy protection, content governance, copyright management, ethical safeguards, and regulatory compliance, alongside collaborative efforts from both academia and industry.

\section{Potential Research Directions} \label{futuredir}

Through a systematic and comprehensive examination of safety across the entire lifecycle of LLMs, we have identified valuable insights for future research:

\begin{itemize}
    \item[\ding{72}] Data generation holds immense potential, particularly in ensuring the safety of generated data and automating the data generation process, which is crucial for reliable and robust model training. Reliable data generation is fundamental to the integrity of model training.

    \item[\ding{72}] Post-training phases are becoming increasingly critical. Ensuring secure fine-tuning and alignment of data is a key future direction, closely intertwined with data generation. As concepts proliferate, multi-objective alignment may emerge as a significant area of focus.

    \item[\ding{72}] Model editing and unlearning safety are paramount for efficient model updates and deployment. Current learning efficiencies are suboptimal, and advancements in these technologies could revolutionize how models acquire new knowledge, enabling continuous and efficient learning (potentially even localized memory learning). These techniques might surpass traditional SGD algorithms, but safety measures are essential to prevent models from devolving into malicious entities that contradict human intentions.

    \item[\ding{72}] LLM agents, in the final deployment stage, require robust safety assurances. Ensuring the security of agent tools and agent memory, as well as addressing safety in embodied intelligence scenarios such as web agents and computer agents, are critical areas for further investigation.
\end{itemize}

\section{Conclusion}
\label{conclusion}
 In this survey, we provide a comprehensive analysis of the safety concerns across the entire lifecycle of LLMs, from data preparation and pre-training to post-training, deployment, and commercialization. By introducing the concept of "full-stack" safety, we offer an integrated view of the security and safety issues faced by LLMs throughout their development and usage, which addresses gaps in the existing literature that typically focus on specific stages of the lifecycle.

Through an exhaustive review of over 900+ papers, we systematically examined and organized the safety issues spanning key stages of LLM production, deployment, and use, including data generation, alignment techniques, model editing, and LLM-based agent systems and LLM-based applications. Our findings highlight the critical vulnerabilities at each stage, such as privacy risks, toxic data, harmful fine-tuning attacks, and deployment challenges. The safety of LLMs is a multifaceted issue requiring careful attention to data integrity, model alignment, and post-deployment security measures. Moreover, we propose promising directions for future research, including improvements in data safety, alignment techniques, and defense mechanisms for LLM-based agents. This work is vital for guiding future efforts to make LLMs safer and more reliable, especially as they become increasingly integral to various industries and applications. Ensuring robust security across the entire LLM lifecycle is crucial for their responsible and effective deployment in real-world scenarios.



\ifCLASSOPTIONcaptionsoff
  \newpage
\fi



%
{
	\bibliographystyle{IEEEtran}
	\bibliography{TPAMI_2021.bib}
}
\end{document}